\documentclass[12pt,preprint]{aastex}

\shorttitle{LOW LUMINOSITY COMPANIONS TO WHITE DWARFS}
\shortauthors{J.  Farihi}

\begin{document}

\title{LOW LUMINOSITY COMPANIONS TO WHITE DWARFS}

\author{J. Farihi\altaffilmark{1,2},
	 E. E. Becklin\altaffilmark{2}, \& 
	 B. Zuckerman\altaffilmark{2}}

\altaffiltext{1}{Gemini Observatory,
			Northern Operations,
			670 North A'ohoku Place,
			Hilo, HI 96720}
\altaffiltext{2}{Department of Physics \& Astronomy,
			University of California,
			430 Portola Plaza,
			Los Angeles, CA 90095}


\email{jfarihi@gemini.edu,becklin@astro.ucla.edu,ben@astro.ucla.edu}

\begin{abstract}

This paper presents results of a near-infrared
imaging survey for low mass stellar and substellar 
companions to white dwarfs.  A wide field proper
motion survey of 261 white dwarfs was capable of
directly detecting companions at orbital separations
between $\sim100$ and 5000 AU with masses as low as
0.05 $M_{\odot}$, while a deep near field search of 
86 white dwarfs was capable of directly detecting
companions at separations between $\sim50$ and 1100 AU
with masses as low as 0.02 $M_{\odot}$.  Additionally, 
all white dwarf targets were examined for near-infrared
excess emission, a technique capable of detecting
companions at arbitrarily close separations down 
to masses of 0.05 $M_{\odot}$.

No brown dwarf candidates were detected, which
implies a brown dwarf companion fraction of $<0.5$\%
for white dwarfs.  In contrast, the stellar companion 
fraction of white dwarfs as measured by this survey is
22\%, uncorrected for bias.  Moreover, most of the known
and suspected stellar companions to white dwarfs are low
mass stars whose masses are only slightly greater than 
the masses of brown dwarfs.  Twenty previously unknown
stellar companions were detected, five of which are
confirmed or likely white dwarfs themselves, while
fifteen are confirmed or likely low mass stars.

Similar to the distribution of cool field dwarfs
as a function of spectral type, the number of cool 
unevolved dwarf companions peaks at mid-M type.  Based
on the present work, relative to this peak, field L dwarfs
appear to be roughly $2-3$ times more abundant than companion 
L dwarfs.  Additionally, there is no evidence that the initial
companion masses have been altered by post main sequence binary
interactions

\end{abstract}

\keywords{binaries: general ---
	stars: fundamental parameters ---
	stars: low-mass, brown dwarfs ---
	stars: luminosity function, mass function ---
	stars: formation ---
	stars: evolution ---
	white dwarfs}

\section{INTRODUCTION}

Until a decade ago and despite many searches, scientists
had no definite evidence of astrophysical objects with
masses between those of stars and planets.  Yet the missing
link between these two, dubbed brown dwarfs, are now a field
of study unto themselves.  There remains much to learn about
these failed stars, especially their astrophysical niches, 
origins and destinies.

Searching for brown dwarfs as companions to stars offers
the opportunity to search systems very near to the Sun and
requires less time than field and cluster searches covering
a relatively large portion of the sky.  The very first serious
brown dwarf candidate was discovered as a companion to the
white dwarf GD 165 \citep{bec88}.  GD 165B ($M\sim0.072$
$M_{\odot}$, $T_{\rm{eff}}=1900$ K) remained unique for
a number of years but eventually became the prototype for a
new spectral class of cool stars and brown dwarfs, the L
dwarfs \citep{kir99a,kir99b}.  The first unambiguous brown
dwarf was also discovered as a companion to a star,
Gl 229 \citep{nak95}.  Gl 229B ($M\sim0.040$ $M_{\odot}$,
$T_{\rm{eff}}=950$ K) became the prototype T dwarf 
\citep{mar96,kir99b}, the coolest known spectral class,
all of whose members are brown dwarfs.

Precision radial velocity techniques are sensitive to brown
dwarfs orbiting within $\sim5$ AU but have revealed very few.
\citet{but00} estimate the brown dwarf companion frequency
to be less than 0.5\%.  This figure includes the search of
several hundred stars and is thus a very good measure for
the innermost orbital separations.

Direct imaging searches have also produced a dearth of
brown dwarf companions to main sequence stars (relative to
stellar companions) at wider separations \citep{zuc87,zuc92,
hen90,nak94,opp01,sch00,hin02,mcc04,far04c}.  The separation
range corresponding to the peak in the stellar companion 
distribution for both G and M dwarf primaries is roughly $10-100$
AU \citep{duq91,fis92}.  This range and beyond has been searched
quite extensively by the aforementioned surveys and very few
brown dwarfs have been found.  The most optimistic estimate
for the brown dwarf companion frequency to main sequence stars
is a few percent \citep{low01}, although most studies conclude
that it is less than 1\%.  The only exception being the case 
when the primary star is a very low mass star or brown dwarf 
($M\la0.10$ $M_{\odot}$).  For these low mass primaries, the
binary fraction is estimated to be $\sim20\%$ \citep{rei01,
clo02,sie03,bur03}.  Thus it appears that the brown dwarf
companion frequency is a strong function of primary mass
\citep{far04c}.

This paper presents the details on 371 white dwarfs
which were searched for low luminosity companions using
near-infrared imaging arrays at several telescopes over
the past 18 years.  The particular techniques and analyses
used for each camera data are described along with follow
up observations.  The white dwarf sample is analyzed from
a kinematical perspective in order to make an overall age 
assessment -- critical for any calculation of completeness
as a function of secondary mass.  The measured distribution
of low mass companions is presented as a function of spectral
type, which is then transformed into mass using existing 
empirical and theoretical relations.  The implications for
binary star formation and evolution are discussed.

Within this supplement is a summary of all available
data on known companions to the 371 white dwarfs in the
sample.  Specifically included in this paper are new data 
and analysis of companions to white dwarfs previously
reported in \citet{zuc92} and \citet{sch96}.  Tables include
proper motions, $UVW$ space velocities, spectral types, and
distances for the white dwarf primaries.  Data on previously
known or newly discovered companions (including candidates)
includes optical and near-infrared photometry, low resolution
optical spectra, and proper motion measurements.

\section{THE SEARCH}

\subsection{The Steward Survey}

The core of this work was analysis of images
obtained at Steward Observatory.  Beginning in 1991 and
continuing through 2003, a program to image nearby white
dwarfs in the near-infrared was conducted on the Bok 2.3
meter telescope at Kitt Peak, Arizona.  Images of 273
targets were acquired using the Bok facility near-infrared
camera \citep{rie93}.  The imaging procedure was to acquire
$J$ band (1.2 $\mu$m) images in a 5-point dither pattern
with 90 seconds integration per dither position for a total
exposure time of 7.5 minutes.  Each raw science image was 
dark subtracted, flat fielded, scaled to the central image,
registered, shifted and then averaged.

\begin{figure}
\plotone{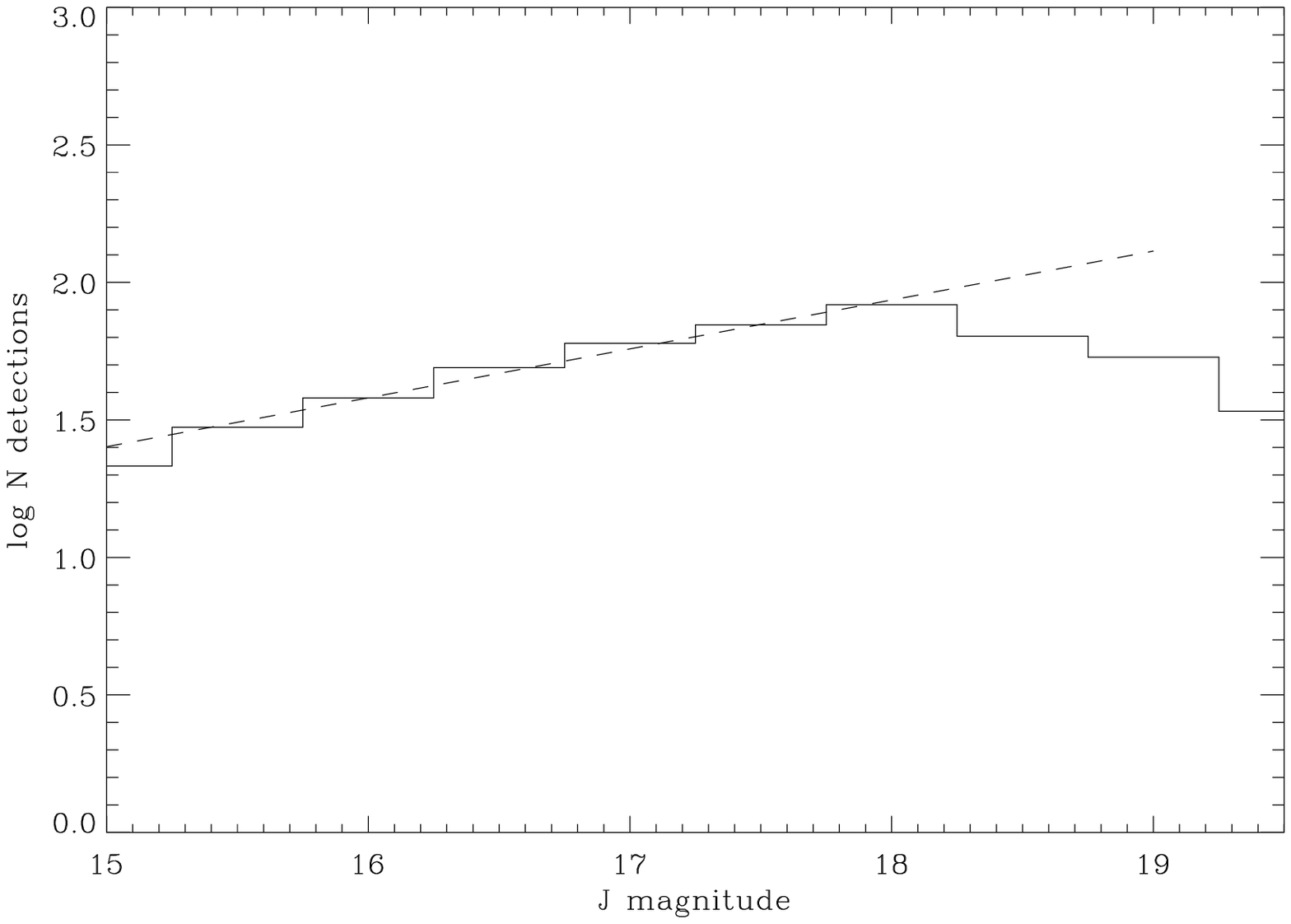}
\caption{Number of objects detected as a function of $J$
magnitude for an ensemble of representative images from all
observing runs at Steward Observatory, indicating the
completeness limit is $J=18$ mag.
\label{fig1}}
\end{figure}

In order to determine the completeness of the Steward
Survey, it was necessary to measure the flux of the faintest
reliably detectable sources in images taken over the entire
length of the study.  Figure \ref{fig1} shows the number of
objects detected with signal-to-noise ratio (S/N) $\geq3$ as
a function of $J$ magnitude in images representative of each
observing run conducted at Steward Observatory.  The survey was
complete to $J=18.0$ mag.  It is estimated that 37\% of
$J=18.5$ mag and 57\% of $J=19.0$ mag objects were
missed.

In the last two years of this survey, supplementary data
were gathered at Lick Observatory on the Shane 3 meter
telescope with the GEMINI camera \citep{mcl93}.  GEMINI
sits behind a telescope that is 70\% larger than the Bok
telescope and employs nearly identical detector technology.
$J$ band data acquired with GEMINI generally went about 1
magnitude deeper, providing greater sensitivity to $J\sim19$
mag objects.  However, the completeness limit of this survey
remains at $J=18.0$ mag and all calculations will be
restricted to this edge.

The Steward survey was a common proper motion companion
search.  By imaging the field around a target white dwarf
at two epochs separated by a sufficient interval of time,
the proper motion of the white dwarf can be measured and 
compared to any motions exhibited by stars in that field.
GEOMAP, a program within the Image Reduction and Analysis
Facility (IRAF) was used for this task.  This program creates
a general transformation between two sets of coordinates
corresponding to sources in the same field at two different
epochs.  Proper motion stars can be identified by their
residuals from this map and their motions measured against
the near zero motion of background stars and galaxies, which
provide a measure of the standard error.

A typical white dwarf with 5 or more field stars
produced a map with a standard deviation in the residuals
of approximately $0.2$ pixels or $0.13''$ on the Steward 
camera.  With a typical time baseline of 5 years, proper 
motions as small as $0.08''$ $\rm{yr}^{-1}$ can be
measured at the $3\sigma$ level.  These are characteristic
values -- the actual measurement values and errors depend
on the proper motion of a particular white dwarf, the
number of field sources with good S/N, the quality of
the images at each epoch, and the time baseline between
epochs (which varied between 2 and 10 years).  Additionally,
many proper motions for white dwarf targets and candidate 
companions were measured using the digitized versions of 
large sky survey photographic plates, such as the Palomar
Observatory Sky Survey I and II.  Although the spatial
resolution of these scans is lower ($1.0-1.7''$ ${\rm{pixel}}
^{-1}$) than the near-infrared data, the epochs are separated
by $\sim40$ years and hence provide better measurements and 
smaller errors.

\subsection{The Keck Survey}

From 1995 to 2001, a program to image
nearby white dwarfs at near-infrared wavelengths
was conducted on the Keck I 10 meter telescope at
Mauna Kea, Hawaii.  Images of 91 targets were acquired
using NIRC \citep{mat94}.  These data were not taken in
an analogous way to the data collected at Steward.  In
general, each target was observed at $J$ and usually one
or more of the following bands in order of decreasing
usage: $z$ (1.0 $\mu$m), $K$ (2.2 $\mu$m), $H$ (1.6 $\mu$m).
A typical total integration time at $J$ was $60-80$ seconds
times 5 dithers.  Data at all wavelengths were reduced in a
manner identical to the Steward data.

It is not possible to establish a completeness limit
in the same way as for the Steward data (Figure \ref{fig1})
because the relatively small NIRC field of view did not contain
sufficient background sources.  An analysis was performed
on many images representative of each observing run conducted
at Keck.  $J$ magnitudes and S/Ns were measured for faint 
but reliably detected objects in the images.  The result
being that $J=21.0$ mag or brighter objects were consistently
detected with ${\rm{S/N}}>10$.  Therefore, the NIRC survey
is likely complete down to $J=21$ mag.

The $z-J$ and/or $J-K$ color of point sources in the
field of the white dwarf, together with their flux relative
to the primary at these wavelengths, were used to filter out
uninteresting background stars and discriminate candidate
companions.  Data were taken at $z$ \& $J$ for about 75\%
of the white dwarfs in this survey.  For the remainder of 
the sample, images were obtained at $K$ for all fields which
contained point sources at $J$.  The $z-J$ color is the most
indicative of low mass stars and brown dwarfs because it is
monotonically increasing with decreasing effective temperature,
unlike $J-K$ \citep{leg02}.  If a candidate could not be ruled
out based on these criteria, then a follow up image was taken
at a later epoch in order to perform astrometry and search for
common proper motion.

\subsection{Companion Data \& Analysis}

For each confirmed or candidate common proper motion
companion, optical photometry and spectroscopy, in addition
to near-infrared photometry, was performed in order to identify
or constrain its temperature and class.  Near-infrared $JHK$
data were acquired with the same instruments used for the wide
field survey, namely the Steward and GEMINI cameras.  Images 
were taken and reduced in a similar manner to the survey
observations.

Optical $BVRI$ data were obtained at Lick Observatory
using the Nickel 1 meter telescope CCD camera.  In general, 
exposures were $1-10$ minutes depending on conditions and
individual target brightness.  Images were cleaned of bad 
pixels in the area of interest, bias subtracted, flat
fielded and averaged if there were multiple frames.

Optical spectroscopic data were acquired at Lick
Observatory using the Kast dual spectrograph on the 
Shane 3 meter telescope.  The exact setup varied between
observing runs but all observations were done at low resolution 
($\sim500$).  Additional spectroscopy was performed at Steward
Observatory using the Boller \& Chivens spectrograph on the Bok 2.3
meter telescope.  All optical spectra were reduced using standard 
IRAF software.  The spectral images were bias subtracted, cleaned 
of bad pixels and cosmic rays, then flat fielded. For each target, 
two spectra of the sky were extracted, averaged and subtracted 
from the extracted target spectrum.  The resulting spectra were
wavelength and flux calibrated by comparison with observed lamp
spectra and standard stars.  No attempt was made to remove 
telluric features.

Optical and near-infrared magnitudes and colors were used as
the main source of constraints for stellar classification and
spectral typing of the discovered companions.  Using a circular 
aperture centered on the target star and an annulus on the 
surrounding sky, both the flux and S/N were calculated for a
range of apertures from one to four full widths at half maximum.
The flux measurement was taken at or near the aperture size which
produced the largest S/N.  In this way the flux of all targets was
measured, including photometric standard stars.  A fairly large
aperture was used for all calibrators, and flux measurements for
science targets were corrected to this standard aperture.
Optical and near-infrared standard stars were taken from
\citet{lan83}, \citet{hun98}, and \citet{haw01}.

\subsection{The IRTF Survey}

As mentioned in \S1, there was a previous phase to the
search for substellar companions to white dwarfs that began
in late 1986 \citep{zuc92}.  Although carried out on several
different telescopes and instruments, the majority of those data
were obtained at the NASA Infrared Telescope Facility (IRTF).
Because some results of the IRTF survey were previously published,
the details will not be discussed here.  However, those results will
be updated below and the white dwarfs surveyed at the IRTF will be
included in the overall sample and statistics.  Of the over 150
white dwarfs observed during this early phase, 66 were later 
reobserved at either Keck or Steward Observatory.  There were
84 white dwarf targets observed at the IRTF and not elsewhere.

\subsection{The Search for Near-Infrared Excess}

Although data at both $J$ and $K$ were taken for about
one fourth to one third of the stars surveyed, all 371 white
dwarfs in the sample were searched for near-infrared excess
emission.  A digital finding chart of each white dwarf was
overlaid with the 2MASS point source catalog data.  The 
measured $JHK_s$ values for the the white dwarf were then 
compared to the model predicted values extrapolated from
optical data based on the effective temperature of the star.
Stars with excess emission and good S/N at one or more of
these near-infrared wavelengths were noted as possible or
probable binaries.

\section{THE SAMPLE}

\subsection{Target Selection}

Nearly every target observed for this project can be
found in either current or earlier versions of the
white dwarf catalog of \citet{mcc87,mcc99}.  The catalog
is mostly composed of stars selected by one of two criteria:
(1) faint proper motion stars or (2) stars with ultraviolet 
excess.  Most of these stars were spectroscopically confirmed
to be white dwarfs, but there remains some contamination by
nondegenerate stars.  Hot subdwarfs, blue horizontal branch
stars, BL Lacertae objects, and Population II stars all
display one of the characteristics above and can have spectra
difficult to differentiate from that of a white dwarf with
older photographic techniques.  

A sample of nearby hot and massive white dwarfs is
ideal to search for substellar companions.  Proximity to
the Sun is desirable for the ability to partially or totally
resolve close companions from the primary star and because 
flux falls off inversely as the square of the distance.  Hot
white dwarfs are ``recently deceased'' and are therefore younger
than their cooler counterparts.  A typical white dwarf with
$M=0.6$ $M_{\odot}$ and $T_{\rm{eff}}=20,000$ K has been
cooling for only 70 Myr \citep{ber95}.  Most massive white
dwarfs ($M\geq0.9$ $M_{\odot}$) are thought to descend
from main sequence progenitors with masses, 6 $M_{\odot}<M<$
8 $M_{\odot}$ \citep{ber91,ber92,bra95,wei87,wei90,wei00,kal05}.
Hence massive white dwarfs have a total age on par with their
cooling age because the main sequence lifetime of the progenitor
would have been relatively short.  Although a large ($N\sim100$)
sample with all three characteristics does not exist, these
attributes were guiding principles in selecting stars for
the survey.

\subsection{Kinematics}

Proper motion can be an indicator of age relative
to the three basic kinematic populations of the Galaxy:
young disk ($\tau\sim1-2$ Gyr), old disk ($\tau\sim5-10$
Gyr), and halo stars ($\tau\sim10-15$ Gyr).  Ages between
$2-5$ Gyr are considered intermediate disk ages.  In
general, the $UVW$ space velocities (and perhaps more
important, velocity dispersions) of stellar populations
increase with increasing age.  This is primarily due to
gravitational upscattering for disk objects, while the 
halo is a distinctly separate kinematical group in every
sense.  Therefore in general, smaller values of $(U,V,W)$
and ($\sigma_{U},\sigma_{V},\sigma_{W})$, for a given
kinematical sample, are correlated with younger
objects \citep{wie74,mih81,leg92,jah97,bin98}.

Although it is three dimensional space motion that
determines kinematic populations and indicates likely
membership for an individual star, proper motion is
often used as a proxy because the radial velocity ($v_r$)
is not known.  In addition, it is particularly challenging
to measure the radial velocity of white dwarfs due to their 
wide, pressure broadened line profiles and intrinsic faintness.
However, two studies have compared the $UVW$ space motions of 
over 100 white dwarfs in wide binaries calculated with and without
the assumption $v_r=0$ \citep{sil01,sil02}.  Accurate radial
velocities obtained from a widely separated main sequence
component in each binary yielded two major conclusions:
(1) the overall sample kinematics were consistent with
the old, metal-poor disk population and (2) the assumption
of $v_r=0$ did {\em not} significantly affect the results
\citep{sil01,sil02}.  This is an important result because,
in the end, the sample of white dwarfs in the present work 
can only be tied together with kinematics.

Table \ref{tbl-1} lists all 395 target stars observed at
all facilities beginning in 1990.  The first column lists 
the white dwarf number from \citet{mcc87,mcc99}, except
where noted, followed by a name of the star in the second
column.  The third column lists the spectral type of the
white dwarf \citep{mcc87,mcc99}.  The integer value in column
three represents the effective temperature index, defined as
$10\times{q_{\rm{eff}}}$, where $q_{\rm{eff}}=5040/T_{\rm{eff}}$
\citep{mcc87,mcc99}.  The fourth column lists the distance, $d$, 
in parsecs as determined photometrically, using the best available
data in the literature and models, or by trigonometric parallax.
The fifth and sixth columns are the proper motion, $\mu$, in 
arcseconds per year, followed by the position angle, $\theta$,
in degrees.  These quantities were taken from the most accurate
and reliable source available.  In decreasing order these are:
the Tycho 2 catalog \citep{hog00}, the UCAC catalogs \citep{zac00,
zac04}, the USNO B1.0 catalog \citep{mon03}, and the white dwarf
catalog of \citet{mcc99} and references therein.  A few proper 
motions were measured for this work.  The seventh column lists 
the Galactic $UVW$ space velocity, corrected for the solar motion
$(U,V,W)=(-9,+12,+7)$ \citep{wie74,mih81} relative to the local 
standard of rest (LSR) in km $\rm{s}^{-1}$.  These quantities were
calculated from the vector ($\alpha,\delta,d,\mu,\theta,v_r$) where
$\alpha$ is the right ascension, $\delta$ the declination, and
$v_{r}=0$ was assumed to provide a uniform treatment of the sample.
$U$ is taken to be positive toward the Galactic anticenter, $V$ 
positive in the direction of Galactic rotation, and $W$ positive 
toward the North Galactic Pole.  The eigth column lists the
facility or facilities at which the white dwarf was observed:
S = Steward, K = Keck, I = IRTF.   

The Galactic $UVW$ space motions and statistics for the
white dwarf sample was calculated in order to evaluate the most
probable range of stellar ages.  As stated above, smaller values 
of $UVW$ and their dispersions, implying more circular Galactic 
orbits, correlate with younger stellar populations that have 
experienced fewer gravitational events since their birth in and
around the spiral arms \citep{mih81,bin98}.  Table \ref{tbl-2} 
contains the kinematical properties calculated for the white 
dwarf sample.  The quantity $T$ is the total space velocity 
with respect to the LSR $(T^2=U^2+V^2+W^2)$, and $\sigma_{T}$
is the total dispersion in space velocity $(\sigma^2_{T}=\sigma
^2_{U}+\sigma^2_{V}+\sigma^2_{W})$.  The sample does not appear
to consist primarily of old, metal-poor disk stars.  It seems 
likely that the sample contains a relatively high fraction of
stars with intermediate and young disk kinematics -- stars with
ages $\tau\la5$ Gyr.

\begin{figure}
\plotone{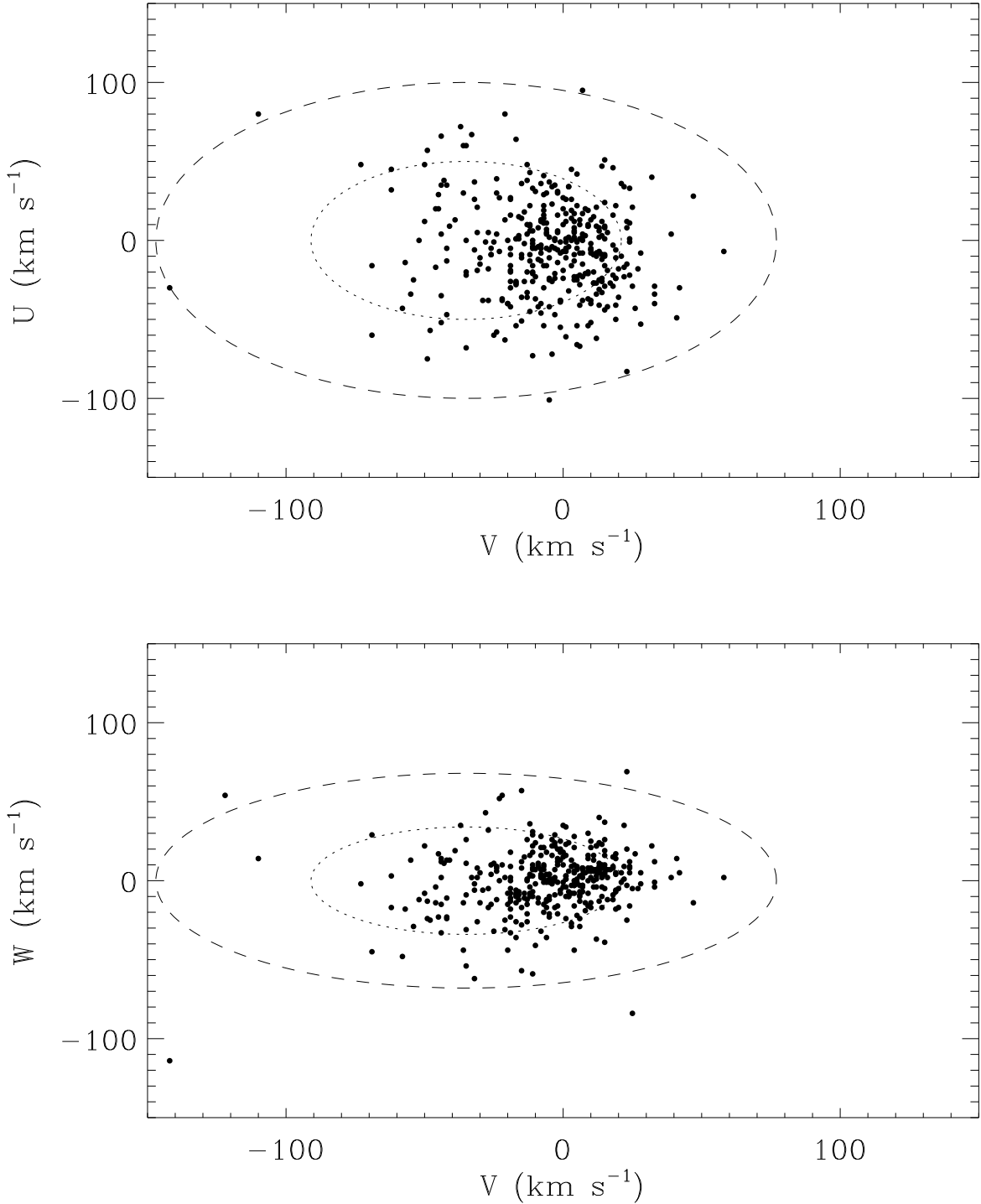}
\caption{Galactic space velocity distribution in the $UV$ and $WV$ planes
for all 371 white dwarfs in the sample, assuming $v_r=0$.  The ellipses 
represent the 1 and 2 $\sigma$ contours for old, metal-poor disk stars
from \citep{bee00}.
\label{fig2}}
\end{figure}

In Figure \ref{fig2}, the white dwarf sample is plotted 
in the $UV$ and $WV$ planes.  Also shown in the figure are
the 1 and 2 $\sigma$ velocity ellipsoids for old, metal-poor
disks stars from \citet{bee00} -- a kinematical study of the
halo and thick disk utilizing a large sample of nonkinematically
selected metal-poor stars.  The ellipsoid parameters in Figure 
\ref{fig2} were taken from the first row of Table 1 in \citet{bee00},
141 stars with $-0.6\leq$ [Fe/H] $\leq-0.8$ and $|Z|<1$ kpc.  $Z$ is
the scale height above the Galactic plane, and hence this old disk
sample is unlikely to be contaminated significantly by halo stars. 
The ellipsoids are centered at $(U,V,W)=(0,-35,0)$ km $\rm{s}^{-1}$
with axes $(\sigma_{U},\sigma_{V},\sigma_{W})=(50,56,34)$ km $\rm{s}
^{-1}$.  From Figure \ref{fig2} and Table \ref{tbl-2} it is clear 
that the white dwarf sample is centered much closer to $(U,V,W)=
(0,0,0)$, values that represent the undisturbed circular Galactic
disk orbits of younger stars \citep{mih81,bin98}.  Older disk stars
lag behind the Galactic rotation of the LSR and hence have
increasingly negative $V$ velocities with increasing age
\citep{bee00}.

Comparison of the values in Table \ref{tbl-2} with the values
for kinematical populations of known ages from $Hipparcos$ 
measurements of nearby stars, yields additional evidence that
the white dwarf sample contains young disk stars.  The average
$UVW$, their dispersions, and the total velocity dispersion
$(\sigma_{T})$ values of the entire sample are consistent with
those of disk stars of intermediate age ($\tau=2-5$ Gyr), but
inconsistent with stars of age $\tau=5$ Gyr due to the relatively
small  negative value of $\langle V \rangle$.  This comes from a
direct  comparison of Table \ref{tbl-2} with Table 5 \& Figures
$3-5$ of \citet{wie74}, and with Table 4 of \citet{jah97}.  In
fact, the subsample in Table \ref{tbl-2}, white dwarfs with 
$\mu<0.50'' \ \rm{yr}^{-1}$, is quite consistent with stars
of age $\tau\sim2$ Gyr \citep{wie74,jah97}.

The white dwarfs surveyed in this work are not similar
to the white dwarf samples of \citet{sil01} and \citet{sil02},
which clearly belong to the old disk kinematical population
($\tau\sim5-10$ Gyr).  Neither is the sample similar to any
of the white dwarf kinematic subgroups in \citet{sio88} with 
the exception of the DH and DP stars (magnetic white dwarfs).  
The sample of magnetic white dwarfs in \citet{sio88} was 
expanded from only 13 stars to 26 stars in \citet{ans99} 
with the same results (both studies assumed $v_r=0$ as in this
work) -- these stars appear to have young
disk kinematics.  In fact, the subsample of moderate proper
motion white dwarfs in Table \ref{fig2} have nearly identical 
kinematical properties as magnetic white dwarfs, implying 
relatively young ages ($\tau\sim2$ Gyr) \citep{sio88,ans99}.

\subsection{Cooling \& Overall Age}

It must be kept in mind that the present sample
consists of a mixture of hot and cool degenerate stars.
The cooling age of a typical hot white dwarf ($T_{\rm{eff}}>
11,000$ K) is less than 500 Myr but the main sequence progenitor 
age is not known.  Hence the total ages of hot white dwarfs in
the sample are potentially consistent with relatively young disk 
objects.  But for cool white dwarfs in the sample, it is more 
likely that they are intermediate age disk stars.  For example,
a white dwarf with $T_{\rm{eff}}<7500$ K is at least 1.5 Gyr
old according to cooling theory \citep{ber95}.

\begin{figure}
\plotone{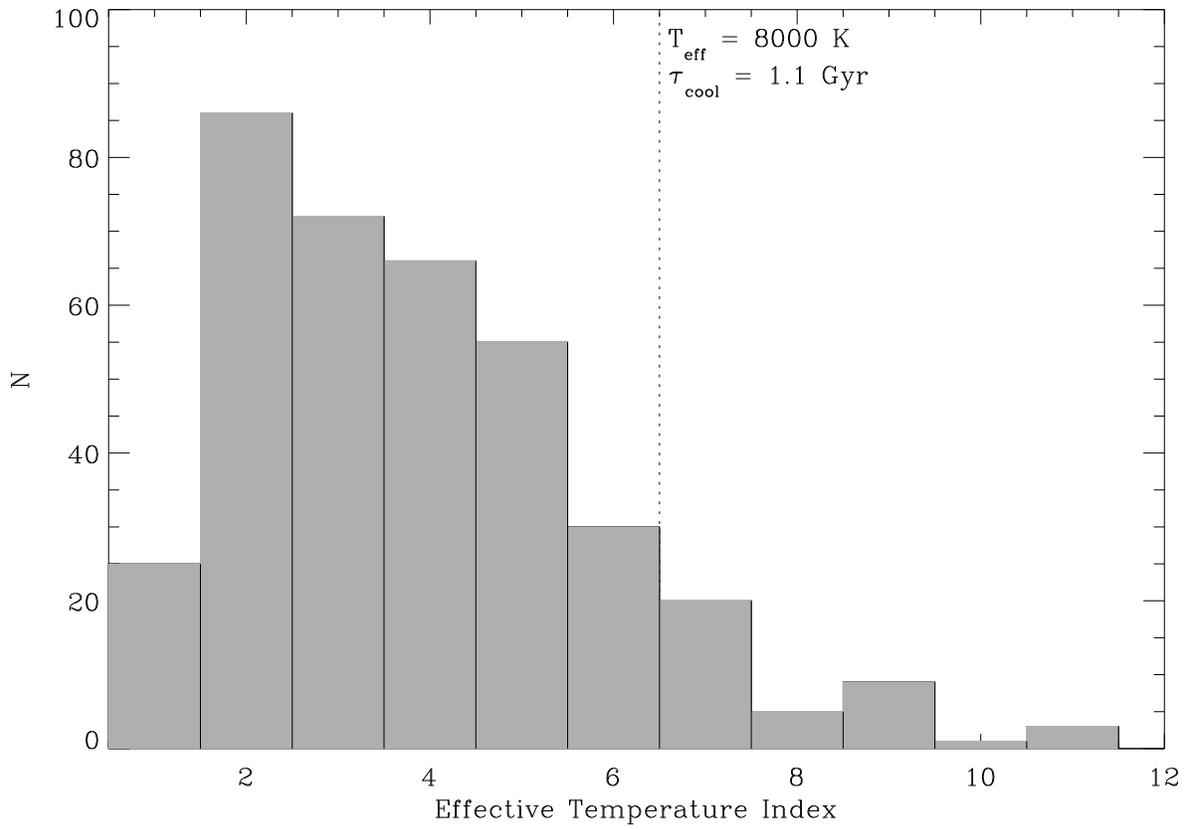}
\caption{Number of sample white dwarfs versus effective temperature index
(as discussed in text \S3.3).  The dotted line represents a cooling age of
1.08 Gyr for a typical DA white dwarf \citep{ber95}.
\label{fig3}}
\end{figure}

In Figure \ref{fig3} is plotted the number of white dwarfs in the 
sample versus effective temperature index.  Exactly 90\% of the 
sample stars have temperatures above 8000 K -- implying cooling 
ages less than 1.1 Gyr for typical hydrogen atmosphere white dwarfs 
\citep{ber95}.  Moreover, 67\% of the sample have temperatures above 
11,500 K and typical cooling ages less than 0.4 Gyr.  Hence the 
cooling ages of the sample stars are consistent with the total 
age estimate inferred from kinematics -- that of a relatively
young disk population.

Since one does not know the main sequence progenitor ages
for the white dwarf sample, caution must be taken not to
over interpret the kinematical results.  In principle, any
individual star of any age can have any velocity.  It is 
possible to estimate total ages for white dwarfs if their
mass is known by using the initial to final mass relation
\citep{wei87,wei90,wei00,bra95}.  However, this is only 
feasible for DA white dwarfs (whose masses can be determined 
spectroscopically), white dwarfs with dynamical mass
measurements, or those with trigonometric parallaxes
\citep{ber92,ber97,ber01}.  The sample in Table \ref{tbl-1}
contains many degenerates with no mass estimate and therefore
no way to confirm or rule out the relatively young total ages
indicated by their kinematics.  While their cooling ages are 
consistent with young disk objects, a conservative approach
would be to explore a range of ages when interpreting the
implications of the survey results.  Realistically, a typical
white dwarf in the sample is likely to be between $\tau=2-5$ 
Gyr old.

\section{RESULTS}

\subsection{All Companions}

Table \ref{tbl-3} lists all companions to white dwarf sample
stars detected in this work or published in the literature.  
Many targets were thought to be single white dwarfs when this 
project began in the late 1980's but subsequently have been 
established to be binaries in various studies.  Although only 
low mass stellar and substellar companions were directly sought
in this study, the overall multiplicity of white dwarfs is of 
astrophysical interest for many reasons.  The first column 
lists the name of the companion.  This is generally the name
of the white dwarf primary plus the letter ``B'' for a secondary,
or ``C'' for a tertiary.  The second column lists the known or
suspected spectral type of the companion, while the third column
lists the primary white dwarf number.  For companions discovered 
in this work, spectral types were estimated from optical and 
near-infrared colors with the longest baselines (such as $V-K$).
The fourth column lists the primary spectral type.  The fifth
and sixth columns list the separation on the sky and position angle
of resolved companions.  If unresolved, an upper limit to the 
separation is given, whereas a designation of ``close'' implies
the system is a known radial velocity variable.  The seventh and
eighth columns list the best distance estimate for the white 
dwarf and the projected separation of the binary.  The ninth
column lists the absolute $V$ magnitude for white dwarf companions
or the absolute $K$ magnitude for low mass stellar and substellar 
companions.  The final column lists references to the initial 
discovery, critical data and analysis of each companion.

In all, there are 83 companion objects in 75 stellar
systems containing at least one white dwarf: 76 doubles,
6 triples, and 1 quadruple system.  Of all the companions,
excepting GD 1400B, there are 18 white dwarfs and 64 main 
sequence stars, and 1 brown dwarf.  There were 24 multiple
systems independently discovered in this work, 20 of which
are reported here for the first time and the remaining 4
previously published \citep{fin97b,mcc99,far04,sch04}.  In addition,
new data and analysis of 32 binaries reported in \citet{zuc92} and
\citet{sch96} have resulted in more accurate descriptions of those
systems.

\subsection{Near-Infrared Excess \& Unresolved Companions}

White dwarfs with $T_{\rm{eff}}\ga10,000$ K have blue or 
zero optical and near-infrared colors.  Cooler white dwarfs 
will have colors that are just slightly red \citep{ber97,
leg98,ber01}.  For example, a typical white dwarf with 
$T_{\rm{eff}}=6750$ K will have $V-K=1$, $J-K=0.2$ 
\citep{ber95}.

Very low mass stars and brown dwarfs have radii that
are approximately 10 times larger than a typical white
dwarf radius, $R\sim1$ $R_{\oplus}$ \citep{bur97}.
Therefore, despite  very low effective temperatures and
luminosities, an unresolved cool companion to a white dwarf
can dominate the spectral energy distribution of the system
at longer wavelengths, especially in the near-infrared
\citep{pro83,zuc87,zuc87b}.  Therefore, a white dwarf with 
red colors in the near-infrared or red portion of the optical
spectrum can indicate the presence of an unresolved cool companion
\citep{gre86a,bec88,zuc92,far04b}.

There are basically two methods for obtaining parameters 
for unresolved low mass stellar or substellar companions 
to white dwarfs -- optical and/or near-infrared photometry 
or optical spectroscopy.  Near-infrared spectroscopy is not 
typically performed because spectral 
types for low mass stars and cool dwarfs in general (M \& L
dwarfs) were established optically \citep{kir94,kir99b}.

Optical spectroscopy can reveal unresolved companions
to white dwarfs for a range of white dwarf to red dwarf 
luminosity ratios.  If the white dwarf is cool enough 
and/or the red dwarf is bright enough, a composite 
spectrum can be seen even in the blue and visual portion
of the optical spectrum \citep{gre86b,fin97}.  Red dwarf
companions which are too dim, relative to their white 
dwarf hosts, in the blue or visual can still be seen at 
red optical wavelengths ($7000-10,000$ \AA; \citealt{max98}).  
To extract information on the companion, one can
visually examine the spectrum and compare it to known spectral
types.  For better accuracy, one can fit the bluest portion
of the spectrum with models and effectively subtract the
contribution of the white dwarf, leaving only the
companion spectrum for analysis \citep{ray03}.

However, the lowest luminosity companions to white 
dwarfs do not contribute a relatively sufficient 
amount of light in the optical for accurate spectral
typing or study if they are unresolved \citep{kir93}.
Near-infrared methods must be used for these companions.
Near-infrared spectroscopy can verify the presence of a
companion, but has only a limited ability to provide a
spectral type for the reason mentioned above.  The most
successful method for doing so uses near-infrared photometry.
With models, one can extrapolate the flux of the white dwarf
into the near-infrared and subtract its expected contribution, 
thereby obtaining photometry for any unresolved, very
low luminosity companion \citep{zuc87,bec88,zuc92,
gre00,far04b}.  The resulting near-infrared colors 
(or near-infrared plus red optical colors or upper limits) 
can be compared with the colors of known isolated low 
luminosity objects such as late M dwarfs and L dwarfs
for determination of spectral type
\citep{kir94,kir99b}.

In this work, both near-infrared and optical colors
resulting from photometry were used to estimate
spectral types for all unresolved companions, while 
optical spectroscopy was used to verify the presence 
of the companion, where possible.  Most of the white
dwarf primaries with unresolved cool companions are 
quite well studied and hence model extrapolation to 
longer wavelengths is likely to be reliable.  The
model grids of P. Bergeron (2002, private communication) 
for pure hydrogen and pure helium atmosphere white dwarfs
were used to predict $RIJHK$ fluxes for white dwarfs in
such systems.  These fluxes, together with the measured 
composite fluxes, were then used to calculate $RIJHK$ 
magnitudes for the unresolved red dwarf component of the
binary.  The resulting optical and near-infrared colors were 
then compared to those of \citet{kir94} to determine spectral
type.  Unlike both \citet{zuc92} and \citet{gre00}, absolute
$K$ magnitudes of the companions were generally not used to
estimate spectral type.

\subsection{New Companions}

\begin{figure} 
\plotone{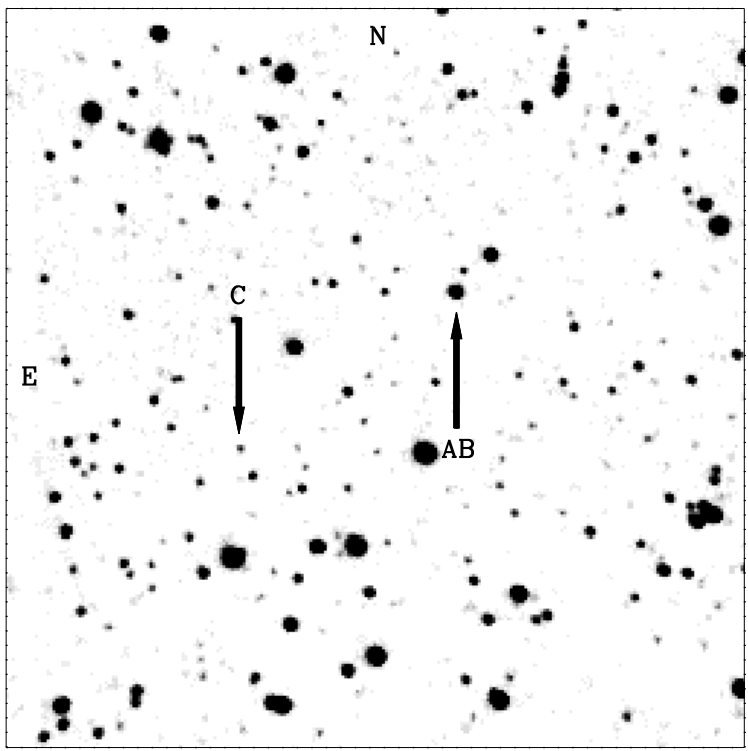}
\caption{Near near-infrared finding chart for G21-15C, taken at $J$ band with
the Bok 2.3 meter telescope in July 2001.  The image is $166''$ square
with $0.65''$ pixels.  The coordinates for the companion are $18^{\rm h}
27^{\rm m} 16.4^{\rm s}$, $+04^{\circ} 04' 09''$ J2000.
\label{fig4}}
\end{figure}

\begin{figure} 
\plotone{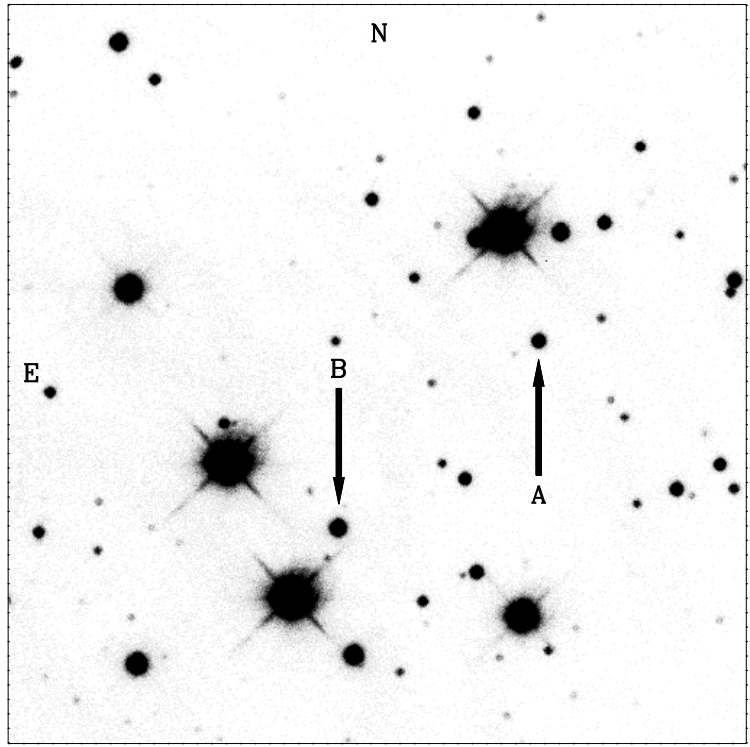}
\caption{Optical finding chart for GD 60B, taken at $I$ band with
the Nickel 1 meter telescope in January 2003.  The image is $184''$ square
with $0.36''$ pixels.  The coordinates for the companion are $04^{\rm h}
20^{\rm m} 15.2^{\rm s}$, $+33^{\circ} 34' 48''$ J2000.
\label{fig5}}
\end{figure}

\begin{figure} 
\plotone{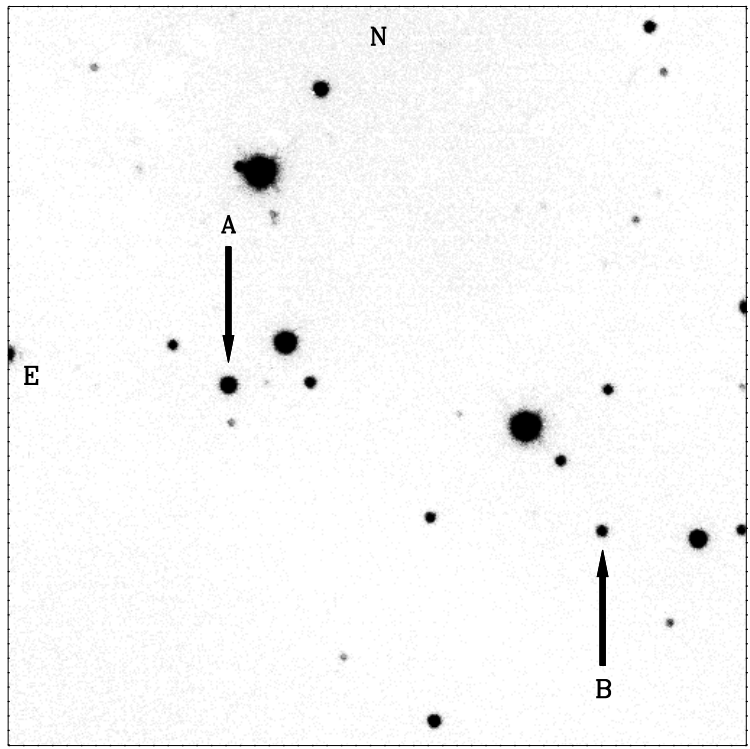}
\caption{Optical finding chart for GD 74B, taken at $R$ band with
the Nickel 1 meter telescope in January 2003.  The image is $184''$ square
with $0.36''$ pixels.  The coordinates for the companion are $06^{\rm h}
28^{\rm m} 55.8^{\rm s}$, $+41^{\circ} 30' 11''$ J2000.
\label{fig6}}
\end{figure}

\begin{figure} 
\plotone{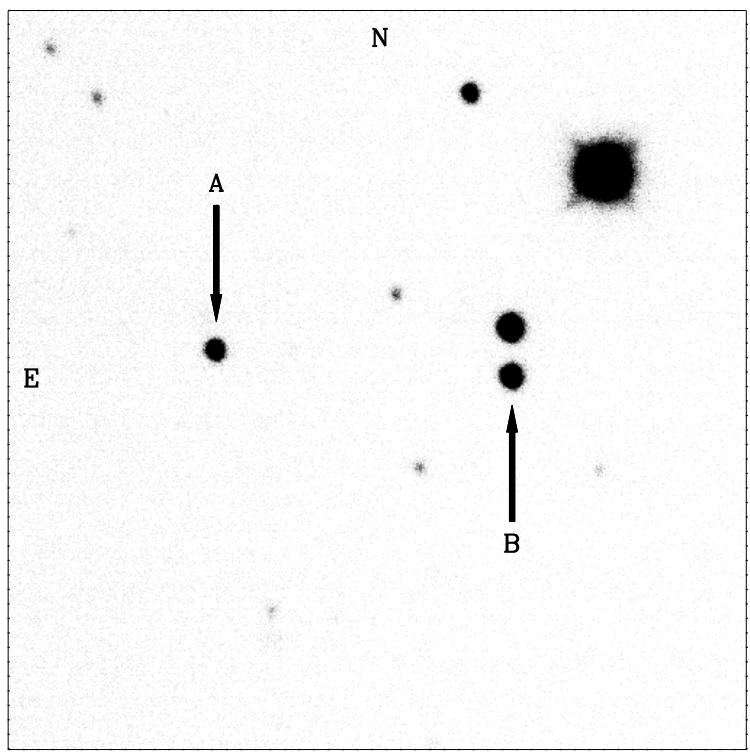}
\caption{Optical finding chart for GD 84B, taken at $R$ band with
the Nickel 1 meter telescope in October 2001.  The image is $184''$ square
with $0.36''$ pixels.  The coordinates for the companion are $07^{\rm h}
17^{\rm m} 54.6^{\rm s}$, $+45^{\circ} 47' 48''$ J2000.
\label{fig7}}
\end{figure}

\begin{figure} 
\plotone{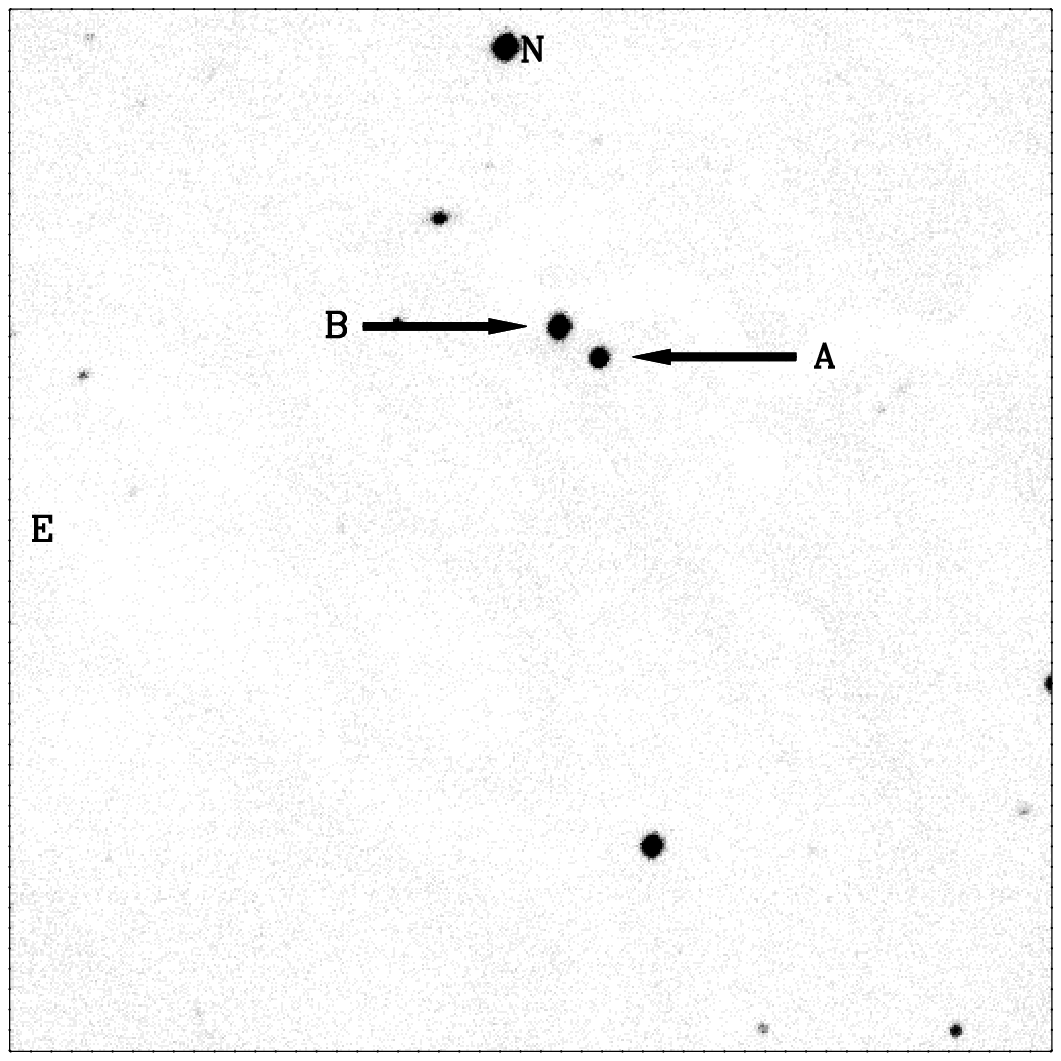}
\caption{Optical finding chart for GD 267B, taken at $I$ band with
the Nickel 1 meter telescope in March 2003.  The image is $184''$ square
with $0.36''$ pixels.
\label{fig8}}
\end{figure}

\begin{figure} 
\plotone{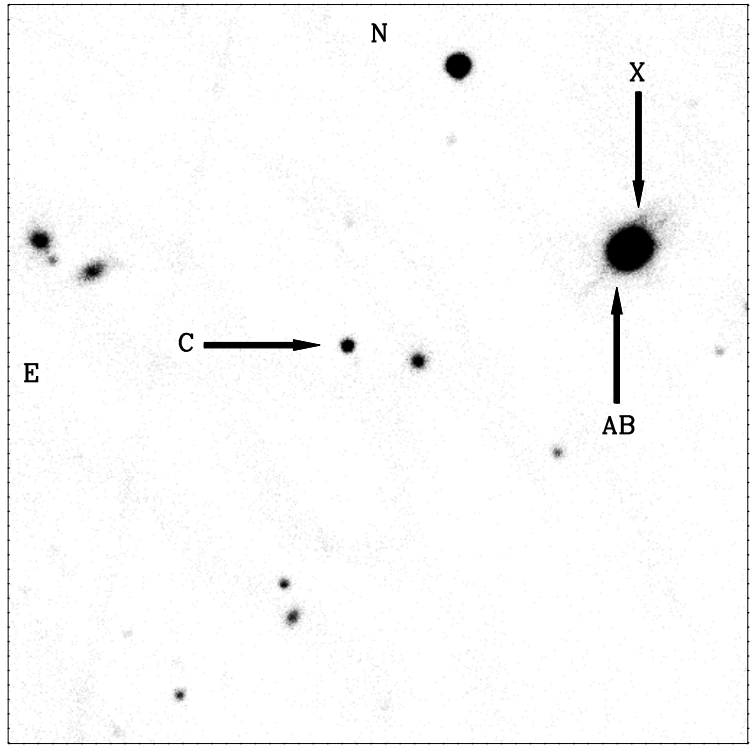}
\caption{Optical finding chart for GD 319C, taken at $I$ band with
the Nickel 1 meter telescope in March 2002.  The image is $184''$ square
with $0.36''$ pixels.  The coordinates for the companion are $12^{\rm h}
50^{\rm m} 12.7^{\rm s}$, $+55^{\circ} 05' 36''$ J2000.  The object
labelled `X' is a foreground K dwarf located $\sim2.5''$ away from GD 319AB.
\label{fig9}}
\end{figure}

\begin{figure} 
\plotone{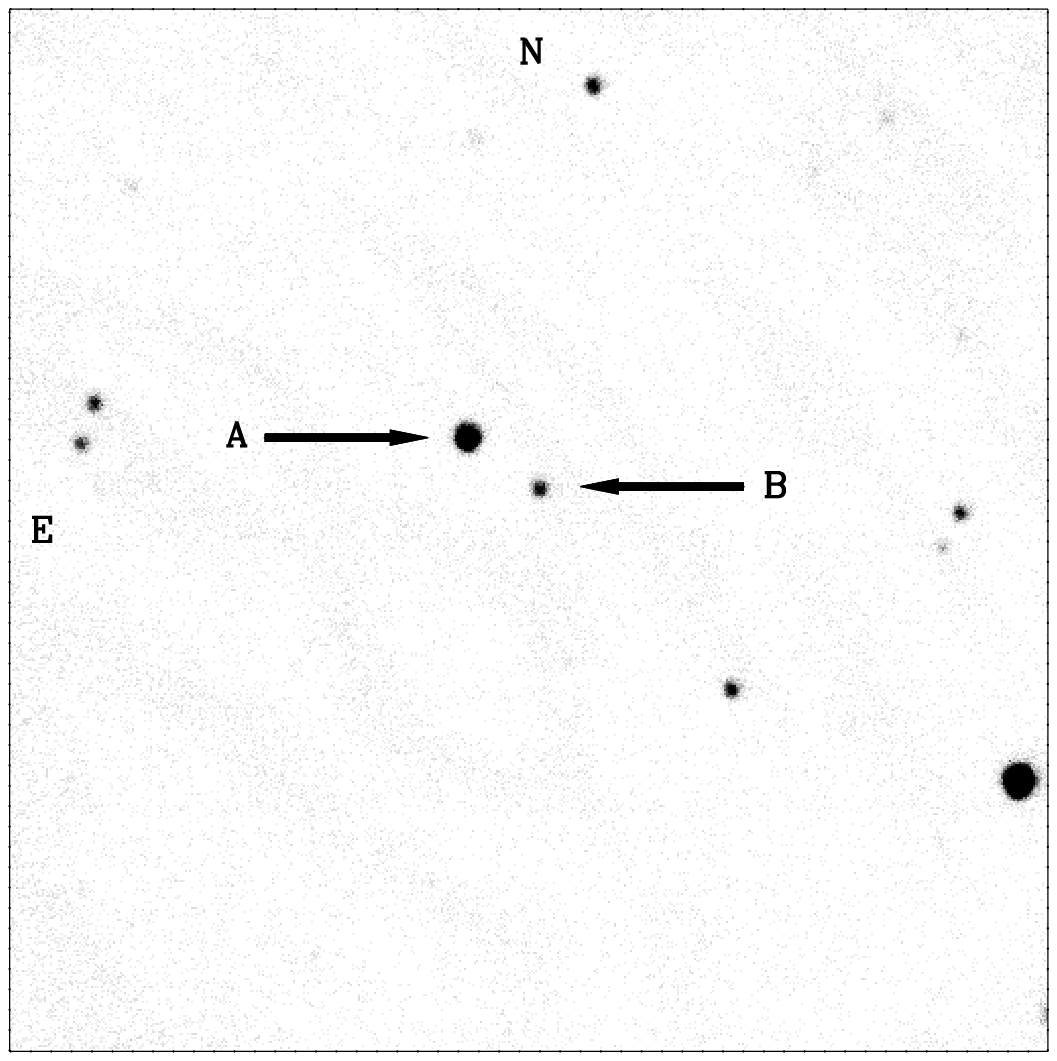}
\caption{Optical finding chart for GD 322B, taken at $I$ band with
the Nickel 1 meter telescope in March 2002.  The image is $184''$ square
with $0.36''$ pixels.
\label{fig10}}
\end{figure}

\begin{figure} 
\plotone{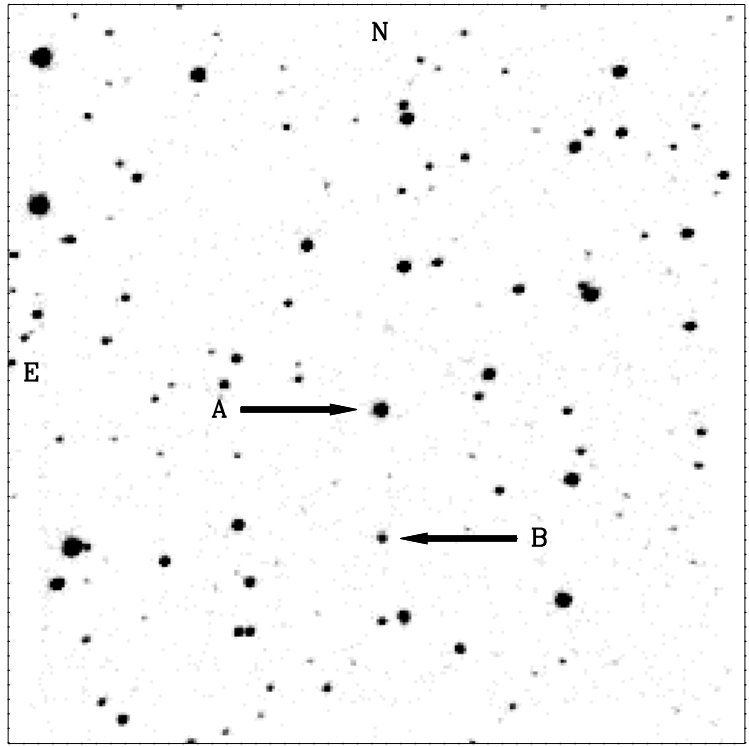}
\caption{Near near-infrared finding chart for GD 559B, taken at $J$ band with
the Bok 2.3 meter telescope in October 1996.  The image is $166''$ square
with $0.65''$ pixels.  The coordinates for the companion are $23^{\rm h}
21^{\rm m} 17.2^{\rm s}$, $+69^{\circ} 25' 54''$ J2000.
\label{fig11}}
\end{figure}

\begin{figure} 
\plotone{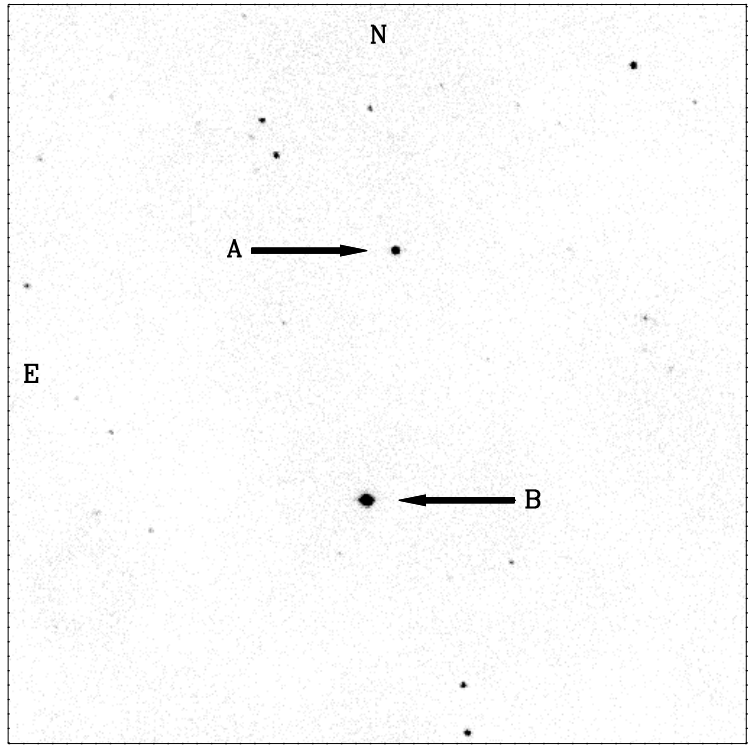}
\caption{Optical finding chart for GD 683B, taken at $I$ band with the
Swope 1 meter telescope in November 2003.  The image is $328''$ square
with $0.44''$ pixels.  The coordinates for the companion are $01^{\rm h}
08^{\rm m} 21.6^{\rm s}$, $-35^{\circ} 36' 33''$ J2000.
\label{fig12}}
\end{figure}

\begin{figure} 
\plotone{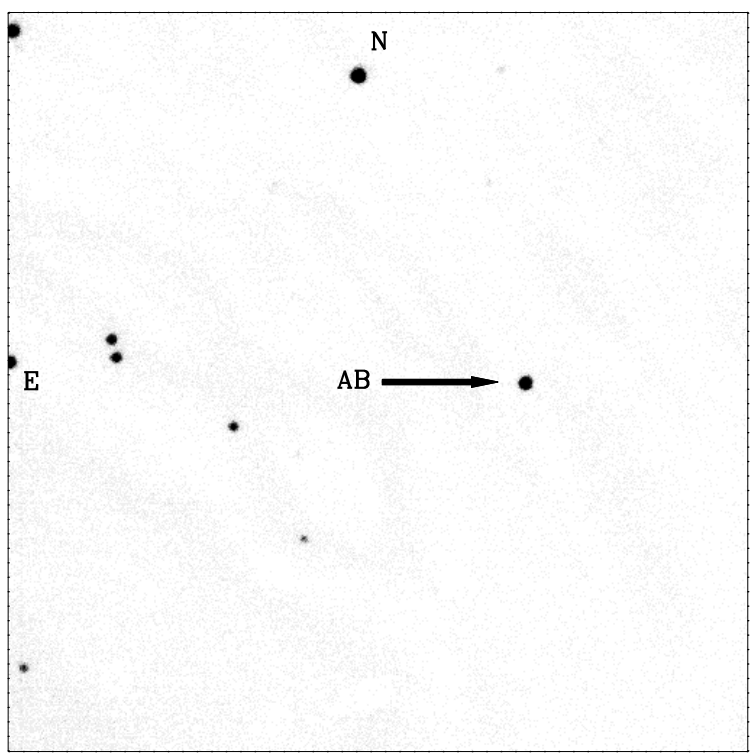}
\caption{Optical finding chart for LP 618-14, taken at $I$ band with
the Nickel 1 meter telescope in June 2002.  The image is $184''$ square with
$0.36''$ pixels.  The coordinates for the composite binary are $13^{\rm h}
36^{\rm m} 16.1^{\rm s}$, $+00^{\circ} 17' 33''$ J2000.
\label{fig13}}
\end{figure}

\begin{figure} 
\plotone{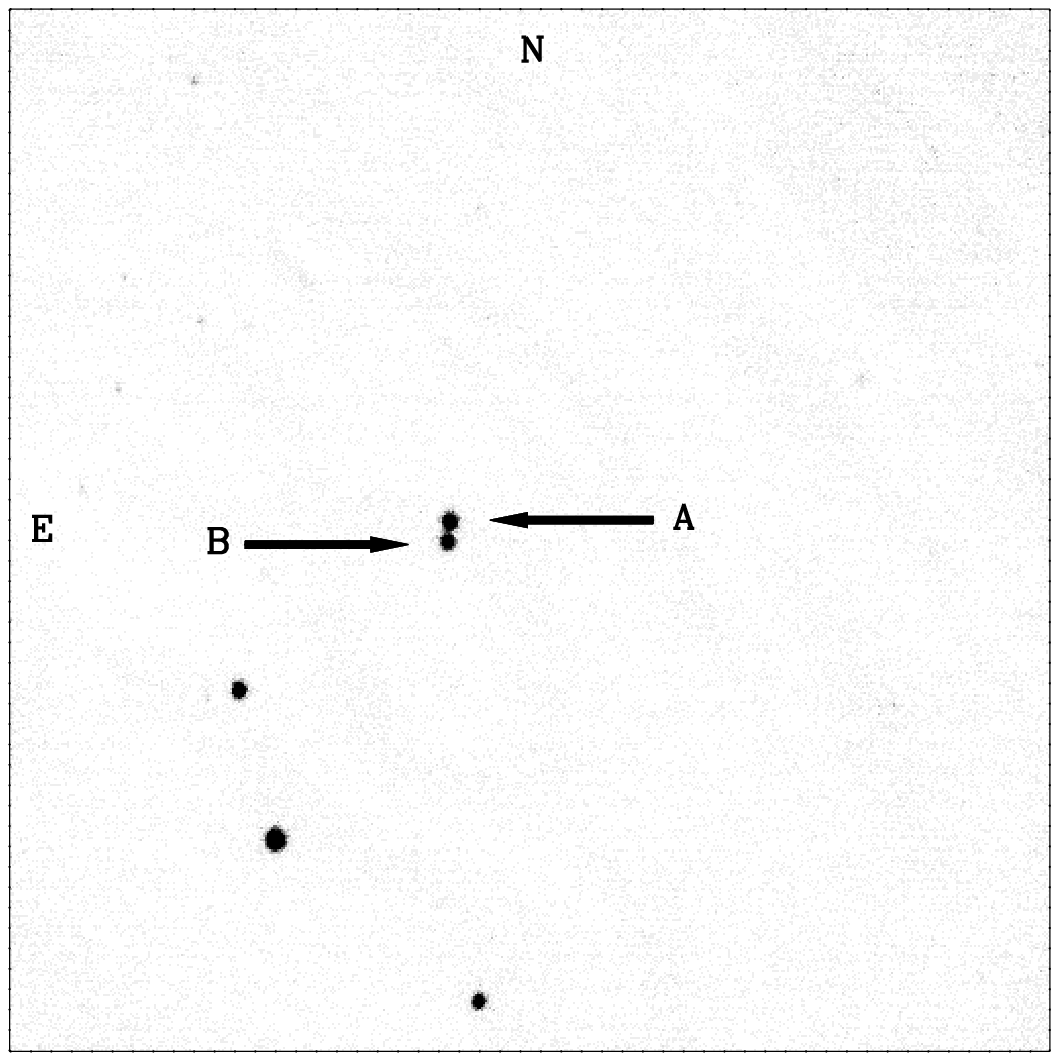}
\caption{Optical finding chart for PG 0901+140B, taken at $I$ band with
the Nickel 1 meter telescope in April 2003.  The image is $184''$ square
with $0.36''$ pixels.
\label{fig14}}
\end{figure}

\begin{figure} 
\plotone{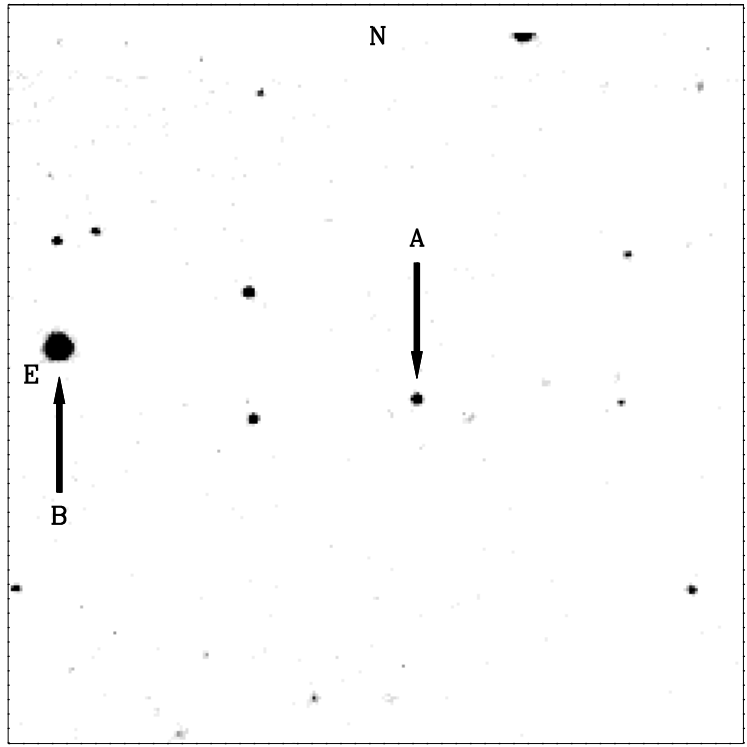}
\caption{Near near-infrared finding chart for PG 0933+729B, taken at $J$ band with
the Bok 2.3 meter telescope in December 2002.  The image is $166''$ square
with $0.65''$ pixels.  The coordinates for the companion are $09^{\rm h}
38^{\rm m} 39.8^{\rm s}$, $+72^{\circ} 42' 31''$ J2000.
\label{fig15}}
\end{figure}

\begin{figure} 
\plotone{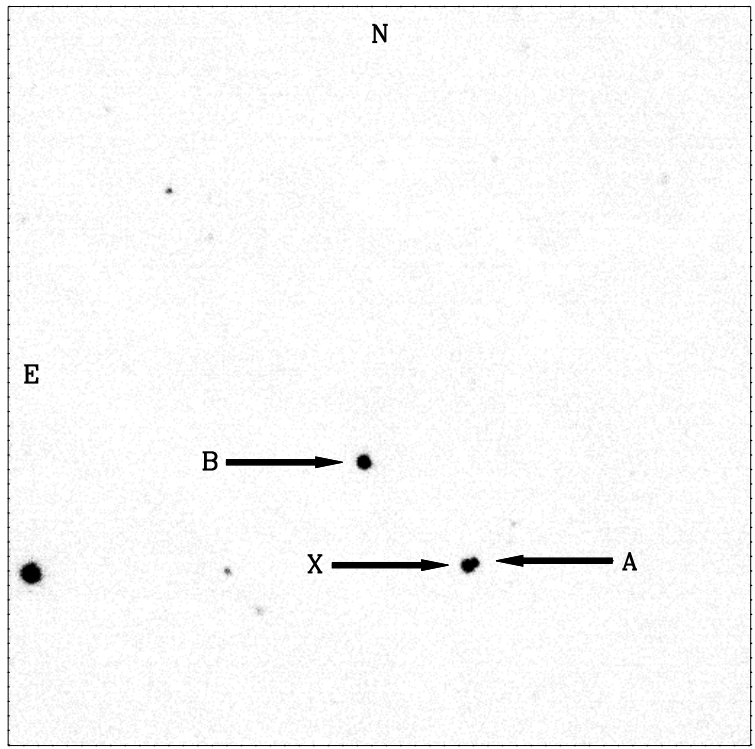}
\caption{Optical finding chart for PG 1015+076B, taken at $I$ band with
the Nickel 1 meter telescope in March 2003.  The image is $184''$ square
with $0.36''$ pixels.  The coordinates for the companion are $10^{\rm h}
18^{\rm m} 03.5^{\rm s}$, $+07^{\circ} 21' 50''$ J2000.  The object
labelled `X' is a background G dwarf located $2.0''$ away from
PG 1015+076A.
\label{fig16}}
\end{figure}

\begin{figure} 
\plotone{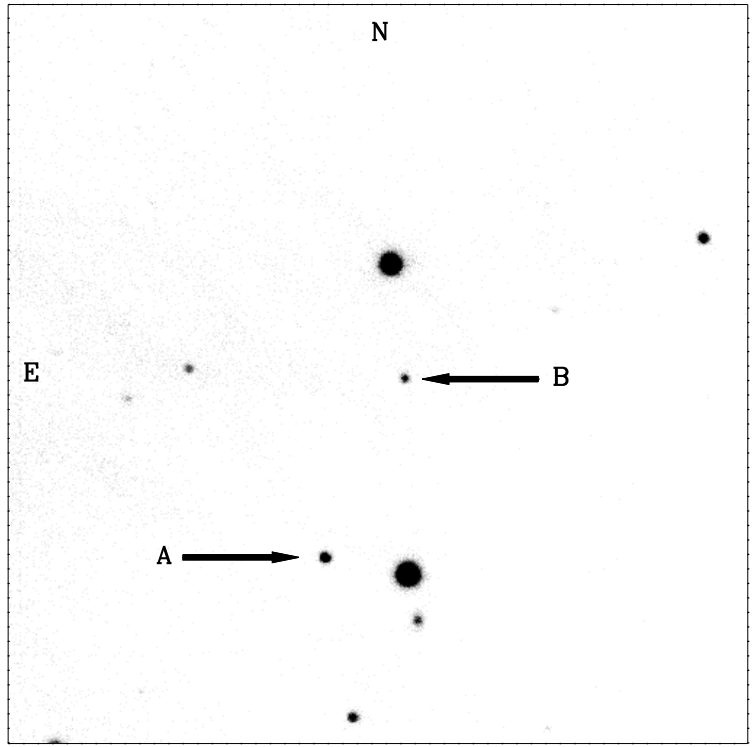}
\caption{Optical finding chart for PG 1017+125B, taken at $I$ band with
the Nickel 1 meter telescope in January 2003.  The image is $184''$ square
with $0.36''$ pixels.  The coordinates for the companion are $10^{\rm h}
19^{\rm m} 54.6^{\rm s}$, $+12^{\circ} 17' 18''$ J2000.
\label{fig17}}
\end{figure}

\begin{figure} 
\plotone{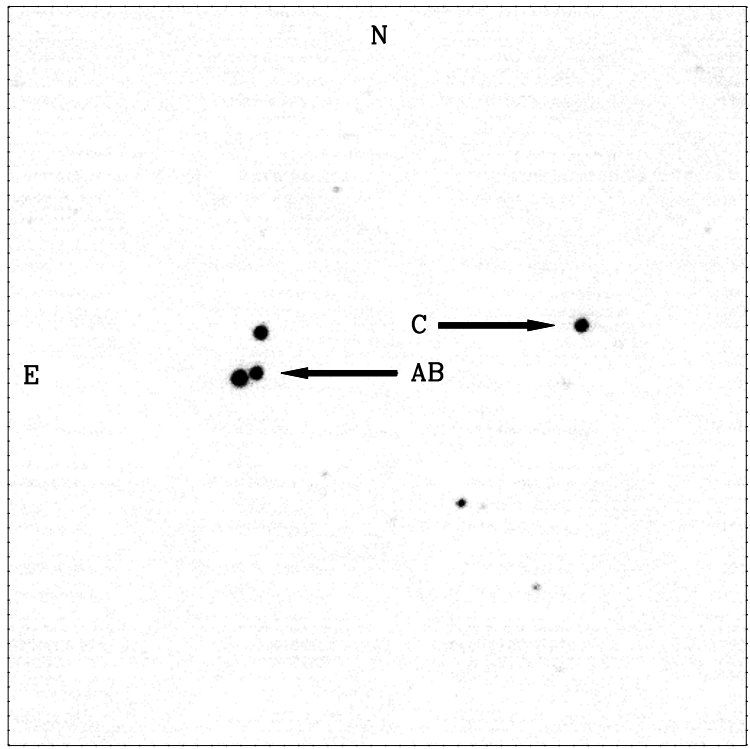}
\caption{Optical finding chart for PG 1204+450C, taken at $I$ band with
the Nickel 1 meter telescope in April 2003.  The image is $184''$ square
with $0.36''$ pixels.  The coordinates for the companion are $12^{\rm h}
06^{\rm m} 39.8^{\rm s}$, $+44^{\circ} 50' 09''$ J2000.
\label{fig18}}
\end{figure}

\begin{figure} 
\plotone{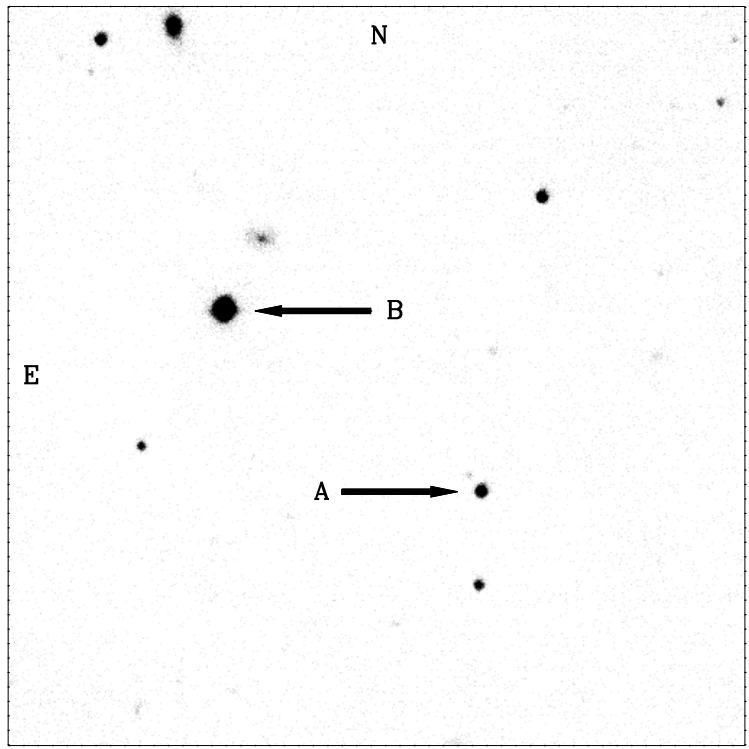}
\caption{Optical finding chart for PG 1449+168B, taken at $I$ band with
the Nickel 1 meter telescope in April 2003.  The image is $184''$ square
with $0.36''$ pixels.  The coordinates for the companion are $14^{\rm h}
52^{\rm m} 16.1^{\rm s}$, $+16^{\circ} 38' 48''$ J2000.
\label{fig19}}
\end{figure}

\begin{figure} 
\plotone{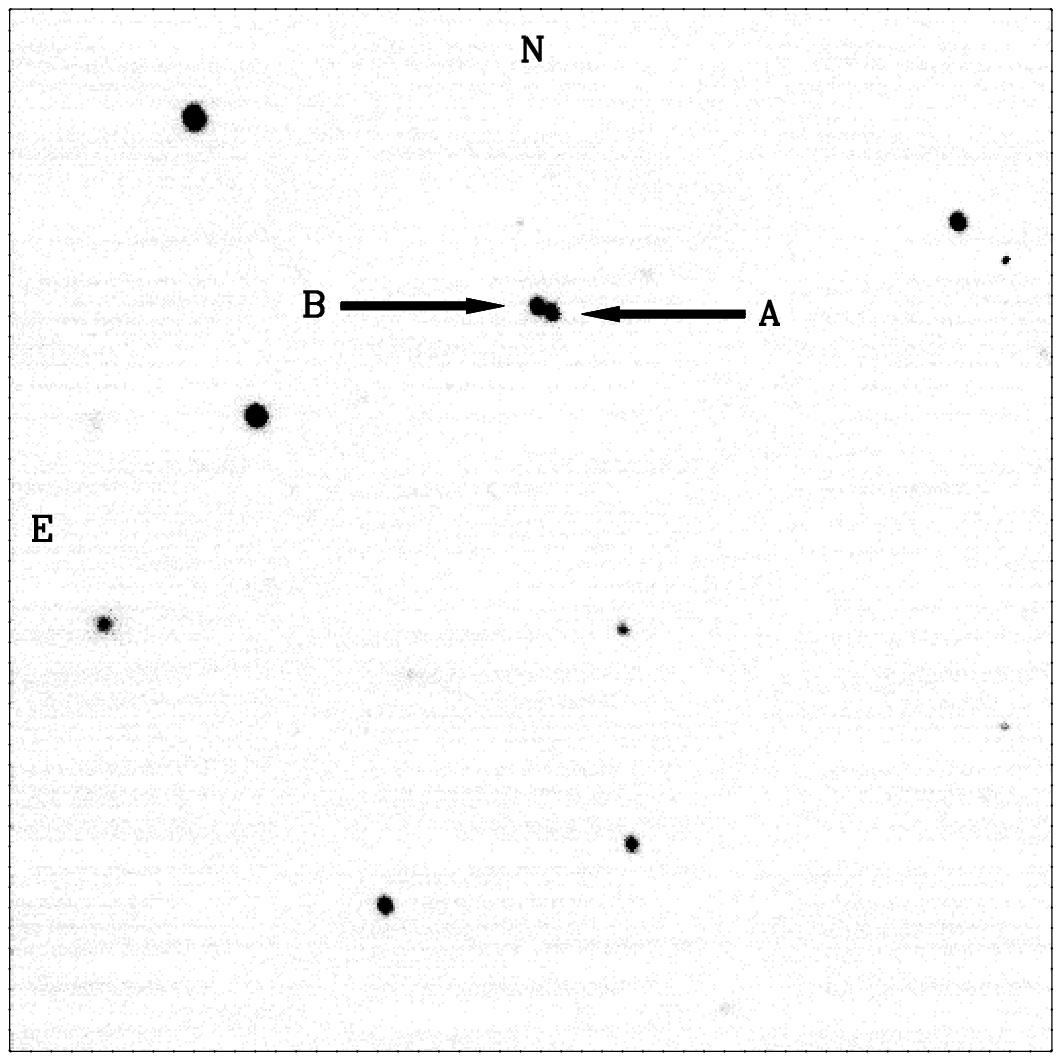}
\caption{Optical finding chart for PG 1539+530B, taken at $V$ band with
the Nickel 1 meter telescope in April 2003.  The image is $184''$ square
with $0.36''$ pixels.
\label{fig20}}
\end{figure}

\begin{figure} 
\plotone{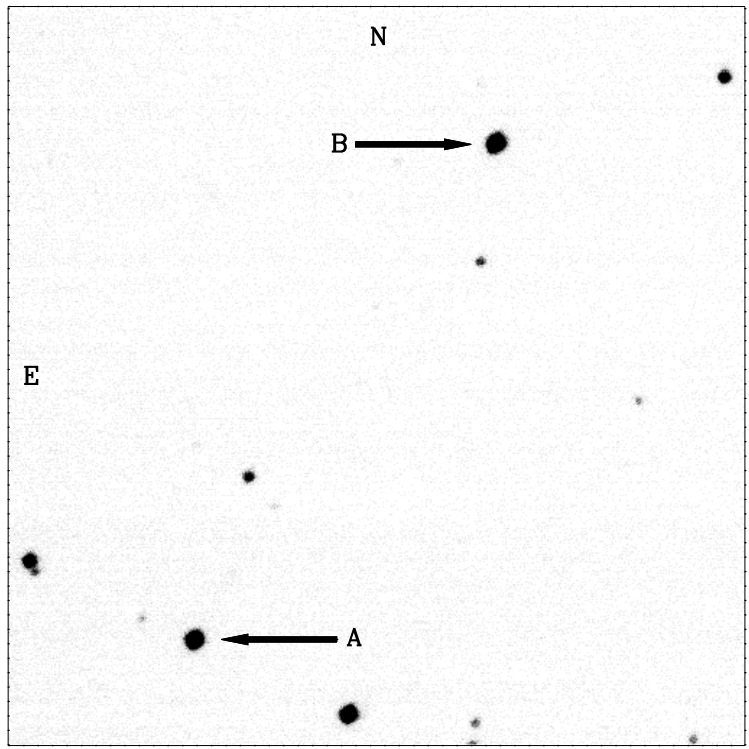}
\caption{Optical finding chart for PG 1659+303B, taken at $V$ band with
the Nickel 1 meter telescope in April 2003.  The image is $184''$ square
with $0.36''$ pixels.  The coordinates for the companion are $17^{\rm h}
01^{\rm m} 02.3^{\rm s}$, $+30^{\circ} 17' 45''$ J2000.
\label{fig21}}
\end{figure}

\begin{figure} 
\plotone{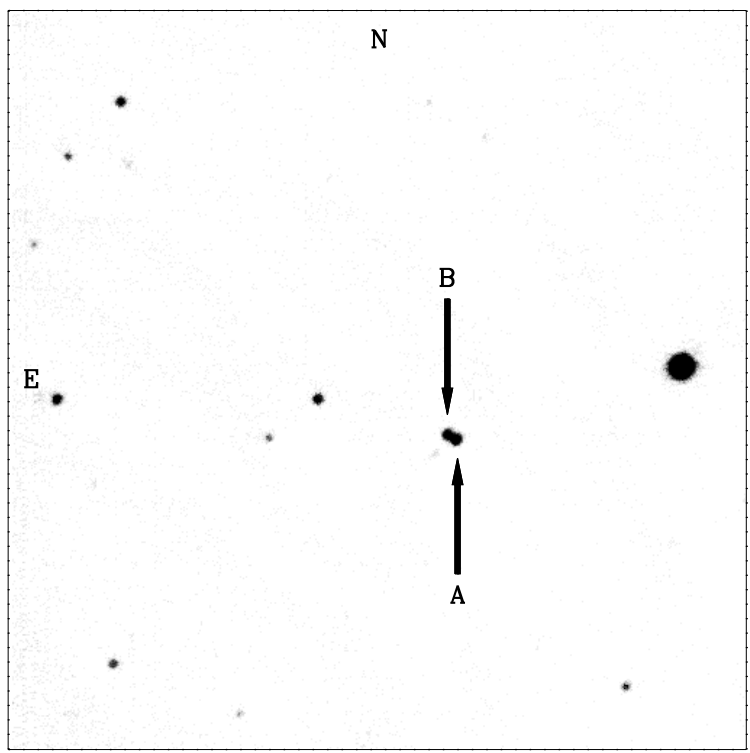}
\caption{Optical finding chart for PG 2244+031B, taken at $I$ band with
the Nickel 1 meter telescope in July 2003.  The image is $184''$ square
with $0.36''$ pixels.
\label{fig22}}
\end{figure}

Figures \ref{fig4}--\ref{fig23} are finding charts
for companions reported here for the first time, including
candidate companions.  In the case of PG 1619+123, its
newly identified common proper motion companion, HD 147528,
is already known and hence no chart is provided here.  The
objects GD 392B, LDS 826C, \& PG 0922+162B were discovered
independently in the course of this survey but are previously
published with finding charts \citep{fin97b,sch04,far04}.  
GD 559B (Figure \ref{fig11}) is reported only in \citet{mcc99}
with no other available journal reference.

For ease of use at the telescope, the charts are given
at optical wavelengths when possible.  In a few cases, 
the quality of a near-infrared image is superior and used
instead.  Generally, these are $\sim3'$ square field of
view CCD or near-infrared array images taken at Lick 
Observatory or Steward Observatory.  Coordinates are
given for the companion if: (1) it is separated from
the white dwarf primary by more than $20''$, (2) 
coordinates in the literature are inaccurate or 
difficult to find, (3) a finder chart is not 
published or difficult to find.

Table \ref{tbl-4} lists measured proper motions 
for all confirmed and candidate common proper motion 
pairs discovered in this work.  These values are the
result of the mapping process discussed in \S2.1.  
The map residuals were generally $\sim0.01''\ 
{\rm{yr}}^{-1}$, but never greater than $0.02''\
{\rm{yr}}^{-1}$.

An important item to note is that the uncertainty
in the measured proper motions is not a total measurement 
error.  This is because there was no independent astrometric 
calibration apart from the point sources in each individual 
mapped field.  In essence only {\em relative proper motions} 
were measured, not the absolute positions of the stars.  The
uncertainties reported by GEOMAP are the root mean square of
the map residuals and do not take into account the following 
factors: (1) any overall nonzero motion of objects in the map,
(2) the number of objects used in the map, (3) the S/N for
individual point sources and their measured coordinate centroids
in the map.  The fields used to measure proper motions were 
between $166''$ and $300''$ in size, hence the number of field
stars was limited, especially at higher galactic latitudes.
Saturated stars are also unreliable because they can skew the
centroiding process, as are faint field stars due to low S/N.
Therefore, the Table \ref{tbl-4} measurements should be
considered of only limited accuracy.  This is also the main
reason that a few candidate companions have been retained for
further investigation despite apparently discrepant measured
proper motions (\S6).

Ironically, there was only a single common proper motion
companion detected solely in the near-infrared and not also in 
the optical.  All other pairs were essentially detectable by
``blinking'' the first and second epoch Digitized Sky Survey
scans (e.g. in the northern hemisphere, the first and second 
epoch Palomar Observatory Sky Survey plates).  These digitized 
scans were also used to measure proper motions when possible
because the longer time baselines provide higher accuracy and 
the ability to measure smaller proper motions.

There are 3 new and 11 previously known visual 
binaries $(a<10'')$ studied here for which no proper 
motion measurement was made.  In a few cases, there 
exists insufficient time baseline between available 
images in which the pair is resolved to measure proper 
motions, or the data available in proper motion catalogs 
are for the composite pair (whether unresolved or extended)
or absent.  But for most, the companionship of the visual
pair is highly probable due to one or more of the following:
(1) an unchanging visual separation and position angle or 
elongation axis between the pair over $15-50$ years (this 
is essentially equivalent to a common proper motion 
determination because most if not all of these pairs 
have $\mu>0.05''\ {\rm{yr}}^{-1}$ and can be clearly
seen moving with respect to background stars by ``blinking''
two DSS epochs), (2) the common photometric distance implied
by the spectral energy distributions of both components 
together with the statistical likelihood of companionship 
based on proximity in the sky, (3) spectroscopic evidence 
presented here or elsewhere, (4) astrometric evidence 
presented elsewhere.

\subsection{Known Companions}

The majority of the objects in \citet{zuc92}
and \citet{sch96} are unresolved white dwarf plus red 
dwarf binaries.  As discussed in \S4.2, the parameters of 
the two components must be deconvolved from one another.
Most of these binaries were investigated with a thorough
and updated literature search, optical photometry and 
spectroscopy to both confirm the identity and further 
constrain the properties of the low mass companion.
This resulted in a higher confidence in their spectral 
classifications.  For those partially or completely
resolved pairs previously reported, all with
$1''<a<9''$, \S4.3 applies.

Generally speaking, the updated analysis of
known low mass stellar companions has shown they
have spectral types which are earlier than previous
estimates.  This is because early M dwarfs can
contribute a significant amount of flux at optical
wavelengths and cause a white dwarf to appear redder 
(in $B-V$ for example) and more luminous (at $V$ for 
example) than it would as a solitary star.  The 
effective temperature inferred for the white dwarf
will be too low, and the inferred distance modulus
will be too close (because $M_V$ will be too dim
and $V$ will be too bright; \citealt{far04c}).

\subsection{Photometry}

Circular aperture photometry was used to determine
instrumental fluxes and magnitudes for all unresolved
and resolved binary stars in this work (\S2.3).  
Comparison with one or more standard stars yielded
the true magnitudes listed in Tables \ref{tbl-5} \& 
\ref{tbl-6}.  $BVRI$ photometry is on the Johnson-Cousins
system and $JHK$ photometry is on the Johnson-Glass system,
collectively known as the Johnson-Cousins-Glass system 
\citep{bes88,bes90b}.

For binary pairs that were spatially well resolved 
from each other $(a>3'')$ and from neighboring stars,
the flux measurement error was generally 5\% or less
for $m<m_c$, where $m_c\approx(19,18,17)$ mag for $(BVRI,JH,K)$,
and $\sim10$\% or greater otherwise.  For separations smaller
than $\sim3''$ between target star and neighbor or companion, 
overlapping point spread functions (PSFs) effectively
contaminate flux measurements even in small apertures 
of $1-2$ pixels in radii.  In these cases, the IRAF program
DAOPHOT was used to simultaneously fit two or more PSFs
within a given area, deconvolve and extract their individual
fluxes.  This method works quite well for pairs with 
$\Delta m\approx3$ mag or less and generates errors equivalent
to those quoted above.  Table \ref{tbl-5} lists all photometry
for resolved binary components, including those stars 
requiring PSF deconvolution from neighbors or companions.

Close binaries consisting of a white dwarf plus red dwarf
which were indistinguishable from a single point source were
treated as a single star and aperture photometry performed
accordingly.  This is true also for those pairs with separations
$(a<2'')$ too small to be accurately fit with two PSFs due to
pixel scale, seeing conditions, and/or $\Delta m>3$ mag.

In Table \ref{tbl-6} are the measured optical and near
infrared magnitudes for all composite binaries.  For each
system, the table has three entries.  The first line is the 
composite photometry itself, with all measurement errors for
this work being 5\% or less in this range of magnitudes. The
second line gives the predicted magnitudes for the white 
dwarf (WD) component based on the most current hydrogen and
helium atmosphere model grids of P. Bergeron (2002, private 
communication), which are considered more accurate than previous
generations \citep{ber95,ber95c}.  The predicted white dwarf 
magnitudes are calculated by adding model colors (appropriate
for its $T_{\rm{eff}}$, and log $g$ if known) to a photometric
bandpass that is essentially uncontaminated by its cool red 
dwarf companion -- either $U$ or $B$ (or $V$ in a few rare
cases).  If the calculation was done from $U$, a reference
is given for the photometry.  The temperature and surface
gravity used as input for the models are taken from the most
reliable sources available with the reference provided.
The third line gives the deconvolved magnitudes for the
red dwarf (RD), generally only $IJHK$ due to
large uncertainties at shorter wavelengths.

Based on comparisons with \citet{kir94}, spectral types 
were estimated from $I-K$.  Unlike $J-K$, which is highly
degenerate across most of the M type dwarf spectral class,
$I-K$ is essentially monotonically increasing from M2 until
well into the L spectral class \citep{kir94,kir99b}.

Some binary systems lack data at one or multiple 
wavelengths due to unavoidable circumstances including,
but not limited to: time constraints, poor weather, 
instrument problems, telescope pointing limits, and 
telescope size.

\subsection{Spectroscopy}

The purpose of the spectroscopy was
to identify the spectral class of companions.  
Standard stars and spectral flats were taken to
ensure the target spectra were free of both detector
and instrument response.  None of the spectra were 
corrected for telluric features or extinction.  In 
the case of white dwarfs, the purpose was to look 
for the presence or absence of highly pressure broadened 
hydrogen or helium lines (implying spectral classes DA, 
DB, or DC for no lines).  In the case of M dwarfs, the
search was for the characteristic TiO and CaH bands.

The Kast Dual Spectrograph sits atop Mount Hamilton, 
which is close to the city lights of San Jose. Sodium
at 5880 \AA \ can be seen very brightly in Kast spectra
and can be difficult or impossible to remove completely
in low S/N observations.  Hence positive and negative residuals
often remain.  During one observing run with the Kast, the red
side of the spectrograph was used without any dichroic or blocking
filter on the blue side.  Hence second order blue light was 
present in the red spectra of all objects excepting very red 
objects such as single M dwarfs.  This effect was mostly,
but not completely, removed by calibrating with a standard star observed in 
the same arrangement.

The one spectroscopic observing run with the Boller \&
Chivens Spectrograph at Steward Observatory was over 3
nights with a very bright moon.  The solar spectrum reflected 
from the moon can be seen quite brightly over the entire chip
and was generally stable and removable.  However, the regions
around the Balmer lines were problematic in a few instances and
some lower S/N spectra still contain residuals in the region
around H$\alpha$, H$\beta$, and H$\gamma$.

A few miscellaneous stars were observed
with LRIS \citep{oke95} at the Keck telescope because they
were considered important yet too faint to obtain reliable
spectra with a 3 meter class telescope.  The observations
were kindly performed by colleagues at other institutions.
In a few cases, calibration stars were not observed and
hence the flux calibration is not perfect and had to
be adjusted as best possible.

In several of the binary systems reported here, the primary
white dwarf is poorly documented in the literature, or missing
all together.  Spectra are displayed in Figures \ref{fig24}--\ref{fig31}
in order of decreasing temperature for those stars which do not have
published spectra or are misclassified or missing from the literature,
for new white dwarf identifications, and for those systems where
binarity or other issues have precluded proper analysis.  Also shown
are the optical spectra of all unpublished white dwarf wide binary
companions discovered uniquely in this study.

Figures \ref{fig32}--\ref{fig44} present the spectra of
unpublished resolved M dwarf secondaries and tertiaries,
displayed in order of decreasing temperature.  Figures
\ref{fig45}--\ref{fig53} contain the composite spectra of white
dwarf plus red dwarf pairs (at least one of which has been resolved
photometrically but not spectroscopically), displayed in order
of decreasing red dwarf temperature.

\section{ANALYSIS}

\subsection{Companion Spectral Type Frequency}

In Figure \ref{fig54} is plotted the number of
unevolved low mass companions versus spectral type
for objects studied in this work.  Despite excellent
sensitivity to late M dwarfs and early L dwarfs in all
survey phases, very few were detected. 

For comparison, Figure \ref{fig55} shows similar
statistics for cool field dwarfs within 20 pc of
Earth taken from \citet{rei00,cru03}.  The data 
plotted in Figure \ref{fig55} have been corrected
for volume, sky coverage, and estimated completeness.
Can one reconcile Figure \ref{fig55} with the common
notion that there are at least as many brown dwarfs
as low mass stars \citep{rei99}?  To resolve this 
possible discrepancy, most field brown dwarfs would
have to be of spectral type T or later, since it is
clear from the figure that, in the field, L dwarfs
are much less common than stars.

However, there are several things to keep in mind
regarding the relative number of field brown dwarfs
versus stars.  There should be be a relative dearth
of L dwarfs compared to T type and cooler brown dwarfs
in the field because cooling brown dwarfs pass through
the L dwarf stage relatively rapidly.  The lower end of
the substellar mass function is poorly constrained at 
present \citep{bur04} and the relative number of substellar
objects versus low mass stars in the field depends on the
shape of the mass function in addition to the unknown
minimum mass for the formation of self-gravitating substellar objects
\citep{low76,rei99,bur04}.  Furthermore, even for only moderately
rising mass functions, such as those measured for substellar
objects in open clusters \citep{hil00,luh00,ham99,bou98},
there will be more brown dwarfs than stars if the minimum
formed, self-gravitating substellar mass is $<0.010$ $M_{\odot}$.
Ongoing and future measurements of the local T dwarf
space density will constrain the substellar field mass
function.

Figures \ref{fig54} \& \ref{fig55} are quite
similar.  Clearly, the peak frequency in spectral
type occurs around M3.5 for both field dwarfs and
companions to white dwarfs.  In fact, the peak is
identical; 25.6\% for both populations.  By itself,
this could imply a common formation mechanism, a
companion mass function similar to the field mass
function in this mass range.  But, relative to the
peak, there are $\sim2-3$ times more L dwarfs and 
$\sim4-5$ times more M6$-$M9 dwarfs in the field
than companions.  For the T dwarf regime, uncertainty
remains because only the Keck portion of the white 
dwarf survey was sensitive to such cool brown dwarfs
(and only for certain separations) plus the current
incomplete determination of the field population density.

Hence, binary systems with small mass ratios 
$(q=M_2/M_1<0.05)$ are rare for white dwarf
progenitors (which typically have main sequence
masses $\sim2$ $M_{\odot}$).  Although there exists
some speculation regarding the possibility that brown
dwarfs are ejected in the early stages of multiple
system or cluster formation, there is currently no
evidence of this occurring.  It is conceivable that
low mass companions in very wide orbits may be lost
to gravitational encounters in the Galactic disk
over a few billion years, but given the fact that
there are a dozen or so known L and T dwarfs in wide
binaries, this seems like a rare mechanism, if it 
occurs at all.

It is possible that very low mass companions to
intermediate mass stars undergo major or catastrophic
alteration during the red giant or asymptotic giant branch
(AGB).  It has been calculated that there is a critical mass,
$(M_c\sim0.02$ $M_{\odot})$, below which low mass companions are
completely evaporated or cannibalized within the AGB envelope
\citep{liv84,ibe93}.  Above $M_c$ companions may accrete a 
significant amount of material during inspiral, perhaps enough
to transform into low mass stars \citep{liv84}.  It is not yet
empirically known whether any of these scenarios actually occur
in nature.  If they do, then the possibility of secondary
evaporation should be less likely for the companion mass range
in question here $(M>0.04$ $M_{\odot})$, but it is not certain.
In \S5.4, the secondary spectral types in binaries which may have
experienced a common envelope phase will be compared to those
which did not.

In a way, the relative dearth of late M dwarfs alleviates a
potential interpretation problem.  Had it been the case that
many late M dwarfs were detected but only one or two L dwarfs,
it might have been argued that the L dwarfs were cooling beyond
the sensitivity of the search.  Since all M dwarfs (and the first
few L dwarf subclasses) at $\tau\geq1$ Gyr are stellar according
to theory, this concern does not exist.  The measured dearth
is real and is not caused by brown dwarf cooling and the
resulting lower sensitivity.

\subsection{The Companion Mass Function}

Dynamical masses measurements do not exist for
any of the companions described in this work.  There
are a few systems -- close white dwarf plus red dwarf
spectroscopic binaries -- whose secondary masses have been 
estimated \citep{saf93,marsh96,max98}.  This is not a mass
measurement as it ultimately relies on models, and what is
really measured in these systems is the mass ratio (hence
the need for a white dwarf mass from models).  But this
method has been used successfully to estimate red dwarf
masses that are consistent with both theory and existing 
dynamical mass measurements for low mass stars in the same
range of spectral types, temperatures, and ages.

For M spectral types, the works of \citet{kir91,hen93,
kir94,dah02} contain: (1) absolute magnitudes as a function 
of spectral type, (2) mass versus luminosity relations, (3)
spectral type as a function of mass based on all available 
dynamical measurements of very low mass stars.  These
empirical and semiempirical relationships are for disk stars
of intermediate age, which is appropriate for the sample of
white dwarfs in this work.  These relations have been used
to provide masses for spectral types M1 through M9.  It is
unnecessary to extrapolate these empirical relations into 
the L dwarf regime, because only two white dwarf plus L dwarf
systems are known and both companions have published mass
estimates from models, based on likely age ranges
\citep{kir99a,far04b}.

Figure \ref{fig56} shows the first step in
the construction process -- mass versus $K$ band
luminosity relations from model, empirical and
semiempirical relations.  The models used are from 
\citet{cha00} and show tracks for ages of 1 and 5 Gyr, 
appropriate for young to intermediate disk ages.  The
minimum mass for hydrogen burning (HBMM) in these models
is $M_{\rm{HBMM}}=0.072$ $M_{\odot}$.  The track for
5 Gyr turns downward (relative to the track for 1 Gyr) 
before the stellar/substellar boundary because the lowest 
mass stars are still contracting onto the main sequence
\citep{bur97,burr01,cha00}.  Below the HBMM, the downturn
is the result of brown dwarf cooling.

For the ages appropriate here, dynamical masses have been
measured down to spectral type M6 $(M=0.10$ $M_{\odot}$;
\citealt{kir94}), but none later.  Hence the empirical 
relation below this spectral type and corresponding mass
is really semiempirical.  Adjustments had to be made
according to the progress in this field over the past 
decade.  For example, an extrapolation of the strictly
empirical  relation predicts a clearly substellar mass of
0.066 $M_{\odot}$ at spectral type M9.  This is not
currently accepted as correct for intermediate disk
ages \citep{bur97,cha00}.

Figure \ref{fig57} plots the absolute $K$ magnitude
versus spectral type for all the low mass stellar and 
substellar companions discovered in this work.  Also
plotted in the same figure is the combined empirical 
relation of \citet{kir94} and \citet{dah02}, both based
on trigonometric parallax measurements.  This figure
demonstrates the possibility of inaccurate distance estimates
for many of the white dwarf primaries and is the major reason
why absolute magnitude was not used as a proximate for mass
in this work.  Unlike previous work \citep{zuc92} and similar
studies \citep{gre00} -- both of which employed $M_K$ as an
indicator of spectral class -- the present study uses spectral
class itself.  The reasons for this are twofold.  First, in many 
cases photometric distances for white dwarfs are inaccurate 
for a variety of reasons.  For example, because white dwarfs
have varying radii at a given effective temperature, their
distances cannot be estimated with
as much confidence as main sequence stars.  Binarity can also
cause a white dwarf to have an erroneous distance estimate.
\citet{sma03} found that, for a sample of six single white
dwarfs, in general, the published photometric distance is
an overestimate of the distance found by trigonometric
parallax.  Second,
$M_K$ is a proximate for luminosity, not for temperature.  Color
and  spectral type are temperature indicators and do not require
a precise distance determination.  For main sequence stars, the temperature 
can be used with an HR diagram (i.e. an empirical radius 
versus temperature relation) to calculate a mass.  This 
is, in essence, what has been done in the present work.

With perhaps one or two exceptions discussed in the
Appendix, all of the Table \ref{tbl-3} these stars have published spectra (in this
work or elsewhere) which are consistent with solar metallicity.  
Therefore the combined correlation between absolute magnitude
and spectral type of \citet{kir94} and \citet{dah02}, for 
low mass field stars of intermediate disk age and solar
metallicity, will suffice to confidently predict secondary
masses.

The final step is to combine the empirical and semiempirical
relations of Figures \ref{fig56} \& \ref{fig57} into Figure
\ref{fig58}, which shows the resulting correlation between
spectral type and mass.  Figure \ref{fig59} is a histogram of
the number of detections versus companion mass, using the
correlation data in Figure \ref{fig58}.

\subsection{Sensitivities \& Biases}

When considering the overall survey mass sensitivity, for
low mass stellar companions, age is not an issue.  But for
substellar objects, determination of mass sensitivity must
include an age estimate.  In \S3.3 a likely age range for
the sample was estimated to be $2-5$ Gyr based on the overall
kinematics and cooling ages, but owing to model grid availability
and to avoid additional interpolation errors, calculations were
performed for ages of 1 and 5 Gyr.

The average distance for the sample, calculated to be 57
pc in Table \ref{tbl-2}, along with the models of \citet{cha00},
was used for determining overall mass completeness.  Obviously, the
sensitivity differs for objects which are closer or farther and the
standard deviation of the entire sample is significant at $\sigma_d=47$
pc.  Yet the masses, temperatures and spectral classes implied by the
completeness limits for each phase of the survey (at the average 
distance of the sample stars) epitomize what was detectable.
Table \ref{tbl-7} summarizes these completeness limits.

\citet{zuc92} report a completeness down to $K=15$ mag.  
However, this completeness was limited by single detectors prior 
to the availability of near-infrared cameras \citep{zuc87}.  All of
the objects in Table \ref{tbl-1} that were observed in the IRTF 
survey were imaged with arrays.  For spatially resolved objects,
a conservative completeness limit for these observations is $K=16$ mag.  
Applying this limit at $d=57$ pc for the 84 white dwarfs observed in this 
early part of the survey (but not reobserved at Steward or Keck), these
observations were complete to $M_K=12.2$ mag.  This corresponds to a
spectral type near L6, $T_{\rm{eff}}\sim1650$ K, and $M=0.060-0.070$
$M_{\odot}$ for $1-5$ Gyr \citep{rei99,kir00,dah02,cha00,vrb04}.  GD
165B was discovered amongst the first observations in the program at
$K=14.2$ mag and has a mass estimated at $M=0.072$ $M_{\odot}$
\citep{kir99a}.

The Steward survey of 261 white dwarfs was complete to $J=18$ mag,
implying a completeness down to $M_J=14.2$ mag at $d=57$ pc.  This
corresponds to a spectral type around L7, $T_{\rm{eff}}\sim1500$ K,
and $M=0.053-0.068$ $M_{\odot}$ for $1-5$ Gyr \citep{rei99,kir00,
dah02,cha00}.  The Keck survey of 86 white dwarfs was complete to
$J=21$ mag, implying completeness to $M_J=17.2$ mag at $d=57$ pc.
This corresponds to spectral types  later than T8, $T_{\rm{eff}}<750$
K, and $M\sim0.020-0.040$ $M_{\odot}$ for $1-5$ Gyr \citep{vrb04,
leg02,cha00}.

As mentioned in \S2.5, all 371 sample stars were searched for
near-infrared excess emission between $1-2$ $\mu$m using the 2MASS all
sky catalog database \citep{cut03}.  2MASS provides a highly accurate,
uniform and consistent method for this type of search.  Higher sensitivity
to unresolved companions was not gained at Keck or Steward for two
reasons: (1) the white dwarf was sometimes saturated in the attempt
to image faint companions, especially at Keck, and (2) near-infrared
excess detection requires photometric accuracy, not deep imaging.

The average temperature of a white dwarf in the sample is 
$T_{\rm{eff}}=13,000$ K.  This yields $M_H=11.8$ mag, 
$M_K=11.9$ mag for a white dwarf of typical mass (log $g=8.0$,
\citealt{ber95}).  The 2MASS all sky catalog provides reliable
photometry (${\rm{S/N}}>10$) down to $H=15.1$
mag and $K_s=14.3$ mag for 100\% of the sky and to $H=15.6$ mag
and $K_s=14.8$ mag for 50\% of the sky \citep{cut03}.

Taking the average of these $H$ \& $K_s$ limiting magnitudes
at 57 pc, an excess of 170\% above the white flux would be 
detectable at $M_K=10.8$ mag ($K_s=14.5$ mag).  This yields 
$M_K=11.3$ mag for a cool companion, which is around spectral
type L4.  However, the sensitivity to excess emission is much
greater at $H$.  At 57 pc, an excess of 21\% is detectable at
$M_H=11.6$ mag ($H=15.3$ mag), which yields $M_H=13.5$ mag for
a low mass companion.  This corresponds to spectral type L8,
$T_{\rm{eff}}\sim1400$ K, and $M=0.050-0.066$ $M_{\odot}$ 
for $1-5$ Gyr \citep{rei99,kir00,dah02,cha00}.

The case of GD 1400B proves this point.  Estimated at
spectral type L6, it was detected using the 2MASS database
in a manner identical to that performed for all 371 stars in
the entire sample.  Not included in the sample, GD 1400B was first
identified by \citet{wac03} in the initial phase of a search
utilizing the 2MASS point source catalog to survey the entire
sky near the positions of all white dwarfs in \citet{mcc99}.
\citet{far04b} were the first to distinguish GD 1400 from the
bulk of white dwarfs with near-infrared excess emission and
identify its companion.

None of the searches were sensitive to companions beyond
the field of view of the corresponding cameras.  For the IRTF,
only objects within $\sim12''$ of the white dwarf were detectable.
The NIRC and Steward near-infrared camera fields of view are
$19.2''$ and $83.2''$ from center to edge, respectively.  This
implies for targets at $<d>\pm \ \sigma_d$, separations out
to $4700\pm3900$ AU were probed by the Steward survey, but
only out to $1100\pm900$ AU for the Keck survey.  Although
wider binaries may have been missed in the Keck and IRTF
searches, they would have been picked up by the DSS blinks
(\S2.6), unless they were spectral type L or later, roughly
speaking.

Generally speaking, M and L dwarfs were detectable at 
arbitrarily close physical separations (\S4.2).  However, T
dwarfs are generally not detectable by near-infrared excess unless 
the white dwarf is quite cool or quite massive ($T_{\rm{eff}}<
7000$ K for log $g=8.0$, or $T_{\rm{eff}}<9000$ K for log 
$g=8.5$).  There were few stars in the sample meeting this
criteria and therefore T dwarfs were only detectable if resolved;
at Keck this required a separation on the sky of $\sim1''$ (with
typical Mauna Kea seeing) and at Steward $\sim2''$ (typical Kitt
Peak seeing).  However, no T dwarfs were detected.

On the brighter side, unresolved dwarf stellar companions
earlier than around M1 ($M_V=9.3$ mag) were almost certainly 
selected against.  Depending on the luminosity of the white dwarf,
it is possible for a G$-$K dwarf or even an M0 dwarf to mask the 
presence of a nearby degenerate at optical wavelengths.  These 
types of binaries are likely selected against in surveys which 
identify and catalog nearby white dwarfs.  This explains in part
the drop off at the higher mass end of Figures \ref{fig54} \&
\ref{fig59}.  However, the study was not biased against wide
yellow dwarf companions and two of the white dwarfs in the
sample were found to have such secondaries.

\subsection{Current Mass Versus Initial Mass}

In order to measure the initial mass function for companions
to intermediate mass stars, a critical question remains: are 
the red dwarf masses observed today the same as the initial 
masses when the binary was formed?

When intermediate mass main sequence stars, the progenitors
of white dwarfs, ascend the asymptotic giant branch, their 
outer regions expand by a factor of a few hundred.  If such a
star has a relatively close main sequence companion within this
region ($a\la1-2$ AU), the pair is said to share a common envelope.
Generally speaking, such a close binary pair will transfer much of
its orbital energy (via angular momentum) into the common envelope
through friction, resulting in ejection of the envelope from the
system and an  inspiral to a more negative binary binding energy
and smaller separation \citep{pac76}.

It has been theorized that a low mass companion may
accrete  up to 70\% of its final mass during a common
envelope phase or may evaporate completely during the
inspiral, depending on the initial masses and separations
of both components \citep{liv84}.  Hence, there is a
possibility that the masses of red dwarf secondaries in
close binaries are not their initial masses.  Unforunately,
there is no consensus on the topic.  There appears to be
evidence in support of the idea that secondaries do accrete
a significant amount of mass during the common envelope phase
\citep{dra03}.  Yet there also appears to be evidence that
low mass companions emerge unaffected \citep{max98}.

Table \ref{tbl-8} presents the median spectral types in
this survey for several different low mass companion subgroups.
The ``resolved'' subgroup refers to all resolved binary companions.
The ``close'' subgroup is all known radial velocity binary companions
-- these are the post common envelope binaries.  The ``unresolved /
not close'' subgroup is all unresolved binary companions that are not
known to be radial velocity variables.  Hence this subgroup contains
an unknown number of post common envelope binaries.  The fourth subgroup
supposes that subgroup three binaries are all wide ($a>5$ AU), whereas
the fifth subgroup supposes that subgroup three binaries are all close
($a<0.1$ AU).  Do the Table \ref{tbl-8} values provide any evidence that
post common envelope binary secondaries have accreted a significant amount
of mass (correlated here with spectral type)?  If so, it is not
obvious (the interested reader is referred to Farihi 2004b for
further discussion).

\section{CONCLUSIONS}

Together, the various phases of this survey discovered over 40 
previously unrecognized white dwarf binary and multiple systems.
The wide field, common proper motion survey alone discovered at
least 20 new white dwarf multiple systems.  Based on the analysis
of \S5.4, there is no reason why all unevolved companion stars
should not be included in the initial mass function.

It is worthwhile to mention that none of the 371 survey stars
were found to have near-infrared excess similar to G29-38 and
GD 362 -- the only two single white dwarfs known to be orbited
by circumstellar dust \citep{zuc87,bec05,kil05}.

Before making an estimate of the substellar companion
fraction, GD 1400B must be discussed.  Although not included
in the sample, it is an important datum in the overall
statistics of low mass companions to white dwarfs -- a long,
hard, and much sought after datum (16 years passed between the
discoveries of the first and second L dwarf companions to white
dwarfs).  GD 1400 is a white dwarf not unlike the sample white
dwarfs, with $T_{\rm{eff}}=11,600$ K and a moderate proper motion
of $\mu\approx0.05''$ ${\rm{yr}}^{-1}$ \citep{koe01,fon03,zac04}.
Hence its inclusion here is perfectly consistent with the sample
stars, the search methods and resultant sensitivities (\S5.3).
Due to these consistencies and the fact that GD 1400B is a vital
statistic, it has been included in the analyses and conclusions.

The calculated fraction of white dwarfs with substellar
companions, within the range of masses and separations to which
this work was sensitive, is $f_{bd}=0.4\pm0.1$\%.  This represents
the first measurement of the low mass tail of the companion mass
function for intermediate mass stars, main sequence A and F stars
(plus relatively few B stars) with masses in the range 1.2 $M_{\odot}$
$<M<8$ $M_{\odot}$.  This value is consistent with similar searches 
around solar type main sequence stars for comparable sensitivities 
in mass and separation \citep{opp01,mcc04}.  Therefore that the
process of star formation eschews the production of binaries with
$M_2/M_1<0.05$ is clear from the relative dearth of both L and
late M dwarfs discovered in this work.

\acknowledgments

The authors wish especially to thank Steward Observatory for
the use of their facilities over the years -- this research
would not have been possible without their munificent support.
J. Farihi thanks C. McCarthy for numerous useful discussions
and very generous computing support over the years.  P. Bergeron
kindly provided his models for use throughout this paper.  Part
of the data presented herein were obtained at Keck Observatory,
which is operated as a scientific partnership among the 
California Institute of Technology (CIT), the University of
California  and the National Aeronautics and Space Administration
(NASA).  This publication makes use of data acquired at the NASA 
Infrared Telescope Facility, which is operated by the University
of Hawaii under Cooperative Agreement no. NCC 5-538 with NASA,
Office of Space Science, Planetary Astronomy Program.  Some data
used in this paper are part of the Two Micron All Sky Survey, a 
joint project of the University of Massachusetts and the Infrared
Processing and Analysis Center (IPAC)/CIT, funded by NASA and the
National Science Foundation (NSF).  2MASS data were retrieved from
the NASA/IPAC Infrared Science Archive, which is operated by the
Jet Propulsion Laboratory, CIT, under contract with NASA.  The
authors acknowledge the Space Telescope Science Institute for
use of the Digitized Sky Survey.  This research has been
supported in part by grants from NSF and NASA to UCLA.

\appendix

\section{INDIVIDUAL OBJECTS \& SYSTEMS}

In this section are discussions of: misclassified stars,
unrelated proper motion stars, candidate companions, and
noteworthy confirmed multiple systems.  Complete details
may be found in \citet{far04b}.  The following references
are used throughout this section.  For white dwarf model 
colors, absolute magnitudes, masses, surface gravity,
effective temperatures, and ages: \citet{ber95,ber95c};
P. Bergeron (2002, private communication).  For empirical 
M dwarf colors and absolute magnitudes: \citet{kir94,dah02}.

\subsection{Nondegenerates}

Some stars in Table \ref{tbl-1} are noted which are not
white dwarfs and for which no literature reference can be
found explicitly correcting the misclassification.  The
correct classification is presented here.  These are listed
as white dwarfs in the catalog of \citet{mcc99} and also in
the same current online catalog.

\subsubsection{G187-9}

G187-9 is classified as type DC in \citet{mcc99}, but it
was classified more or less correctly as early as 1967 
\citep{wag67}.  It has spectral type M2 and $M_V=11.27$ mag
\citep{rei95}.  A main sequence M2 star has $M_V=10.2$ mag and
hence G187-9 is subluminous, and by definition a subdwarf.  It
has a high proper motion of $\mu=0.7''$ ${\rm{yr}}^{-1}$.

\subsubsection{GD 617 \& PG 0009+191}

GD 617 is classified as type DAB5 \citep{gre84,mcc99} but
has been reclassified as a very hot subdwarf showing Balmer 
lines, neutral helium, and continuum flux from an unresolved
F/G star \citep{lam00}.  Photometry verifies the probability
of an unresolved main sequence companion.  It has $V-K=+0.8$,
which is too red for a single hot helium burning star.  Hence
GD 617 is very likely type sdB+F/G.

PG 0009+191 is classified as type DA \citep{gre86,mcc87} but
does not appear in \citet{mcc99}.  The absence hints at probable
nondegeneracy.  The spectrum of PG 0009+191 appears to be that
of a hot subdwarf, type sdB \citep{far04c}

\subsubsection{PG 1126+185 \& PG 0210+168}

PG 1126+185 is classified as type DC8 \citep{gre86,mcc99}.
It was reclassified as DC+G/K by \citet{put97}.  It is not
possible to detect a DC star (cool helium atmosphere or very
cool hydrogen atmosphere white dwarf) around a main sequence G
or K star, since the difference in brightness at $V$ would be 
at least 6 magnitudes.  Were it not for the blue continuum, the
spectrum in Figure 2 of \citet{put97} makes a great case for a
metal-poor nondegenerate star; weak lines of Mg, Fe, Na can be
seen in addition to Balmer lines and Ca H \& K.  Photometry done
here gives $V-K=1.88$, which is consistent with a G/K type star.
But PG 1126+185 also has $U-B=-1.12$, implying a very hot object
and confirming the steep blue continuum seen in its spectrum.  
\citet{sma03} have measured a zero parallax for PG 1126+185 
-- clearly inconsistent with a white dwarf, given its relative
brightness at $V=14.0$ mag.  

To explain both the blue continuum and the observed
absorption features of PG 1126+185, a composite system consisting 
of one hot star and one cool star is needed. The luminosities 
should be comparable given the spectrum.  A typical G star has
$M_V\approx5$ mag, but is there a hot star that has a comparable
absolute $V$ magnitude?  Subdwarf B stars have $M_V\approx4.5$ 
mag \citep{max00b}.  Therefore, the most likely explanation
for the spectrum of PG 1126+185 is a composite binary consisting
of an sdB+G/K.  This is also consistent with the zero parallax
measurement.

PG 0210+168 shares a story similar to PG 1126+185.  It was
typed DC originally \citep{gre86,mcc99} and then reclassified
as DC+F/G by \citet{put97}.  Again, it is simply not possible to
detect a DC white dwarf in the optical against the flux of an F 
or G star.  In this case the brightness difference would be at 
least 8 magnitudes.  Figure 2 of \citet{put97} clearly shows weak
metal lines, a Balmer series, plus Ca H \& K.  Its spectrum is 
very similar to that of PG 1126+185 and shows a strong blue
continuum as well.  PG 0210+168 was not observed for optical 
photometry in the course of this work, so its $UBV$ colors are
not known, but with $V-K\approx1.5$, it is quite certain that
PG 0210+168 contains a cool star together with a hot star.  
Following the same reasoning as for PG 1126+185, the conclusion
is that it is very likely an sdB+F/G composite binary.

\subsection{Uncommon Proper Motion}

Common proper motion is a necessary but insufficient
condition to establish physical companionship for wide
binaries.  In this section, evidence is presented which
illustrates this.

Measuring proper motion accurately is a nontrivial task
requiring good S/N on dozens if not hundreds of background,
near zero motion sources.  And still, what is almost always
measured is relative proper motion -- that is, the motion of 
an object relative to nearby stationary sources \citep{mon03,
lep03}.  In identification of common proper motion companions,
this issue is not really important.  A wide binary pair should
have the same absolute and relative proper motion, assuming zero
orbital motion.  However, the issues raised in \S4.3 are valid
and do contribute to measurement uncertainties.  Given the limited 
accuracy of the proper motions measured for this work, many
candidates were flagged which turned out to be background
stars.  This is desireable -- the error was on the side of 
caution and real candidates were not thrown out.

In the end, only reliable trigonometric parallaxes and/or
orbital motion can determine binarity in widely separated pairs
with absolute certainty.  This is illustrated by three white dwarf
target primaries in the sample: G47-18, G66-36, \& G116-16.  All
three of these stars are reported to be members of a wide binary
containing a white dwarf in \citet{mcc99}.  However, further 
investigation proves these cases to be false positives.  None
of the following pairs are physically associated.

\subsubsection{G47-18 \& G116-16}

G47-18 was reported as a common proper motion binary with
the F6 dwarf HD 77408 \citep{egg67}.  The white dwarf has 
$\mu=0.324''$ ${\rm{yr}}^{-1}$ at $\theta=270.2^{\circ}$ 
and $\pi=0.049''$ (three measurements; \citealt{mcc99}).  
However, HD 77408 has $\mu=0.420''$ ${\rm{yr}}^{-1}$
at $\theta=269.5^{\circ}$ and $\pi=0.0199''$ \citep{per97}.

G116-16 was reported as a common proper motion binary with
the G0 dwarf G116-14 \citep{egg65}.  The white dwarf has 
$\mu=0.252''$ ${\rm{yr}}^{-1}$ at $\theta=180.1^{\circ}$ 
and $\pi=0.035''$ \citep{dah82}, while G116-14 has 
$\mu=0.281''$ ${\rm{yr}}^{-1}$ at $\theta=173.1^{\circ}$
and $\pi=0.0194''$ \citep{per97}.

\subsubsection{G66-36}

G66-36 was reported as a common proper motion binary
with the G5 dwarf G66-35 \citep{osw81}.  However,
G66-36 is not a white dwarf but a metal-poor M2 star at
$d=25$ pc \citep{rei95}.  Its proper motion is $\mu=0.32''$
${\rm{yr}}^{-1}$ at $\theta=173^{\circ}$ \citep{mcc99},
while G66-35 has $\mu=0.297''$ ${\rm{yr}}^{-1}$ at
$\theta=187.1^{\circ}$ and $\pi=0.0152''$ \citep{per97}.
It is noteworthy that this is a case of a subdwarf being
mistaken for a white dwarf and two stars with different 
proper motion being identified as companions.

\subsection{Proper Motion Confusion}

Motions in the Galactic disk are primarily responsible
for the proper motion of nearby stars \citep{bin87,bin98}.
Therefore, in principle, two unrelated stars can have nearly
identical proper motions, both in magnitude and direction.
In addition, some proper motion confusion is due to limited
measurement accuracy, as exemplified in \S A.2.  Unlike the
objects in the previous section, the following potential
candidate wide binaries have never appeared in the 
literature.

\subsubsection{GD 248}

In early 2002, a candidate common proper motion
companion to the DC5 white dwarf GD 248 was identified.
A measurement utilizing digitized POSS plates separated
by nearly 40 years seemed to indicate common proper motion.
Another measurement using near-infrared images separated
by almost 5 years gave similar results but with uncertainty
three times as large.

A Keck LRIS spectrum \citep{far04c} covering $4900-10,000$
\AA \ revealed Na at 5880 \AA, a strong MgH band at 5100 \AA,
and a variety of other fairly weak features -- all indicative
of a metal-poor late K or very early M star \citep{rei00}.

Thus, the candidate companion to GD 248 is an
ultrahigh velocity background star.  At an estimated
distance of $\sim800$ pc (assuming $M_V=10.1$ mag, 2
magnitudes below the main sequence for spectral type K7),
its tangential velocity of $v_{\rm{tan}}\approx530$ km
${\rm{s}}^{-1}$ is just under the escape velocity of the
Galaxy \citep{bin87}.  A precise proper motion measurement
has been performed on the nonphysical pair and their error
ellipses differ by exactly $1\sigma=0.008''$ ${\rm{yr}}^{-1}$
(S. L\'epine 2002, private communication).

\subsubsection{GD 304, PG 1026+002, \& PG 1038+633}

Candidate common proper motion companions to the white
dwarfs GD 304, PG 1026+002, \& PG 1038+633 were identified
during the course of the survey.  Table \ref{tbl-9} lists
the data.

Ultimately, spectra revealed \citep{far04c} that the 
candidate companions to PG 1026+002, \& PG 1038+633 were
high velocity background stars of similar temperature to
the candidate companion to GD 248.  The candidate companion
to GD 304 has a much redder $V-K$ color than the other three
objects in Table \ref{tbl-9}.

Assuming $M_V=9.8$ mag (1 magnitude below the main 
sequence for spectral type M0) for the candidate companion
to PG 1026+002, its tangential velocity is $v_{\rm{tan}}
\approx390$ km ${\rm{s}}^{-1}$.  Using $M_V=9.1$ mag for
the candidate companion to PG 1038+633 (1 magnitude below
the main sequence for a K7 star), its tangential velocity 
is $v_{\rm{tan}}\approx400$ km ${\rm{s}}^{-1}$.  These
values are less than that of the candidate companion to 
GD 248, but still quite high.  For the candidate companion
to GD 304 however, a slightly more modest tangential velocity
of $v_{\rm{tan}}\approx210$ km ${\rm{s}}^{-1}$ is calculated
assuming a main sequence absolute magnitude of $M_V=10.8$ mag 
for spectral type M3.  For all four high velocity background
stars, a spectrum was a critical discriminant.

\subsection{Candidate Companions}

These objects show common proper motion of varying degree
of agreement along with colors and/or spectra consistent with
companionship.  Data for candidate companions are listed in
Tables \ref{tbl-3}, \ref{tbl-4}, \ref{tbl-5}, \& \ref{tbl-10}.

\subsubsection{GD 84}

GD 84B is an especially tentative companion due 
to the relative lack of agreement in the measured proper
motions.  A more precise measurement could not be made for
the candidate companion because it is passing in front of
another star in the digitized POSS I scans.

This candidate is retained because it and GD 84 have
similar photometric distances.  \citet{wei95} give
spectral type DQ5.6 for GD 84 ($T_{\rm{eff}}=9000$
K), which yields $M_V=12.59$ mag and a photometric distance
of $d=33$ pc for log $g=8.0$.  The M dwarf, verified by its
optical spectrum in Figures \ref{fig38} \& \ref{fig39}, 
appears to be a normal solar metallicity star.  Using 
$K=10.55$ mag from Table \ref{tbl-5} and $M_K=7.80$ mag
for spectral type M4, its photometric distance is $d=35$ pc.

\subsubsection{GD 683}

The UCAC1 catalog (Table \ref{tbl-4}; \citealt{zac00})
lists nearly identical proper motion measurements for GD
683 and GD 683B.  But for some unknown reason in the UCAC2
catalog, the proper motion of GD 683B changes to approximately
half of its former value and GD 683 is completely absent
\citep{zac04}.  Because of this discrepancy, a (less accurate)
measurement was made for this work (Table \ref{tbl-10}).

Furthermore, with $K=11.28$ mag from Table \ref{tbl-4}
and $M_K=6.00$ mag for spectral type M2 (solar metallicity),
the photometric distance to GD 683B is $d=114$ pc.  There
exist two temperature and surface gravity determinations for
the DA white dwarf GD 683; their average gives $T_{\rm{eff}}=
30,000$ K and log $g=7.79$, which imply $M_V=9.31$ mag and a
photometric distance of $d=121$ pc, consistent with
companionship between the pair.

\subsubsection{PG 0933+729}

PG 0933+729B is retained as a candidate because there
currently exists no optical photometric nor spectroscopic 
data.  The photographic magnitudes, $B_{pg}=16.6$ mag,
$R_{pg}=14.4$ mag in the USNO catalog, together with the
near-infrared magnitudes in Table \ref{tbl-5}, imply a
$V-K$ color of an early M dwarf (type M2 or earlier) and 
a photometric distance $\sim120$ pc.  The white dwarf is 
type DA3, $T_{\rm{eff}}=17,400$ K at $d\approx90$ pc
\citep{lie05}.

\subsection{Outstanding Doubles \& Triples}

New information gleaned in this work -- specifically,
data on previously unknown  companions -- provides more
accurate descriptions and  parameterizations of many single
and a few double degenerates.  In several cases of confirmed
white dwarf plus red dwarf binaries, the photometric distance
to the white dwarf does not agree with the photometric distance
to the red dwarf.  In cases where the red dwarf appears overluminous,
this may be explained by binarity.  However, there are several cases
where the red dwarf appears significantly underluminous.  Although a
metal-poor M companion would be subluminous with respect to the main
sequence, it is more likely that the distance to the white dwarf is
inaccurate as none of the M dwarf companions appear to have metal-poor
spectra.

\subsubsection{G21-15}

G21-15 is a triple degenerate, only the second known.
Its hierarchy is similar to the only other known triple
degenerate (WD 1704+481; \citealt{max00a}), with one close
double degenerate \citep{saf98} plus the widely separated 
cool white dwarf reported here.  \citet{max99} report a
6.27 d period in the single lined DA spectroscopic binary.

This system has a trigonometric parallax $\pi=0.0182\pm
0.0023''$ and hence distance $d=54.9\pm7.1$ pc \citep{ber01}.
Treating the primary DA as a single star, the photometric
analysis of \citet{ber01} yields $T_{\rm{eff}}=12,240$ K,
log $g=7.00$, $M=0.23$ $M_{\odot}$, and $M_V=10.19\pm0.28$
mag.  The spectroscopic analysis of the primary gives
$T_{\rm{eff}}=14,800$ K, log $g=7.61$, and $M=0.39$
$M_{\odot}$ \citep{bra95,max99,ber01}.  The spectroscopic
method produces the wrong absolute magnitude but more
accurately determines mass, which is an intermediate
value of the masses of the individual double degenerate
components \citep{ber91}.

If one assumes that the two white dwarfs are equally
luminous, the mass of each component ($M\approx0.4$ 
$M_{\odot}$; \citealt{ber01}), is still less than the
cutoff  for CO core white dwarfs.  Hence it is possible
that the close binary consists of two low mass, He core
white dwarfs \citep{ber92,mar95}.  This scenario is
consistent with the spectroscopy, photometry, and parallax.
But this fails to correctly predict the single lined DA
nature of the system, unless one of the stars is a yet 
undetected DB, which is unlikely \citep{ber92,ber02}.

The other possibility is two white dwarfs with different
luminosities -- some combination of different masses and
temperatures.  Is there a scenario which is consistent with
all the observations?  The answer is yes.  In order to display
a single lined DA spectrum with a 6.27 d period, whose Balmer
line profiles yield log $g\approx7.6$, $M\approx0.4$ $M_{\odot}$,
one needs a relatively low mass hot component and a relatively
high mass cool component.  The observational data can be
reproduced by assuming two DA stars at 55 pc, one with 
$M=0.60$ $M_{\odot}$ (log $g=8.0$), $T_{\rm{eff}}=
10,000$ K, $M_V=12.15$ mag, and the other with $M=0.35$ 
$M_{\odot}$ (log $g=7.4$), $T_{\rm{eff}}=15,000$ K,
$M_V=10.38$ mag.  Therefore the most likely nature of
the double degenerate is a He core and a CO core white
dwarf.

The optical and near-infrared colors of G21-15C
(Table \ref{tbl-5}) are consistent with a helium
atmosphere (it is too red for a hydrogen atmosphere)
degenerate with $T_{\rm{eff}}=4750$ K.  The trigonometric
parallax gives $M_V=15.30\pm0.28$ mag and indicates
a radius corresponding to log $g\approx8.0$, or $M=0.57$
$M_{\odot}$.  The cooling age of a helium atmosphere
white dwarf with such parameters is 6.6 Gyr.  Its
kinematics are consistent with a disk star of age
$\tau=5-10$ Gyr (Table \ref{tbl-1}).

It is somewhat surprising that such an old disk white
dwarf is not more massive than $M=0.57$ $M_{\odot}$.  If
correct, this implies the total age of G21-15C, and hence
the entire system, is likely to be closer to 8 Gyr.  The
initial to final mass relation for white dwarfs indicates
that degenerate stars less massive than 0.6 $M_{\odot}$ have
descended from main sequence progenitors less massive than 2
$M_{\odot}$ \citep{wei87,wei90,wei00,bra95}.  Such a progenitor
should have had a main sequence lifetime of more than 1.5 Gyr
\citep{gir00}, yielding a total age for G21-15C of $\tau>8.1$
Gyr.

Designating the brighter, He core white dwarf G21-15A
and the fainter, putative CO core white dwarf G21-15B,
the cooling age of G21-15C can be used as a constraint on the
overall age of the system and place upper limits to the masses
of both main sequence progenitor components of the double
degenerate.  Using $T_{\rm{eff}}=15,000$ K for G21-15A 
and $T_{\rm{eff}}=10,000$ K for G21-15B, they have been
cooling for 0.1 and 0.6 Gyr respectively.  If the above
analysis is correct, the entire system is at least 8.1 Gyr
old and hence the progenitors of G21-15A \& B were very
nearly solar mass stars \citep{gir00}.  

A He core white dwarf is born as it ascends the
red giant branch for the first time, its outer layers
stripped before helium burning can begin \citep{ber92,mar95}.
G21-15A began this ascent after at least 8 Gyr on the main
sequence, corresponding to a star with $M\leq1.10$ $M_{\odot}$
\citep{gir00}.  G21-15B completed its main sequence evolution
0.5 Gyr before component A became a red giant, and hence came
from a slightly more massive progenitor with $M\sim1.15$
$M_{\odot}$ \citep{gir00}.

Now a complete picture, albeit speculative, of the
entire evolution of the double degenerate can emerge.
The progenitors of G21-15A \& B were probably two nearly
solar mass stars with separation $\la1$ AU.  The more massive
component B left the main sequence first, without any mass
transfer in the relatively wide binary during its first red
giant phase.  Upon ascending the asymptotic giant branch, 
component B forms a common envelope around the binary
and the orbit shrinks significantly due to loss of angular
momentum.  The resulting close separation is enough to
cause component A to overfill its Roche lobe as a first
ascent giant and become a He core white dwarf
\citep{nel01}.

\subsubsection{GD 319}

GD 319 is a triple system consisting of an sdB with
a close unseen companion \citep{saf98} plus the widely
separated M3.5 dwarf reported here.  The lower limit
for the mass of the unseen companion is $M\sin i=0.9$
$M_{\odot}$, which is in a 0.6 d orbit around the sdB
\citep{max00b}.  

Although not a white dwarf, GD 319 was included in the
survey due to its presence in \citet{mcc87}.  It was later
reclassified as type sdB by \citet{saf98}, who also first
detected its radial velocity variations.  sdB stars such as
GD 319 are thought to be helium burning stars with very thin
hydrogen envelopes which eventually cool to become white
dwarfs with $M\approx0.5$ $M_{\odot}$ \citep{saf94}.  These
hot subdwarfs may form analogously to low mass white dwarfs,
their surface hydrogen layers mostly removed by a close
companion \citep{ibe93}.  Hence, GD 319 is representative
of a white dwarf system.

There is a K star located $\approx3''$ away from
GD 319, but it was shown to be physically unrelated
\citep{mca96}.  Most photometry reported in the literature
for GD 319 is contaminated by the K star.  The values for
GD 319AB in Table \ref{tbl-5} are measurements from images
in which both stars were well resolved from one another and
there should be no such contamination \citep{mca96,far04c}.
The near-infrared photometry was performed on images taken
at Keck Observatory and the results indicate the K star is
a foreground object.

Although the mass of the close companion GD 319B is
unknown, the lower limit indicates that it is a likely
M dwarf, unless $i<9^{\circ}$.  It is difficult to say
whether the near-infrared photometry of the sdB+dM reveals
a near-infrared excess.  The measured $V-K=-0.80$ is slightly
less blue than one would expect for a 30,000 K star but an
effective temperature for GD 319 has not been established.
In any case, were the unseen companion a dK and not a dM star,
one would expect a definite excess at $K$ band.  To illustrate,
an M0V star has $M_K\approx5.0$ mag, while an sdB has $M_V
\approx4.5$ mag \citep{max00b} plus $V-K\approx-1$ for a
30,000 K object.  Their combined contribution at 2.2 $\mu$m
would be $M_K\approx4.5$ mag -- an excess of $\sim1$ mag, but
this excess is not seen, so the unseen companion cannot be a
K dwarf.

The wide tertiary component, GD 319C, can be used to
constrain the distance to the system.  This is helpful
because at $d\sim400$ pc, GD 319 is too far away for
a trigonometric parallax measurement.  Using empirical
optical and near-infrared absolute magnitudes for an
M3.5 dwarf, the distance modulus for GD 319C is
$m-M=7.39\pm0.10$ or $d=300\pm14$ pc.  This yields 
$M_V=5.3$ mag for GD 319A.  This further constrains
the spectral type of GD 319B to later than $\sim$ M3.5
(because such a companion would imply $V-K\approx-0.5$ for
GD 319AB), implying a mass in the range 0.3 $M_{\odot}>M
\geq0.09$ $M_{\odot}$.  This in turn limits the inclination
of the orbit by $\sin i>0.09/0.3=0.3$, which yields
$i>17.5^{\circ}$.

\subsubsection{LDS 826}

LDS 826 is a triple system consisting of a white
dwarf plus red dwarf visual pair together with the widely
separated M8 dwarf reported here and in \citet{sch04}.
The data analyzed for this work indicate the system is
a DA5.5+M3.5+dM8 which differs slightly from that given
in \citet{sch04}.  There is a significant amount of
published data on this system and its components which
may be inaccurate according to the uncontaminated (\S4.5)
photometric data presented here (see \citet{far04c} for 
details).

\subsubsection{PG 0824+288}

PG 0824+288 is a triple system consisting of a DA1
plus dwarf carbon star double lined spectroscopic
binary \citep{heb93} in a visual pair with the M3.5 dwarf
reported here.  The spectroscopic binary has been searched
unsuccessfully for radial velocity variability \citep{max00}.
An astrometric measurement of the blended ($a\sim3''$)
visual double between POSS I \& II epochs reveals the
pair moving together with the proper motion listed in Table
\ref{tbl-4}.  The visual binary was first suspected by
\citet{gre94}, but remained unconfirmed until now.

One can estimate the distance by noting that the
visual magnitude of the composite spectroscopic binary,
$V=14.22$ mag, is estimated to have a 25\% contribution 
from the dwarf carbon star, PG 0824+288B \citep{heb93},
yielding $V=14.53$ mag for the white dwarf.  Two spectroscopic
analyses of PG 0824+288A give significantly different mass
estimates but similar effective temperatures; \citet{mar97}
give $T_{\rm{eff}}=51,900$ K and log $g=8.00$, while \citet{fin97}
give $T_{\rm{eff}}=50,500$ K and log $g=7.43$ \citep{mar97}.  The
absolute magnitudes predicted by models for these two spectroscopic
parameter fits are quite different -- $M_V=9.09$ mag ($d=122$ pc)
versus $M_V=8.12$ mag ($d=191$ pc).

The nearby visual M dwarf companion, PG 0824+288C, has
an estimated spectral type of M3.5 \citep{gre94}.  This is
consistent with its estimated color, $V-K\approx4.8$, calculated
from the $\Delta g=3.0$ mag difference between the components of
the visual pair \citep{gre94}, plus the measured $K$ magnitude.  
Comparing the $J$ \& $K$ values in Table \ref{tbl-5} with the
absolute magnitudes of an M3.5 dwarf, the distance modulus is
$m-M=5.31$, or $d=115$ pc, which is inconsistent with the mass
and radius determined by \citet{fin97}.  Therefore, it appears
likely that PG 0824+288A is a log $g=8.0$ white dwarf with a
mass of $M=0.70$ $M_{\odot}$ at $d\approx120$ pc.

\subsubsection{PG 1204+450}

PG 1204+450 is a triple system consisting of a double
lined DA spectroscopic binary plus the widely separated
M4 dwarf companion reported here \citep{saf98,max02}.
The spectroscopic binary consists of an approximate DA2+DA3
with a 1.6 d period and a mass ratio of $M_A/M_B=0.87$
\citep{max02}.  They resolved the H$\alpha$ line core
into two components and give $T_{\rm{eff}}=31,000$ K,
$M=0.46$ $M_{\odot}$ for PG 1204+450A and $T_{\rm{eff}}=
16,000$ K, $M=0.52$ $M_{\odot}$ for PG 1204+450B.  One
should keep in mind that these are estimates and that it
is not possible to perform  a spectroscopic fit of the
Balmer lines for the individual components.

\citet{lie05} give $T_{\rm{eff}}=22,600$ K,
$M=0.52$ $M_{\odot}$ in a recent update of previous
work \citep{ber92}, which treats the double degenerate
as a single star.  If these newer parameters are accurate,
then the individual component masses inferred by
\citet{max02} should change similarly to 0.50 and
0.57 $M_{\odot}$ for components A \& B respectively.

The wide M4 companion, PG 1241+450C, should be able
to constrain the effective temperatures and absolute 
magnitudes of PG 1241+450 A \& B by constraining the
distance to the system.  Using empirical optical and 
near-infrared absolute magnitudes for an M4 dwarf, the
distance modulus for PG 1241+450C is $m-M=5.30\pm0.11$
or $d=115\pm6$ pc.  This implies an absolute magnitude
for PG 1241+450AB of $M_V=9.74$ mag.  If the flux ratio
at $V$ is similar to the estimated luminosity ratio from
\citet{max02}, this implies $M_V=10.17$ mag and $M_V=10.95$
mag for components A \& B respectively.  Using these absolute
magnitudes and estimated masses from the previous paragraph,
models predict $T_{\rm{eff}}=22,000$ K for PG 1204+450A and
$T_{\rm{eff}}=16,500$ K for PG 1204+450B.

\subsubsection{PG 1241$-$010}

PG 1241$-$010 is a triple system consisting of a
close double degenerate plus the visual M9 companion
reported here.  The close binary is a single lined
DA in a 3.3 d orbit \citep{mar95}.

\citet{lie05} give $T_{\rm{eff}}=23,800$ K, $M=0.40$
$M_{\odot}$ in a recent update of previous work \citep{ber92},
treating the double degenerate as a single star.  As in the
case of G21-15 above, this close binary is likely to be composed
of one relatively hot He core white dwarf plus another relatively
cool CO core white dwarf.  The spectroscopically determined
absolute magnitude for PG 1241$-$010, $M_V=9.3$ mag \citep{ber92,
lie05}, is almost certainly an underestimate due to the presence
of not one but two white dwarfs.  Unfortunately, parameter
estimates for the individual components do not exist in 
the literature

PG 1241$-$010C was first recognized by \citet{zuc92} and
is one of the latest known confirmed companions to a white
dwarf.  The photometric distance to the tertiary may be used 
as a constraint on the  distance to the system.  Its $I$ band
magnitude was very difficult to measure due to the $\Delta I>4$
mag difference between components C \& AB, just $3''$ apart.  In
any case, its colors (Table \ref{tbl-5}) are consistent with
spectral types M8$-$M9.  Comparing its apparent $K$ magnitude 
with absolute magnitudes of M8 and M9 spectral standards, the
resulting distance modulus is bounded by $4.14<m-M<4.44$ or
$d=72\pm5$ pc.

This distance yields $M_V=9.71\pm0.15$ mag for PG
1241$-$010AB, but a single 24,000 K DA white dwarf
with log $g=7.5$ ($M=0.41$ $M_{\odot}$) should have
$M_V=9.65$ mag according to models, implying that
PG1241$-$010B contributes almost nothing to the overall
$V$ band flux.  A distance around $d=72$ pc is also
inconsistent with the previous spectroscopic mass
determination of $M=0.31$ $M_{\odot}$ ($M_V=9.2$ mag)
for PG 1241$-$010AB.  Assuming the above distance bounds
and the spectroscopic temperature of PG 1241$-$010A are
correct (with log $g=7.5$), the maximum contribution
from PG 1241$-$010B is $M_V=12.3$ mag, implying
$T_{\rm{eff}}<10,000$ K for a log $g=8.0$ white dwarf.
However, it is possible that PG 1241$-$010C is itself
binary, indicating a distance up to $d=109$ pc (two
identical M8 dwarfs) and more consistent with the
spectroscopically determined parameters of PG
1241$-$010AB.

\subsubsection{G261-43}

G261-43A is a DA3 white dwarf with a trigonometric
parallax of $\pi=0.471''$ \citep{mcc99}.  \citet{mcm89}
reports and compares three independent determinations of
effective temperature and surface gravity, all of which
agree very well at $T_{\rm{eff}}=15,400$ K and log $g=7.9$.
Models predict $M=0.57$ $M_{\odot}$ and $M_V=11.04$ mag,
compared with $M=0.61$ $M_{\odot}$ and $M_V=11.19$ mag
from its parallax.  Hence, G261-43 is a DA white dwarf
with a typical mass \citep{ber92}.

The binary nature of G261-43 was first reported by
\citet{zuc97}.  An almost certain white dwarf itself,
G261-43B is located only $1.4''$ away and quite faint 
relative to its primary ($\Delta V\approx3.5$ mag).  

The roughly estimated effective temperature of
G261-43B ($\sim5000$ K; \citealt{zuc97}) is probably
a bit too low given the photometry.  Although the
near-infrared photometry had good S/N, the optical
measurements were difficult due to the brightness ratio
of primary to secondary (B. Zuckerman 2004, private
communication).  Using the three most reliable magnitudes
available ($I=15.7$ mag, $J=15.34$ mag, $K=15.05$ mag) and
the resulting colors, hydrogen atmosphere models predict
$T_{\rm{eff}}=6000$ K almost exactly, regardless of surface
gravity.  Combining this effective temperature with the
trigonometric parallax gives log $g=8.39$ and $M=0.84$
$M_{\odot}$ for G261-43B.  Therefore, the difference
in radii between components A \& B is likely to be
significant \citep{zuc97}.

If this analysis is correct, the main sequence
progenitor of G261-43A spent about 4.4 Gyr on the main
sequence and as a giant.  This is the difference between
the 4.6 Gyr cooling age of component B and the 0.2 Gyr 
cooling age of component A.  The mass of a main sequence
star with such a lifetime, including post main sequence
evolution, should be just greater than 1.3 $M_{\odot}$
\citep{gir00}.

\subsubsection{PG 0901+140}

PG 0901+140 is a visual double degenerate with a
separation on the sky of $3.6''$.  Despite several
studies and comparable luminosities between its two
components ($\Delta V\approx0.5$ mag), the binarity 
has never been mentioned in the literature
\citep{gre86,ber90,lie05}.

\citet{lie05} give $T_{\rm{eff}}=9200$ K, log
$g=8.29$, and $M_V=12.91$ mag for PG 0901+140 by
spectroscopic analysis, assuming it is a single star.
Comparison of hydrogen atmosphere model colors to the 
optical and near-infrared photometry in Table \ref{tbl-5}
yields $T_{\rm{eff}}=9500$ K for the brighter component
PG 0901+140A, and $T_{\rm{eff}}=8250$ K for the fainter
component PG 0901+140B.  These effective temperatures are
quite consistent with the combined effective temperature
determination, whereas the spectroscopically determined
mass ($M=0.79$ $M_{\odot}$; \citealt{lie05}) is an
intermediate value of the masses of the individual
double degenerate components \citep{ber91}.  Assuming
a luminosity ratio of 1.6:1 between components A \& B
(from the 0.5 mag average of $\Delta V$ and $\Delta R$),
a reasonable guess at the individual masses is $M=0.78$
$M_{\odot}$ for PG 0901+140A and $M=0.81$ $M_{\odot}$
for PG 0901+140B.  If correct, this would place the
system at a little more than $d=40$ pc.

Further speculation is not worthwhile, since
the component parameters of the resolved double
degenerate can be determined from individual spectra.
Such a determination would be very useful for the
white dwarf initial to final mass relation -- analogous
to the analysis of PG 0922+162A \& B by \citet{fin97b}.

\subsubsection{LP 618-14}

LP 618-14 was identified by S. Salim (2002, private
communication) during a survey intended to find previously
unidentified white dwarfs in the new Luyten two tenths
catalog \citep{sal02}.  Its reduced proper motion placed
it within the white dwarf sequence while its Sloan colors 
were a little too red for a single degenerate star.  

Its spectrum in Figure \ref{fig49} reveals the
blue continuum of a cool DA or DC star plus the
TiO bands of a red dwarf.  Unfortunately, the spectrum
has sufficiently low S/N as to preclude a more precise
estimate of the white dwarf effective temperature.  
Also the $UBV$ photometry was performed on images acquired
under less than ideal weather conditions.  The $U-B=-0.45\pm
0.15$ color of LP 618-14 is consistent with a large range of
effective temperatures: $6000-13,000$ K for a hydrogen
atmosphere white dwarf with log $g=8.0$, up to 15,000 K
for log $g\leq7.5$, but only $6000-8000$ K for log 
$g\geq8.5$ or for a helium atmosphere at any value of
log $g$.  Additionally, it is not known whether the
red dwarf contributes significantly at $B$ band.
A good blue spectrum and photometric $UBV$ data
would end this ambiguity.

The luminosity contribution of the two components
is further confused by the scenario described in \S4.4
because the distance to the system is not known.
The reduced proper motion indicates LP 618-14 should be
within 50 pc of the sun ($\mu=.32''$ ${\rm{yr}}^{-1}$),
but this interpretation is thwarted by predicting a spectral
type later than M7 for the red dwarf companion ($M_K\geq10.0$
mag).  This appears to be inconsistent with the spectrum in
Figure \ref{fig49}.  Alternatively, if one constrains the
spectral type of the secondary to be no later than M5, then the 
system is located at $d>100$ pc and has a very high tangential
velocity ($v_{tan}>150$ km ${\rm{s}}^{-1}$).  This results
in a white dwarf with a relatively large radius at $\sim9000$
K -- a He core degenerate with $M<0.4$ $M_{\odot}$.  This is
a real possibility for any white dwarf in a close binary 
\citep{mar95} but further investigation is required before
firm conclusions may be drawn.

\subsubsection{LP 761-114}

LP 761-114 is first mentioned in \citet{osw96}
as a white dwarf in a wide binary system, where 
it is reported as the lowest luminosity star in a
sample used to place a lower limit to the age of the
Galactic disk.  \citet{sil02} and \citet{hol02}
corroborate this interpretation, reporting $V=17.45$ mag,
$V-I=1.75$, $T_{\rm{eff}}=4020$ K, and a photometric 
distance of $d=15.3$ pc.

However, the M2 dwarf common proper motion
companion absolutely rules out a distance $d<30$ pc
(see below).  Also, this bright neighbor can easily
contaminate photometric measurements of the cool 
white dwarf.  The M2 companion $7.7''$ distant
positively dominates the binary flux ratio at all
wavelengths, including $U$ band where it is over
2 magnitudes brighter(!) than the white dwarf.
It gets much worse toward longer wavelengths
with $\Delta V=4.3$ mag, $\Delta I=5.8$ mag.

The photometry presented for LP 761-114 in Table
\ref{tbl-5} was measured with near zero contamination
as described in \S4.5, and tells a slightly different
story than existing analyses \citep{osw96,sil02,hol02}.
The $V$ magnitude measured here is 0.4 mag fainter and
the measured color of $V-I=0.73$ is much bluer -- a
clear sign that prior measurements were contaminated by
the relatively bright M companion.

The resulting optical and near-infrared colors
of LP 761-114 are not consistent with a 4000 K
helium atmosphere white dwarf \citep{hol02}.
Comparing its colors (e.g. $U-B=-0.31$, $V-J=1.17$)
with models of cool helium and hydrogen atmosphere
white dwarfs, one finds very good agreement with a
hydrogen atmosphere degenerate at $T_{\rm{eff}}=6000$
K.  The near-infrared measurements were extremely
difficult due to the $\Delta m=6.5-7$ mag brightness
difference between components and these data possess
the lowest (but still reliable) S/N of all the
photometry performed on this system.

Is this interpretation consistent with the photometric
distance to the M2 dwarf companion?  Comparing the optical
and near-infrared magnitudes of LP 761-113 from Table
\ref{tbl-5} with spectral standard absolute magnitudes,
the distance modulus is $m-M=3.38\pm0.04$ or $d=47.4\pm0.9$
pc.  This yields $M_V=14.45$ mag for the white dwarf, indicating
log $g=8.17$ or $M=0.70$ $M_{\odot}$ -- a reasonable mass for a
white dwarf that has been cooling for more than 3.2 Gyr.  

\subsubsection{PG 1539+530}

PG 1539+530 is listed as a DA2 in \citet{mcc87}
but is mysteriously absent from \citet{mcc99}.  The
PG catalog lists type DA2 for this star and in the 
comments column of Table 5 is the note ``DBL'' 
\citep{gre86}.  The removal of an object from the
white dwarf catalog is usually a sign of misclassification,
but in this case the original classification is the
correct one.

The composite spectrum of PG 1539+530AB is shown
in Figure \ref{fig46}.  It clearly displays the spectral
features of an early M dwarf plus a few pressure broadened 
Balmer lines typical of hot white dwarfs.  The H$\alpha$
line has been completely masked by the brighter M companion,
but at least 3 more hydrogen lines are visible out to a
partial H$\epsilon$.

The pair has been resolved at a separation of $2.7''$
on the sky.  The M dwarf component, PG 1539+530B, has 
colors and a spectrum consistent with spectral type M2.
The white dwarf component, PG1539+530A, has optical and
near-infrared colors consistent with a $T_{\rm{eff}}=
25,000$ K DA star and is actually fainter than its companion
at $V$ band and longward.  Comparison of the photometry in
Table \ref{tbl-5} with the optical and near-infrared absolute
magnitudes for M2 dwarfs yields a photometric distance of
$d=162\pm4$ pc for PG 1539+530B.  At this distance, a 25,000
K DA white dwarf has $M_V=10.47$ mag, log $g=8.09$ and a
mass of $M=0.69$ $M_{\odot}$ according to models.

\subsubsection{PG 2244+031}

PG 2244+031 is noted as a DA1 with a composite 
spectrum in the PG catalog \citep{gre86}.  The
present work designates the system as a visual
binary consisting of the DA1 white dwarf PG 2244+031A
plus an M3.5 dwarf companion at $2.4''$, PG 2244+031B.
The composite optical spectrum of the binary is shown 
in Figure \ref{fig48}.

However, there is significant confusion 
in the literature regarding the coordinates,
identity, finding chart, and designated WD
number for PG 2244+031.  In fact, there are
no less than four objects associated with PG
2244+031; the following paragraphs should 
clarify the situation.

Object 1, PG 2244+031 (WD 2244+031) was first
reported as type DC by \citet{gre80} and this
paper contains the only published finding chart
for any of the four objects.  The coordinates given
in Table 3 of \citet{gre80} for object 1 more or
less correctly identify the position of PG 2244+031AB.
The correct coordinates are $22^{\rm h} 44^{\rm m}
49.7^{\rm s}$, $+03^{\circ} 05' 54''$ B1950 or $22^{\rm
h} 47^{\rm m} 22.3^{\rm s}$, $+03^{\circ} 21' 45''$ J2000.  

Object 2, HS 2244+0305 (WD 2244+030) is designated
as a DA1 \citep{hom98} with coordinates identical to
those of object 1 listed above, but is listed separately
from PG 2244+031 in the current version of the online
white dwarf catalog.  This is an error; WD 2244+031
and WD 2244+030 are the same object.

Object 3 is the star which is identified in the 
finding chart of \citet{gre80} for PG 2244+031.
This chart points to coordinates $22^{\rm h}
44^{\rm m} 16^{\rm s}$, $+03^{\circ} 06'
20''$ B1950 or $22^{\rm h} 46^{\rm m} 
48^{\rm s}$, $+03^{\circ} 22' 10''$ J2000 
and a completely different object from PG 2244+031.
Photometry was performed on object 3 and it has 
$B-V=0.44$ and $V-K=1.37$.  These values are
consistent with the colors of a G0 star.

Object 4 is the star corresponding to the
coordinates in the PG catalog for PG 2244+031.
These coordinates neither match the finding chart
of \citet{gre80} nor the coordinates in Table 3 of
\citet{gre80}.  The PG coordinates are $22^{\rm h}
44^{\rm m} 25^{\rm s}$, $+03^{\circ} 08' 52''$ 
B1950 or $22^{\rm h} 46^{\rm m} 57^{\rm s}$,
$+03^{\circ} 24' 41''$ J2000 \citep{gre86}.  No finding
chart is provided as the identity and coordinates of object
3 in \citet{gre80} and object 4 in \citet{gre86} are
assumed to be one and the same.  However, photometry was
performed on object 4 and it has $B-V=0.50$ and $V-K=1.51$.
These values are consistent with the colors of a G5 star.  

In summary, the objects PG 2244+031, WD 2244+031,
HS 2244+0305, \& WD 2244+030 are identical to the DA1+M3.5
visual binary described here as PG 2244+031AB.  All published
coordinates for these objects are in error with the exception
of those in \citet{gre80}, and more accurately in \citet{hom98}.
The only correct finding chart in the literature is the one
shown here in Figure \ref{fig22}.

\subsubsection{GD 74}

GD 74A is well cited in the literature but
only one study has produced a distance estimate.
\citet{ber92} give $T_{\rm{eff}}=16,900$ K,
log $g=7.99$, $M=0.59$ $M_{\odot}$, and $M_V=11.11$ 
mag for GD 74A.  If correct, this would place the
DA white dwarf at just under $d=60$ pc.

However, the optical and near-infrared colors
inferred from the photometry in Table \ref{tbl-5} 
indicate an effective temperature higher than 17,000
K for GD 74.  With colors such as $V-K=-0.79$, hydrogen
atmosphere white dwarf models predict a temperature much
closer to 25,000 K.  Such a DA2 star with log $g=8.0$
would have a photometric distance closer to 85 pc.
Comparison of the photometry in Table \ref{tbl-5} with
the optical and near-infrared absolute magnitudes for
M4 dwarfs yields a photometric distance of  $d=97\pm6$ pc
for GD 74B.

\subsubsection{GD 123}

GD 123A is a DA white dwarf that has been repeatedly
studied in the literature, with a well corroborated
$T_{\rm {eff}}$ near 30,000 K.  The average of five
independent spectroscopic analyses gives $T_{\rm
{eff}}=29,500$ K, log $g=7.92$, and $M_V=9.86$ mag
\citep{fin97,mar97,ven97,nap99,lie05}.  This would 
place the GD 123 system near $d=81$ pc.

The binarity of GD 123 was first found by
\citet{gre86} who noted a composite spectrum of a DA4+K.
Optical and near-infrared analysis done here indicates 
that GD 123B is an M4.5 dwarf with photometric distance
$d=67\pm2$ pc.

\subsubsection{GD 337}

GD 337A is reported as a DA2 white dwarf in
\citet{mcc99}, recently corroborated by
\citet{lie05} who give $T_{\rm{eff}}=22,400$ K, 
log $g=7.80$, and $M_V=10.23$ mag.  A photometric
distance of $d=151$ pc is inferred from these data.
GD 337 also has three parallax measurements in
\citet{mcc99}, all of which give $\pi\leq0.004''$

\citet{pro83} first detected the still unresolved
companion by its near-infrared excess emission.  The
presence of a late type companion has been confirmed
by its observed composite spectrum \citep{gre86,gre86b}.
The analysis here places GD 337B at spectral type M4.5.
Comparing its inferred optical and near-infrared
magnitudes with the absolute magnitudes of M4.5 dwarfs,
its photometric distance is $d=180\pm5$ pc.

\subsubsection{GD 984}

GD 984A is a well studied hot DA white dwarf.
At least five independent spectroscopic studies
have been carried out, yielding average parameters
of $T_{\rm{eff}}=49,000$ K, log $g=7.85$, and 
$M_V=8.95$ mag \citep{fin97,mar97,ven97,nap99,koe01}.
If correct, this would place the system near $d=108$
pc.

There is some spread in the derived effective
temperatures and surface gravities for GD 984A, perhaps
due to its unresolved companion, GD 984B.  The M2 dwarf
contaminates the Balmer lines of the white dwarf as blue
as H$\gamma$, making spectroscopic parameter fits 
less certain \citep{fin97}.  The spread in temperature
estimates ranges from $43,000-57,000$ K and surface
gravity from log $g=7.7-8.2$ \citep{mar97,koe01}.

What does the photometric distance to GD 984B say?
Comparing its optical and near-infrared photometry in
Table \ref{tbl-6} with the absolute magnitudes for an
M2 dwarf, the distance is $d=185\pm2$ pc.  There is a
significant difference between this distance for GD 984B
and $d=108$ pc derived for GD 984A.  Since there is a fair
amount of room for uncertainty in parameters of the white 
dwarf, it is quite possible the farther distance is the
correct one.  At $d=185$ pc, GD 984A would have $M_V=7.77$ mag.
Even with $T_{\rm{eff}}=60,000$ K, this would imply a radius
too large for a single white dwarf with $M\geq0.5$ $M_{\odot}$.
At $T_{\rm{eff}}=50,000$ K, this implies log $g=7.23$ or $M=0.43$
$M_{\odot}$.  Therefore, if GD 984A is 185 pc away, then it is
almost certainly a double degenerate or low mass, He core white
dwarf.

\subsubsection{LTT 8747}

LTT 8747A is a nearby cool DA star ripe for a
trigonometric parallax measurement.  The ARICNS
database \citep{gle91} gives a {\em photometric} parallax
of $0.051\pm0.006''$ but it is not included in a recent
list of white dwarfs within 20 pc \citep{hol02}.  The
reason stated in \citet{hol02} for exclusion is that it
belongs to a group of white dwarfs which ``all have
trigonometric parallaxes smaller than $0.05''$'' while
citing \citet{mcc99}.  This may be an error because the
ARICNS value of $0.051''$ is cited in \citet{mcc99} by
mistake as a {\em trigonometric} parallax and no other
value or measurement exists in the literature.  

The photometric distance of 19.6 pc may be
reliable as it is based only on $UBV$ photometry
(ARICNS).  \citet{zuc03} report $T_{\rm{eff}}=7660$
K and log $g=7.80$ from $UBVRI$ photometry and parallax
(which must be the same {\em photometric} parallax
mentioned above).  This effective temperature is
likely a bit too low as LTT 8747B contributes to
the $I$ band flux, causing the white dwarf to 
appear slightly cooler.

The analysis here finds $T_{\rm{eff}}=8500$ 
K is quite consistent with both the $UBV$ and Stromgren
colors of LTT 8747A (both published and those measured 
in this work; \citealt{egg65,mcc99}).  This results in a
photometric distance of $d=22.6$ pc for log $g=8.0$.

The possible radial velocity companion \citep{sch96,
max00}, LTT 8747B is certainly a late M dwarf by both
its contribution to the composite spectrum in Figure
\ref{fig53} (at least one band of VO can be seen) and
by the $J-K=1.10$ composite color of the binary \citep{kir91,
kir94,kir99b}.  The deconvolved colors of LTT 8747B indicate
an M8 dwarf with an uncertainty of 1 spectral type.  If it
has an absolute $K$ magnitude typical of an M8 dwarf, then
$d=19.4$ pc is inferred. At this distance LTT 8747A would 
have $M_V=13.10$ mag, log $g=8.20$, and $M=0.73$ $M_{\odot}$
for $T_{\rm{eff}}=8500$ K.

\subsubsection{PG 0308+096}

PG 0308+096 is a post common envelope binary
consisting of a DA2 white dwarf and the M4.5 dwarf
reported here in a 0.3 d orbit \citep{saf93}.  A recent
spectroscopic analysis gives $T_{\rm{eff}}=25,900$ K,
log $g=8.08$, and $M_V=10.37$ mag for PG 0308+096A
\citep{lie05}.  The distance inferred from these
white dwarf parameters is $d=101$ pc.  Comparison of
the optical and near-infrared magnitudes of PG 0308+096B
with the absolute magnitudes for M4.5 dwarfs, yields a
photometric distance $d=114$ pc.

\subsubsection{PG 0950+185}

PG 0950+185 is a visual double containing a hot
DA white dwarf plus the M2 dwarf reported here $1.1''$
distant \citep{gre86,gre86b}.  The only spectroscopic 
analysis in the literature gives $T_{\rm{eff}}=31,800$
K, log $g=7.68$, $M=0.50$ $M_{\odot}$, and $M_V=9.29$ mag
for PG 0950+185A \citep{lie05}.  The distance inferred
from these parameters is $d=201$ pc.

If this is correct, the implied absolute magnitude of
$M_K=5.28$ mag for PG 0950+185B is significantly brighter
than the expected value of $M_K=6.00$ mag for a single M2
dwarf.  The difference between the standard optical and
near-infrared magnitudes of an M2 dwarf at $d=201$ pc and
the apparent magnitudes of PG 0950+185B are 0.59 mag at
$I$, 0.60 mag at $J$, 0.70 mag at $H$, and 0.72 mag at $K$.
Hence, if the $d=201$ pc distance is accurate, it is
likely that PG 0950+185B is a binary M2 dwarf consisting
of two nearly equal luminosity stars.

\subsubsection{PG 0956+045}

PG 0956+045 is a visual double containing
a DA3 white dwarf plus the M4.5 dwarf reported here
$2.0''$ distant \citep{mcc99}.  The only spectroscopic
analysis in the literature gives $T_{\rm{eff}}=18,200$
K, log $g=7.81$, $M=0.52$ $M_{\odot}$, and $M_V=10.62$
mag for PG 0956+045A \citep{lie05}.  The distance inferred
from these parameters is $d=113$ pc.  With empirical optical
and near-infrared absolute magnitudes for an M4.5 dwarf, the
distance modulus for PG 0956+045B is $m-M=6.25\pm0.17$ or
$d=178\pm15$ pc.

The $V$ \& $R$ magnitudes for PG 0956+045B are
unusual and do not seem to match what might be
expected from a mid M dwarf.  Specifically, they
are significantly fainter than an extrapolation
from the $IJHK$ magnitudes would predict and color
indices involving $V$ \& $R$ appear to be inconsistent
with color indices involving only the other photometric
bands.  It is possible that the $V$ band magnitude for
PG 0956+045B is unreliable due to both low S/N plus
additional uncertainty from deconvolving its PSF from 
the much brighter PG 0956+045A. However, the $R$ band 
measurement has ${\rm S/N}>200$, and is both reliable
and robust -- the measurement was repeated in numerous
ways, always with near zero residuals after PSF subtraction
and predicting the correct relative instrumental magnitude
for PG 0956+045A (certainly within the typical 0.05 mag 
standard error).  As an example of the resulting discrepancy,
the $I-K=2.34$ color is consistent with a spectral type
of M4 and one might expect something near $R-I\approx1.6$
for this spectral type.  But PG 0956+045B has $R-I=2.24$
and, by itself, predicts a spectral type between M6$-$M6.5.
A spatially resolved optical spectrum of PG 0956+045B might
shed light on this issue.

Concerning the mismatch of photometric distances,
there are three distinct possibilities.  One is that
the flux of the red secondary may have contaminated
the spectroscopic analysis of \citet{lie05}, causing PG
0956+045A to appear cooler that it actually is.  A hotter,
more luminous white dwarf would be more consistent with
a distance near $d=180$ pc.  But its measured colors,
such as $V-K=-0.58$, are consistent with a white dwarf of
$T_{\rm{eff}}\approx17,000$ K.  The second  possibility is
that PG 0956+045B is an M dwarf at $d=113$ pc with $M_K=8.66$
mag, corresponding to a spectral type near M5.5 for the main 
sequence.  This seems unlikely since such an M dwarf should
have $I-K\approx3.0$.  A variation of this second prospect
could be made by invoking a subdwarf M star to explain the
discrepancy between the colors of the secondary and the 
absolute magnitude implied by $d=113$ pc.  While this is 
certainly possible, it is  unlikely based on the disk like
kinematics of the PG 0956+045 system (Table \ref{tbl-1}).
The third possible explanation is simply that the white dwarf
distance has been underestimated due to binarity.  A combination
of the above factors could explain the discrepancy between the
inferred distances of PG 0956+045A \& B.  In any case, the 
difference between their photometric distance moduli is
greater than 0.9 and is worthy of further exploration.

\subsubsection{PG 1015+076}

PG 1015+076A is a DA2 star that has been
repeatedly misclassified as a much cooler white
dwarf due to the presence of a background main
sequence star $2.0''$ distant \citep{gre86,zuc03}.
The background star has $V-K=1.61$ and is probably
a G type star.  The spectrum of PG 1015+076A in Figure
\ref{fig24} is almost certainly contaminated at some
level by the continuum light of the nearby G star.
This is why H$\beta$ \& H$\gamma$ are diluted while
H$\alpha$ is nearly absent.  The optical photometry
for PG 1015+076A and background G star was performed
without mutual contamination.  Its optical colors from
Table  \ref{tbl-5} (plus $U-B=-0.98$) indicate 
$T_{\rm{eff}}\approx25,000$ K for the white dwarf.

PG 1015+076B is the M3 dwarf common proper
motion companion to PG 1015+076A.  Comparing 
its optical and near-infrared magnitudes in Table 
\ref{tbl-5} with the expected absolute magnitudes
of a typical M3, the distance modulus is $m-M=6.44
\pm0.06$, implying $d=194\pm6$ pc.  At this distance
PG 1015+076A should have $M_V=10.16$ mag, which 
corresponds to log $g=7.89$, or $M=0.59$
$M_{\odot}$ for $T_{\rm{eff}}=25,000$ K.

\subsubsection{PG 1210+464}

PG 1210+464 is an unresolved binary DA+dM,
evidenced by its composite spectrum and near
infrared  excess \citep{gre86,zuc92,sch96}.
The white dwarf, PG 1210+464A, has spectroscopic
parameters $T_{\rm{eff}}=27,700$ K, log $g=7.85$,
and $M_V=9.87$ mag \citep{lie05}.  Hence, its
photometric distance is around $d=139$ pc.
The companion, PG 1210+464B, is estimated to be
spectral type M2 based on its deconvolved $R-I$ 
and $I-K$ colors.  Comparing its optical and near
near-infrared magnitudes in Table \ref{tbl-6} with
the expected absolute magnitudes of a typical M2 dwarf,
the distance is $d=111\pm4$ pc.

\subsubsection{PG 1654+160}

PG 1654+160 is a relatively rare DB+dM system --
the only such system out of 61 white dwarf plus M
dwarf pairs discovered or described in this work.
PG 1654+160A itself is an uncommon object, a DBV
\citep{win84,bea99}.

Spectroscopic investigation indicates the pulsating
helium atmosphere white dwarf has parameters in the
range $T_{\rm{eff}}=24,300-27,800$ K, log $g=7.95-8.00$,
depending on whether there is a trace amount of hydrogen
in its atmosphere or none.  Helium atmosphere models 
predict that a white dwarf with $T_{\rm{eff}}=26,000$ 
K and log $g=8.0$ would have $M_V=10.38$ mag and $M=0.61$
$M_{\odot}$.  If correct, this would place the PG 1654+160
system at a distance near $d=171$ pc.  

Recent astroseismological analysis of data acquired
using the Whole Earth Telescope may corroborate the 
spectroscopically determined parameters of PG 1654+160A.
Detection of nearly equal spacing between pulsation 
periods was found to be consistent with the expected
mean period spacing of a normal mass  ($M\approx0.6$ 
$M_{\odot}$) DB white dwarf pulsating in nonradial
$\ell=1$ modes \citep{han03}.  

However, the photometric distance to the M4.5
dwarf companion is positively inconsistent with
this interpretation.  Comparing the optical and
near-infrared magnitudes of PG 1654+160B with 
absolute magnitudes expected of an M4.5 dwarf 
yields a distance modulus of $m-M=4.49\pm0.11$
or $d=79\pm4$ pc.

This serious discrepancy calls into question the
binary nature of the apparent common proper motion
pair.  It is quite easy to see the elongated pair 
moving together while blinking the digitized POSS
I \& II red sensitive plates.  Both the relative
proper motion and the elongation are readily visible
using the blue and near-infrared sensitive scans, but the
contrast is best for the red plates where the components
have nearly equal brightness.  However, the pair is not
well resolved in either of the two POSS epochs, and 
therefore accurate photocenters (centroids) are not 
possible for any of these digitized scans.  The position
angle corresponding to the elongation axis appears constant
over the 46 year baseline between POSS epochs, consistent
with the $131.0^{\circ}$ value (epoch 2003.3) in Table 
\ref{tbl-3}.  Using a single centroid for the elongated
pair in the POSS I \& II red sensitive plate scans, a
proper motion of $\mu=0.085''$ ${\rm{yr}}^{-1}$ at 
$\theta=137^{\circ}$ is obtained over a 39 year baseline.
This value is somewhat greater than the USNO B1.0 catalog
value in Table \ref{tbl-1} but consistent with an object
whose photocenter is biased depending on whether blue 
(white dwarf dominated centroid), red (maximum elongation)
or near-infrared sensitive (red dwarf dominated centroid)
POSS plates are used for astrometry.  If the two stars are
unrelated, and one is a stationary background star, the pair
would have separated by as much as $4.5''$ since the epoch 
1950.4 POSS I red plate was observed.  The conclusion based
on this analysis is that the visual pair is likely a
physical binary based on its common proper motion.  

Still there remains the $\Delta m=1.75$ mag discrepancy
between the expected magnitude of an M4.5 dwarf at $d=79$
pc versus $d=171$ pc.  Clearly, the PG 1654+160 system
requires follow up observations and analysis in order 
to constrain both its distance and the parameters of 
its components.

\subsubsection{PG 1659+303}

PG 1659+303A is a DA white dwarf with spectroscopically
determined parameters $T_{\rm{eff}}=13,600$ K, log $g=7.95$,
and $M_V=11.35$ mag \citep{lie05}, corresponding to $d=53$
pc.  The $V-K=4.28$ color of PG 1659+303B places it between
spectral types M2$-$M2.5.  Its measured spectrum in Figure
\ref{fig33} is consistent with this interpretation and gives
no indication of subsolar metallicity (i.e. it is not a
subdwarf).  Comparing the expected optical and near-infrared
absolute magnitudes with the apparent magnitudes in Table
\ref{tbl-5}, the distance falls in the range $d=67-82$ pc.

\subsubsection{Rubin 80}

Rubin 80A is listed as a DA6 in \citet{mcc99} while
referencing an unpublished spectrum taken around 1979.
Caution is warranted because an unresolved low mass
stellar companion can contaminate both spectra and colors,
thus causing a white dwarf to appear cooler than it is 
\citep{far04c}.  \citet{gre86a} first noted the companion
to Rubin 80A in its composite spectrum as well as commenting
that the companion affected his measured colors of the white
dwarf.  \citet{zuc03} give $T_{\rm{eff}}=7765$ K based
on $UBVRI$ colors, but this might be too cool due to
the inclusion of the $RI$ bands in the determination.
Examining only the $U-B=-0.57$ and $B-V=+0.28$ colors 
(this work finds $U-B=-0.55$, $B-V=+0.29$), an effective
temperature near 8000 K (log $g=8$) is implied by both 
color indices independently.  Hence it appears the DA6
type is likely to be accurate, assuming the flux of
Rubin 80B does not contribute at $B$ (a safe assumption).

However, the blue continuum slope of Rubin 80
(Figure \ref{fig52}) is slightly steeper than that
of LTT 8747 ($T_{\rm{eff}}=8500$ K, Figure \ref{fig53}),
which is consistent with Rubin 80A having a higher effective
temperature than LTT 8747.  Both stars were observed
on the same night with the same instrument, setup and
calibration.  Possible errors in flux calibration were
searched for unsuccessfully and a variety of standard
star sensitivity functions were used, all producing 
similar results.  In order to deconvolve the $IJHK$
magnitudes of the companion, an effective temperature 
of 9000 K was used for the values in Table \ref{tbl-6}.
While this may be too high, it turns out to be more 
consistent with the resulting parameters for the red 
dwarf companion, Rubin 80B.  Some representative,
plausible white dwarf parameters are given in Table
\ref{tbl-11}.  Details can be found in \citet{far04c}.

\subsubsection{Ton S 392}

\begin{figure} 
\plotone{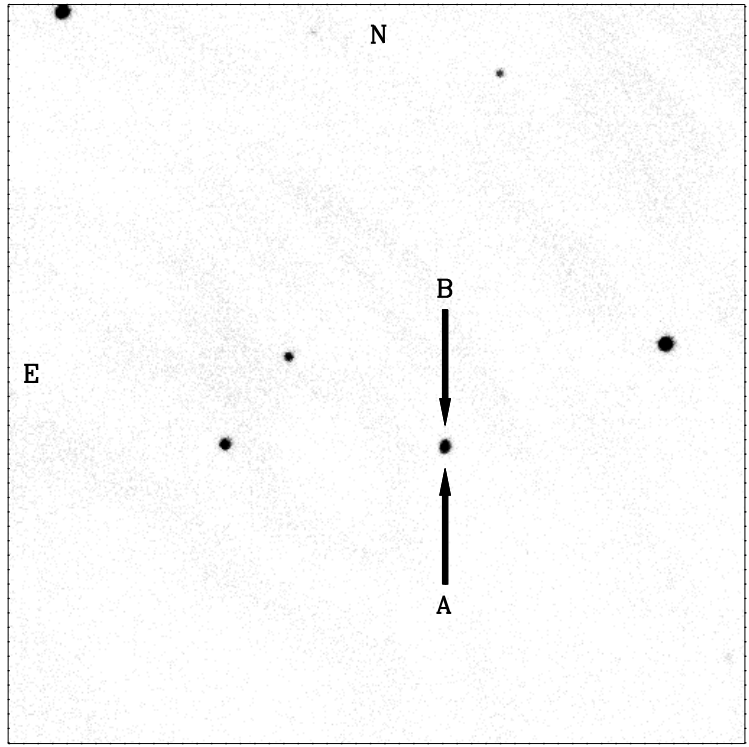}
\caption{Optical finding chart for Ton S 392B, taken at $I$ band with the
Nickel 1 meter telescope in October 2003.  The image is $184''$ square
with $0.36''$ pixels.  The coordinates for the binary are $03^{\rm h}
59^{\rm m} 04.9^{\rm s}$, $-23^{\circ} 12' 25''$ J2000.  At a
separation of $~\sim1.2''$, the companion is just barely resolved in this
image.
\label{fig23}}
\end{figure}

\begin{figure}
\plotone{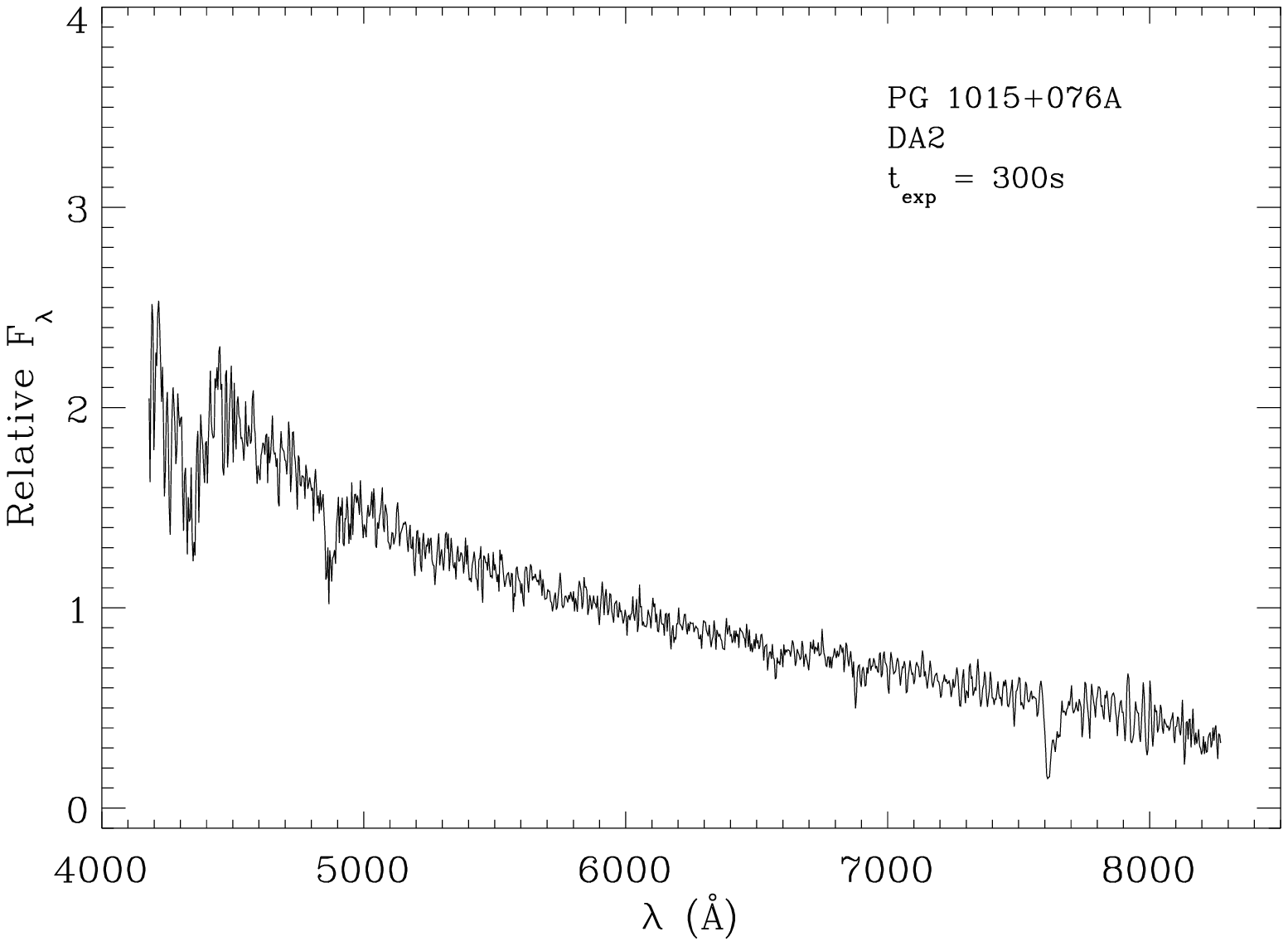}
\caption{Optical spectrum of PG 1015+076A taken with the Boller \&
Chivens Spectrograph on the Bok 2.3 meter in April 2003.
\label{fig24}}
\end{figure}

\begin{figure}
\plotone{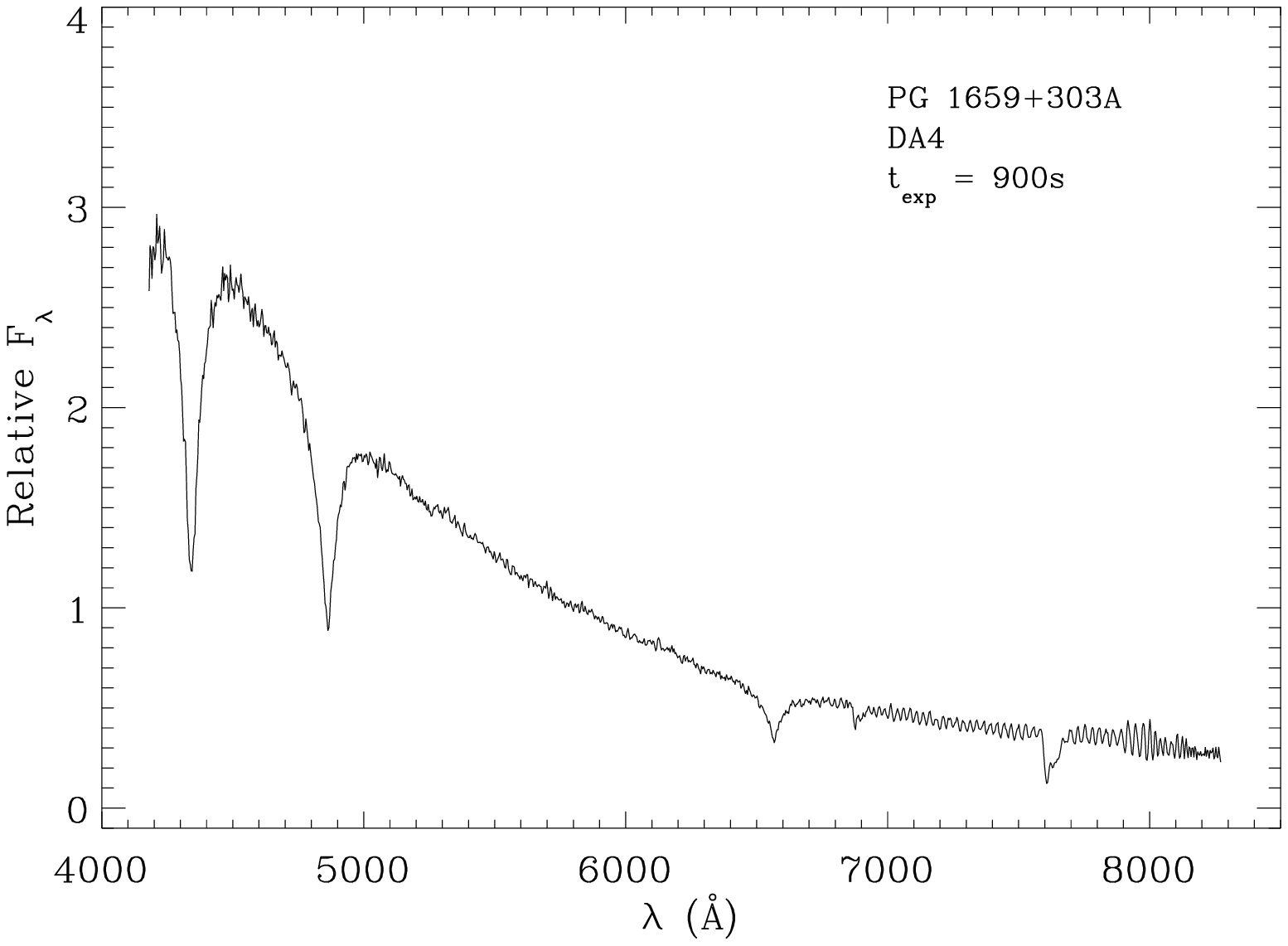}
\caption{Optical spectrum of PG 1659+303A taken with the Boller \&
Chivens Spectrograph on the Bok 2.3 meter in April 2003.
\label{fig25}}
\end{figure}

\begin{figure}
\plotone{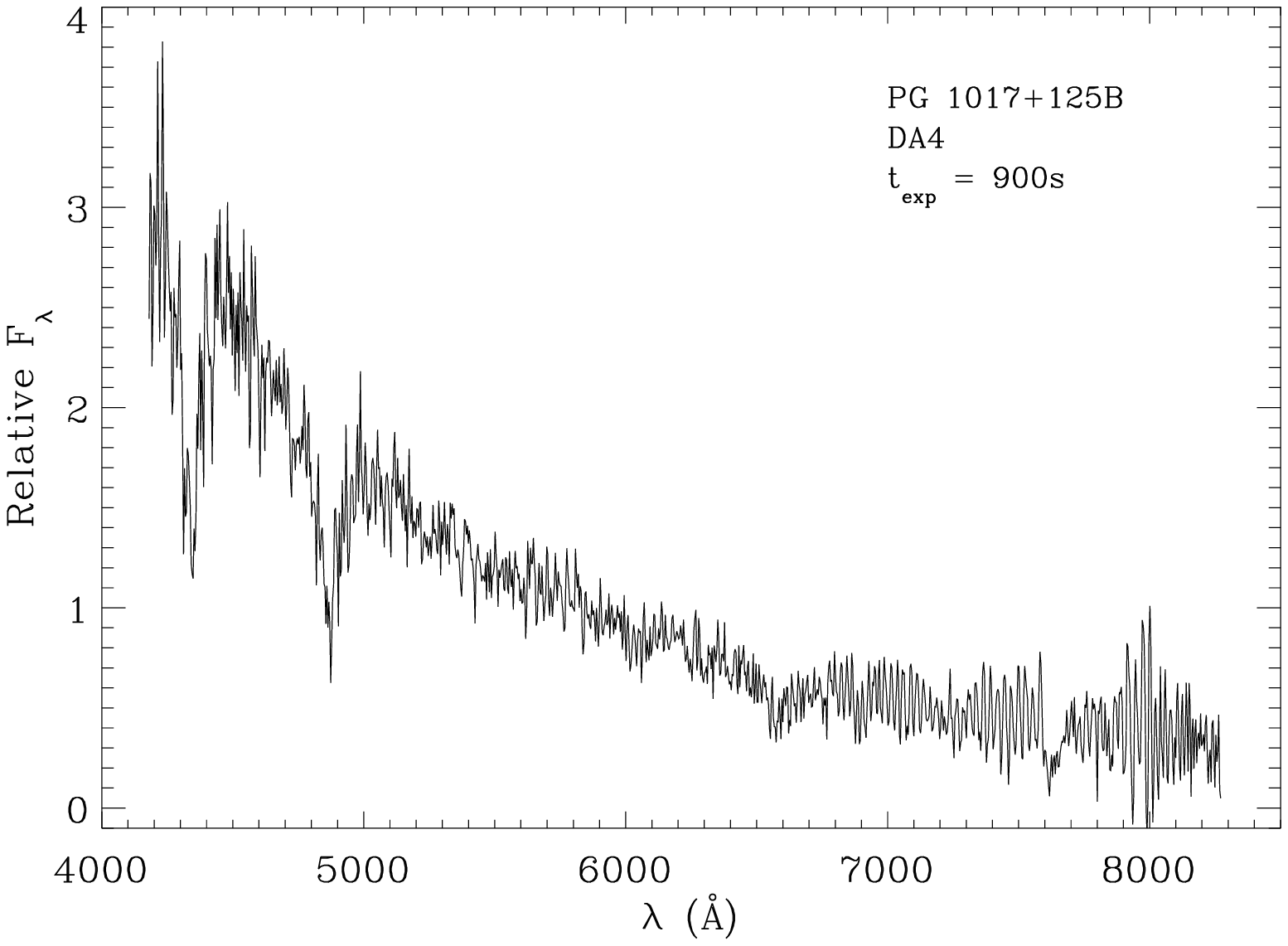}
\caption{Optical spectrum of PG 1017+125B taken with the Boller \&
Chivens Spectrograph on the Bok 2.3 meter in April 2003.
\label{fig26}}
\end{figure}

\begin{figure}
\plotone{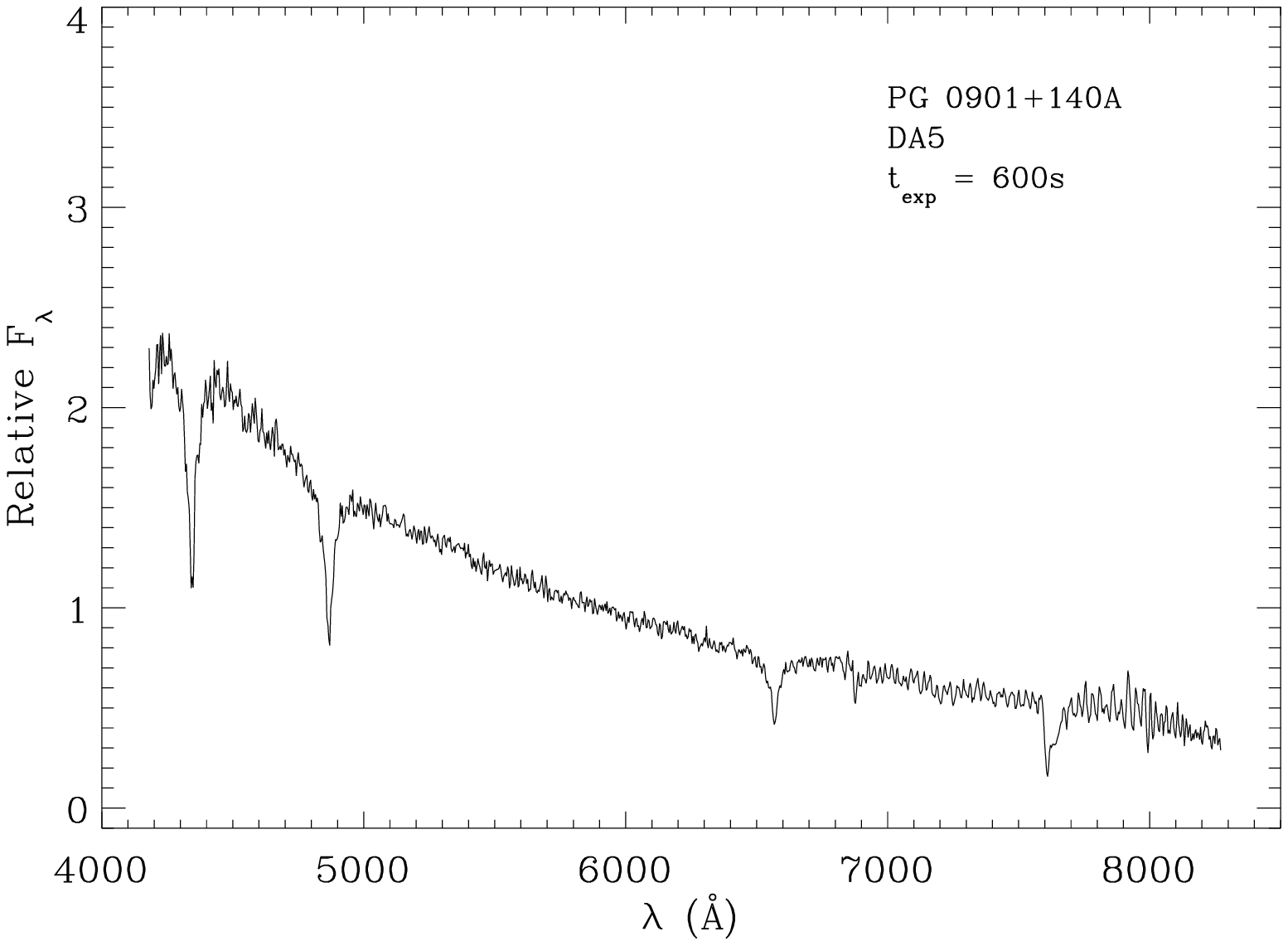}
\caption{Optical spectrum of PG 0901+140A taken with the Boller \&
Chivens Spectrograph on the Bok 2.3 meter in April 2003.
\label{fig27}}
\end{figure}

\begin{figure}
\plotone{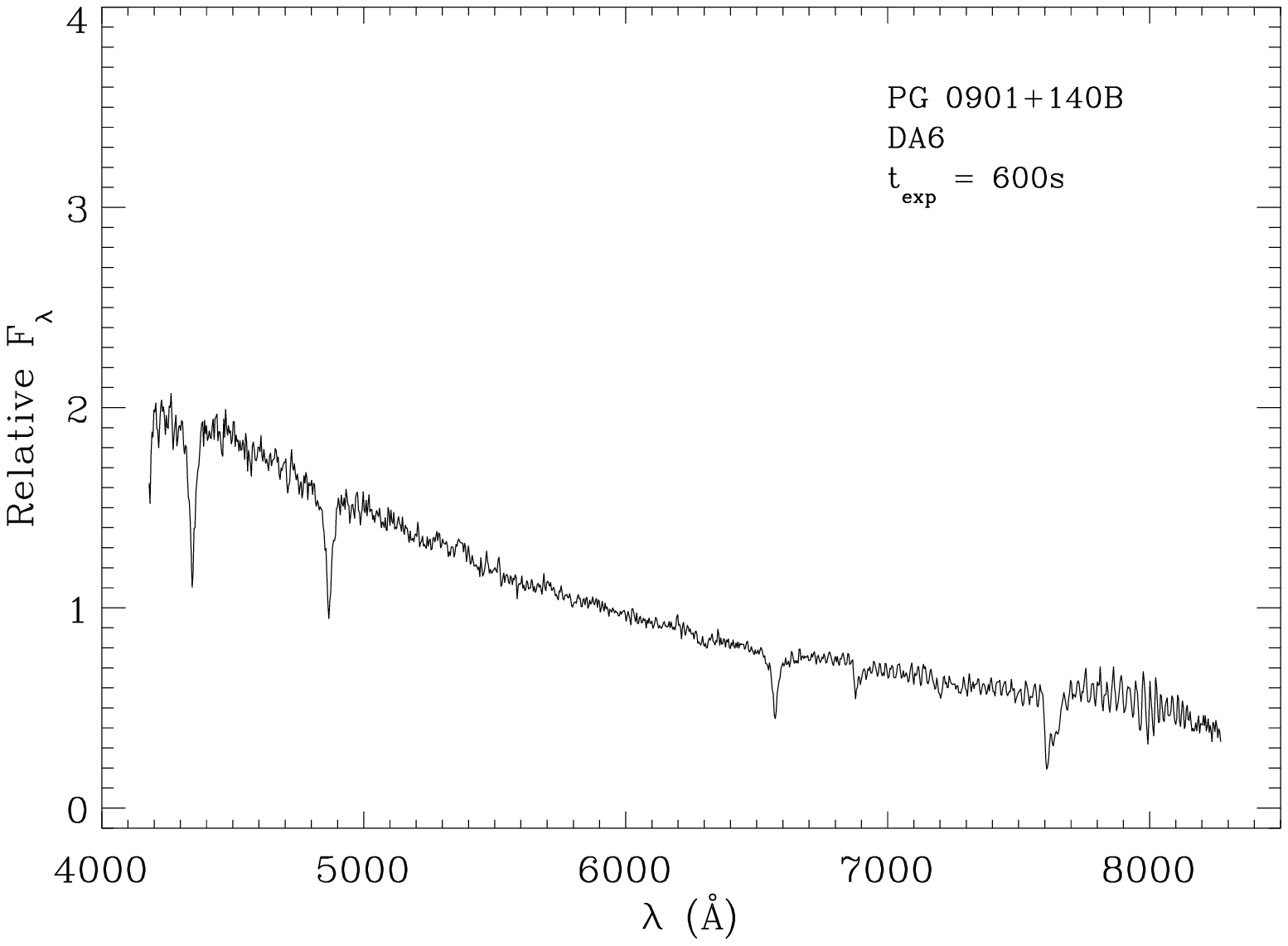}
\caption{Optical spectrum of PG 0901+140B taken with the Boller \&
Chivens Spectrograph on the Bok 2.3 meter in April 2003.
\label{fig28}}
\end{figure}

\begin{figure}
\plotone{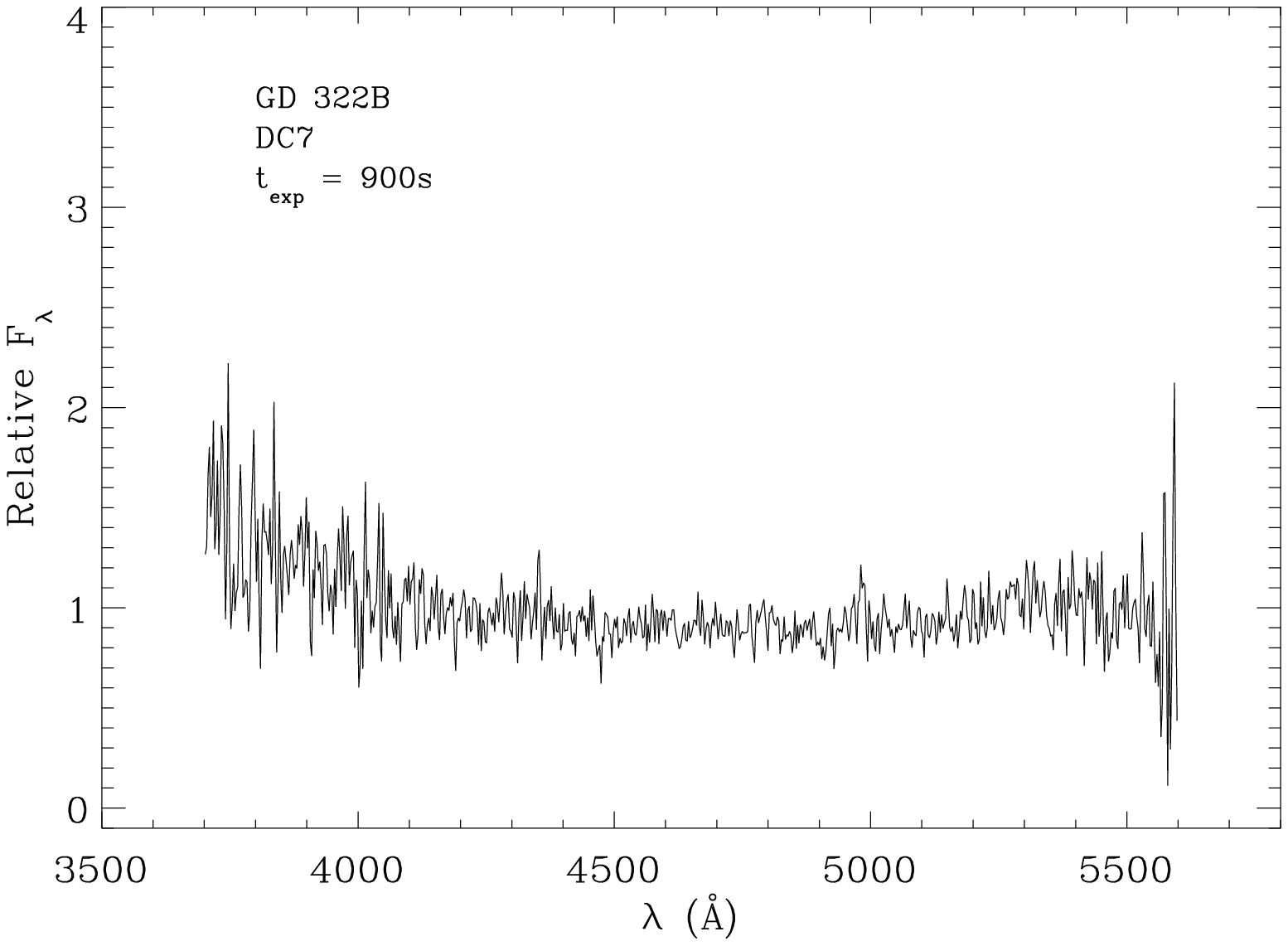}
\caption{Blue optical spectrum of GD 322B taken with the Kast Spectrograph
on the Shane 3 meter telescope in August 2002.
\label{fig29}}
\end{figure}

\begin{figure}
\plotone{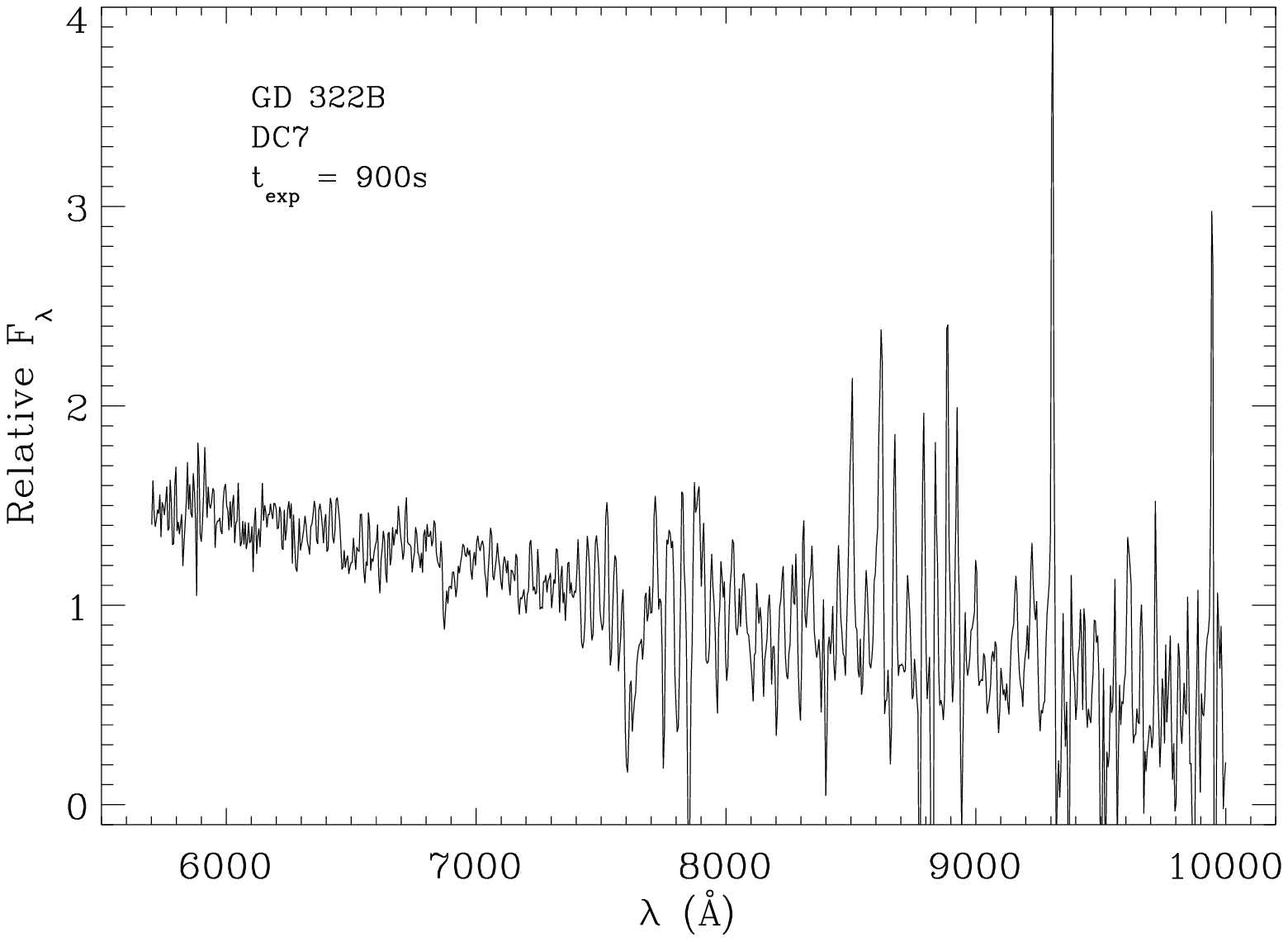}
\caption{Red optical spectrum of GD 322B taken with the Kast Spectrograph
on the Shane 3 meter telescope in August 2002.
\label{fig30}}
\end{figure}

\begin{figure}
\plotone{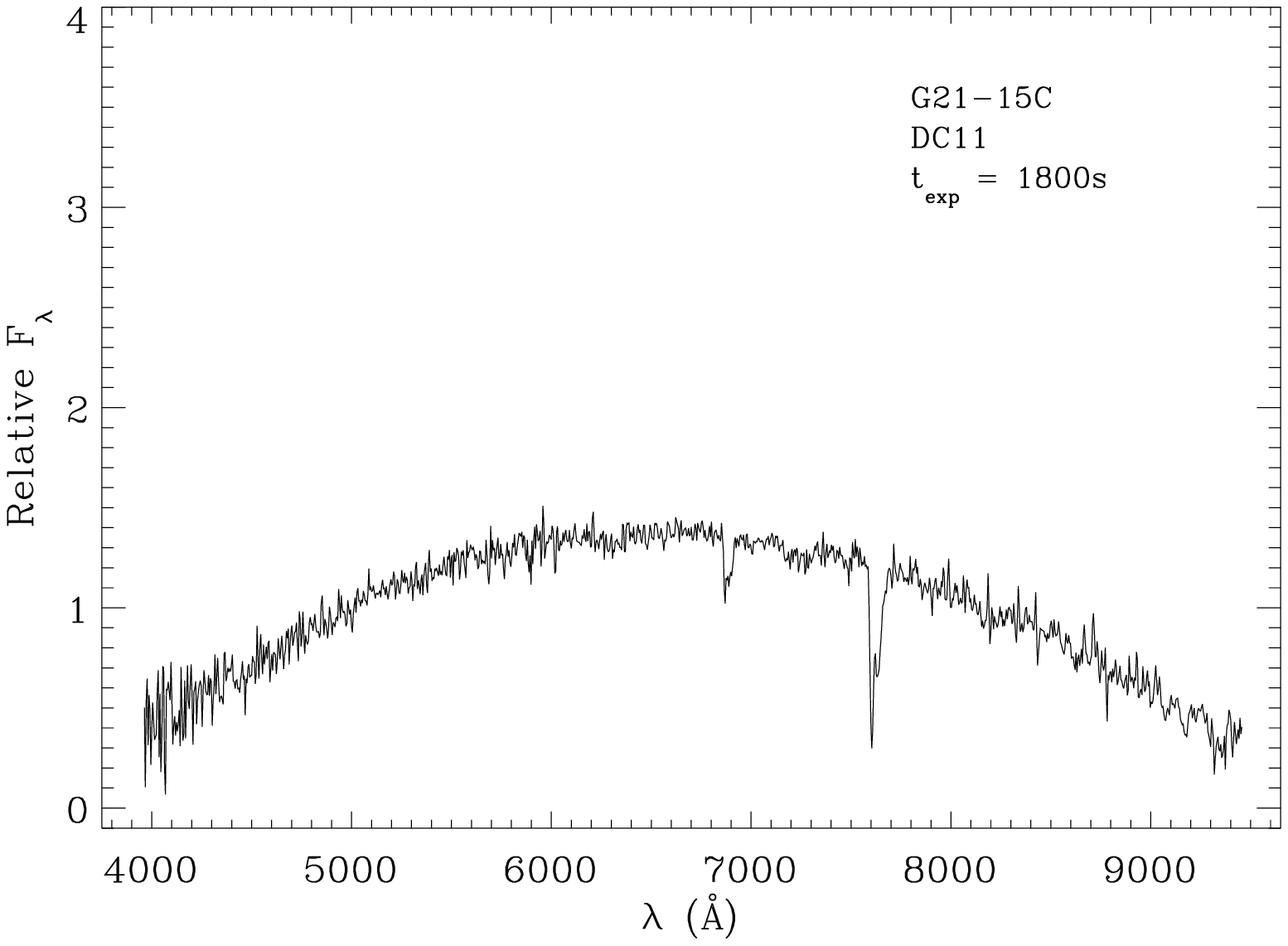}
\caption{Optical spectrum of G21-15C taken with the Kast Spectrograph
on the Shane 3 meter telescope in August 2003.
\label{fig31}}
\end{figure}

\begin{figure}
\plotone{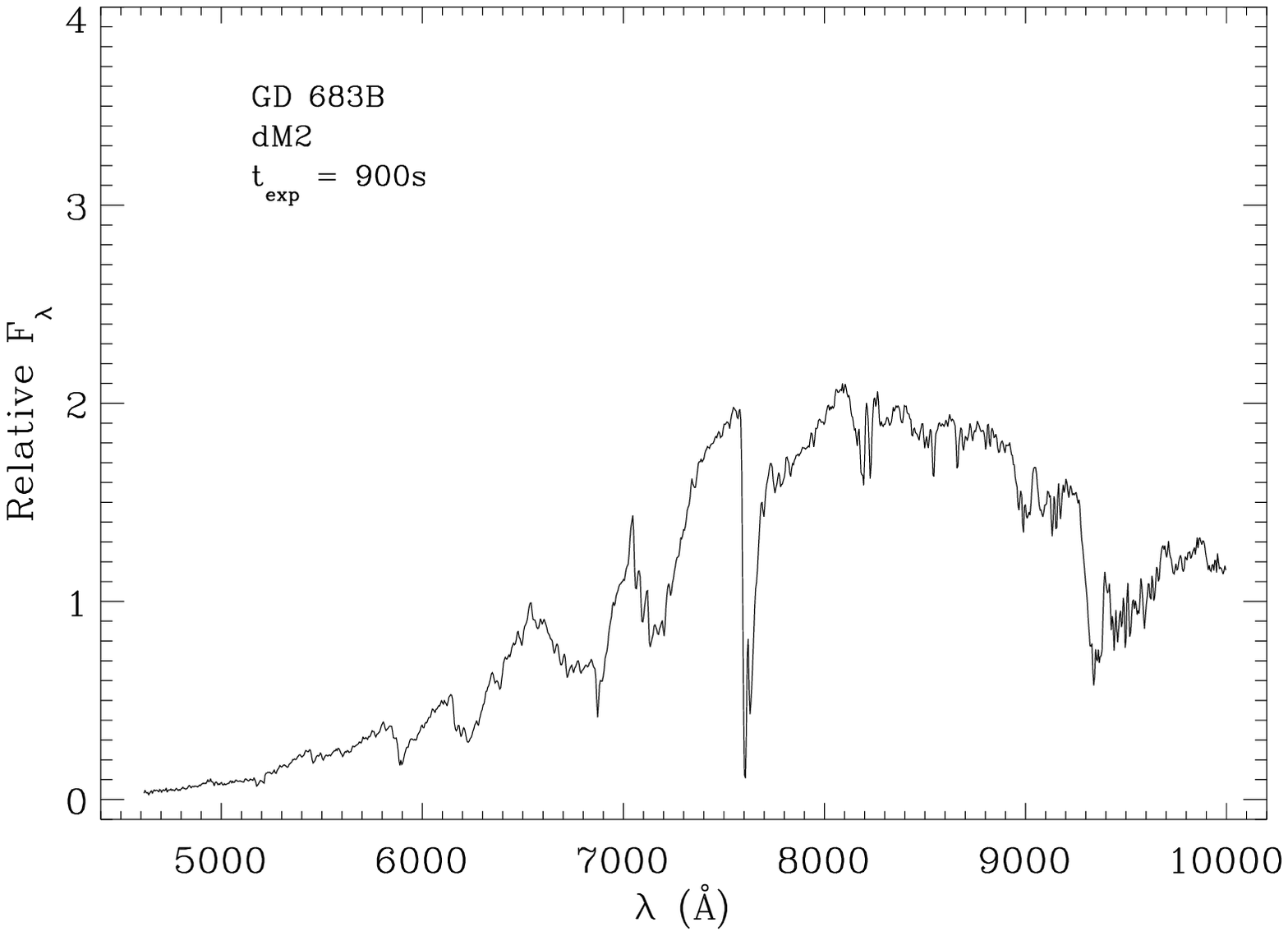}
\caption{Optical spectrum of GD 683B taken with the Kast Spectrograph
on the Shane 3 meter telescope in August 2003.
\label{fig32}}
\end{figure}

\begin{figure}
\plotone{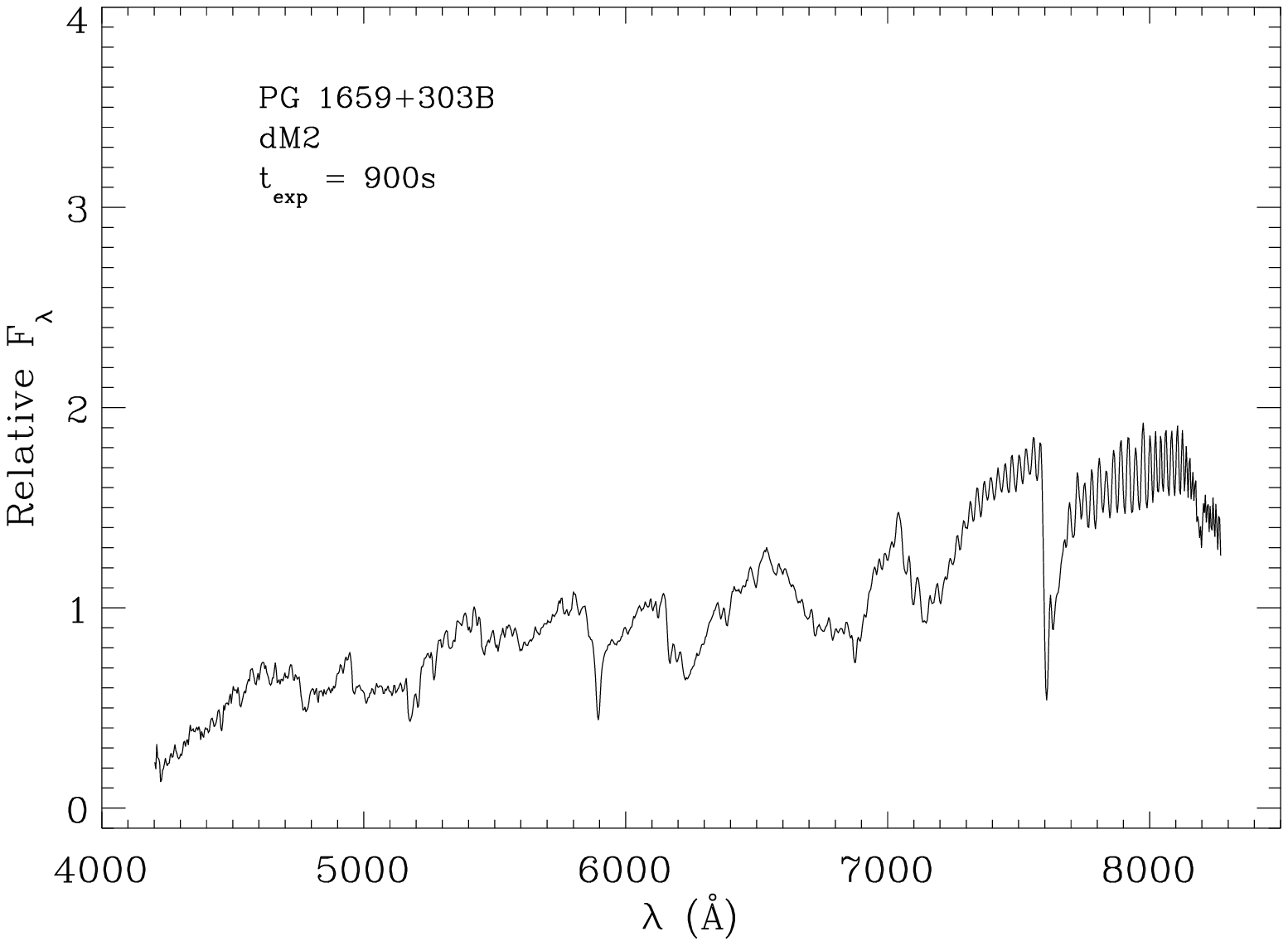}
\caption{Optical spectrum of PG 1659+303B taken with the Boller \&
Chivens Spectrograph on the Bok 2.3 meter in April 2003.
\label{fig33}}
\end{figure}

\begin{figure}
\plotone{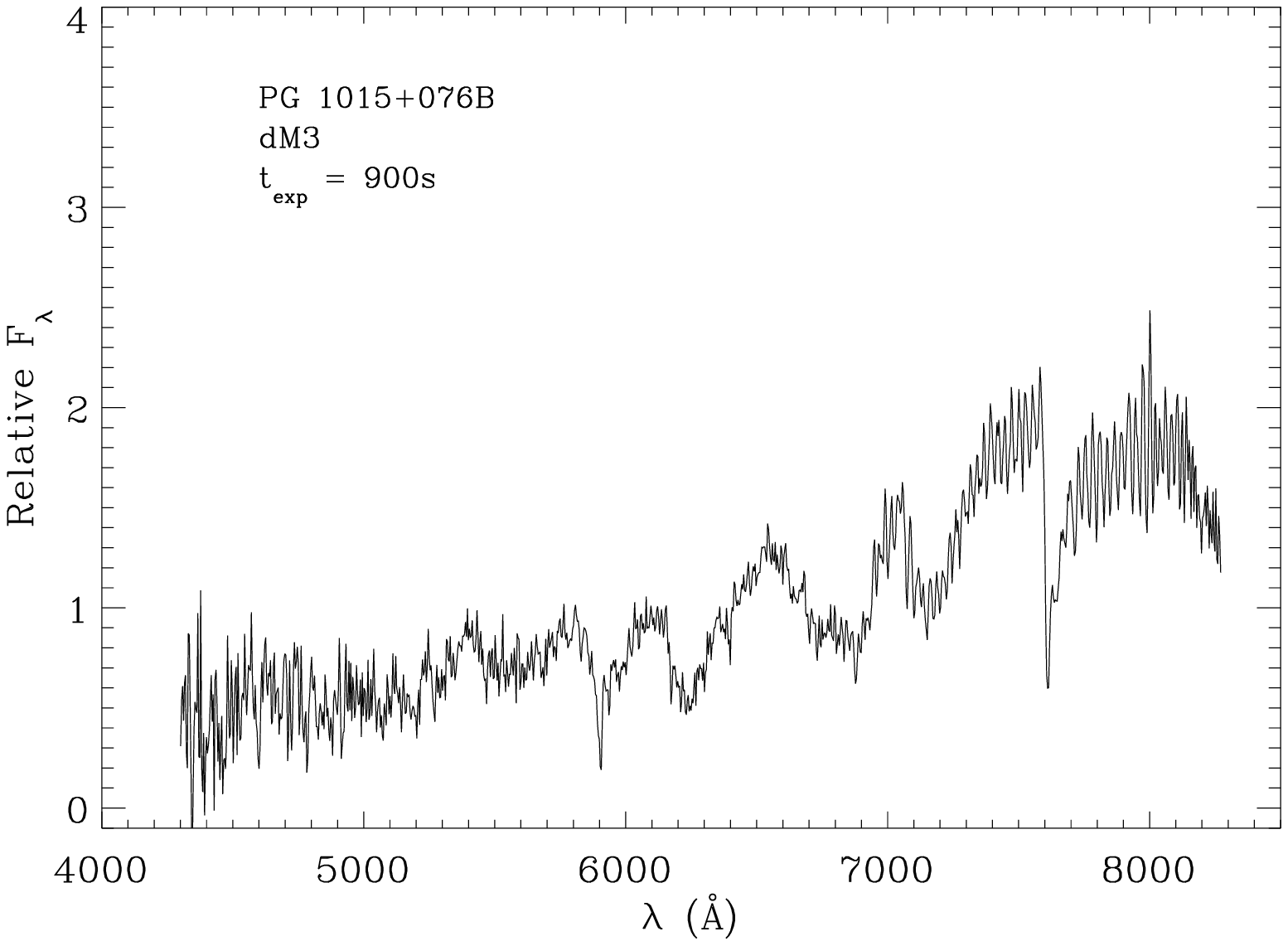}
\caption{Optical spectrum of PG 1015+076B taken with the Boller \&
Chivens Spectrograph on the Bok 2.3 meter in April 2003.
\label{fig34}}
\end{figure}

\begin{figure}
\plotone{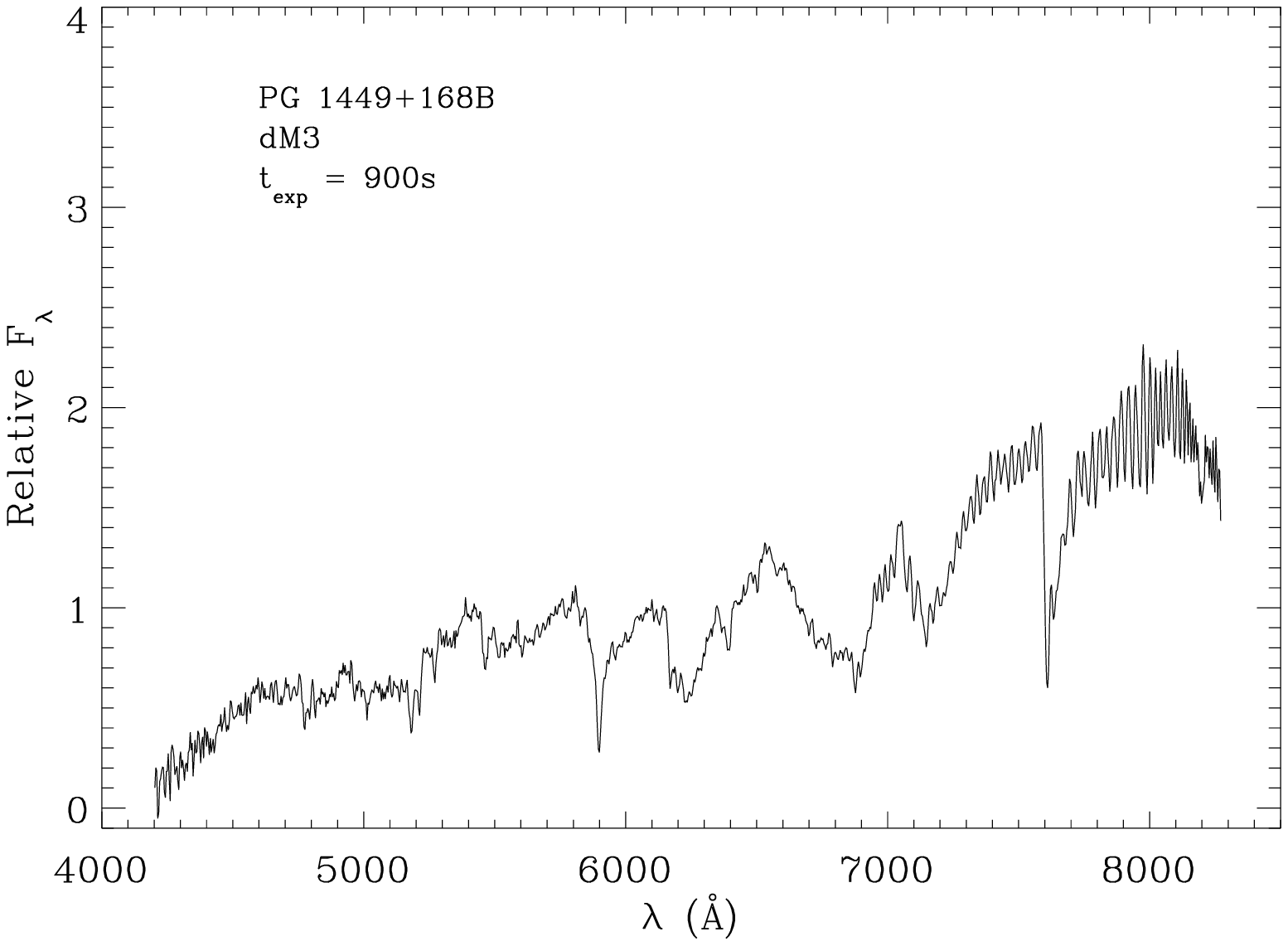}
\caption{Optical spectrum of PG 1449+168B taken with the Boller \&
Chivens Spectrograph on the Bok 2.3 meter in April 2003.
\label{fig35}}
\end{figure}

\begin{figure}
\plotone{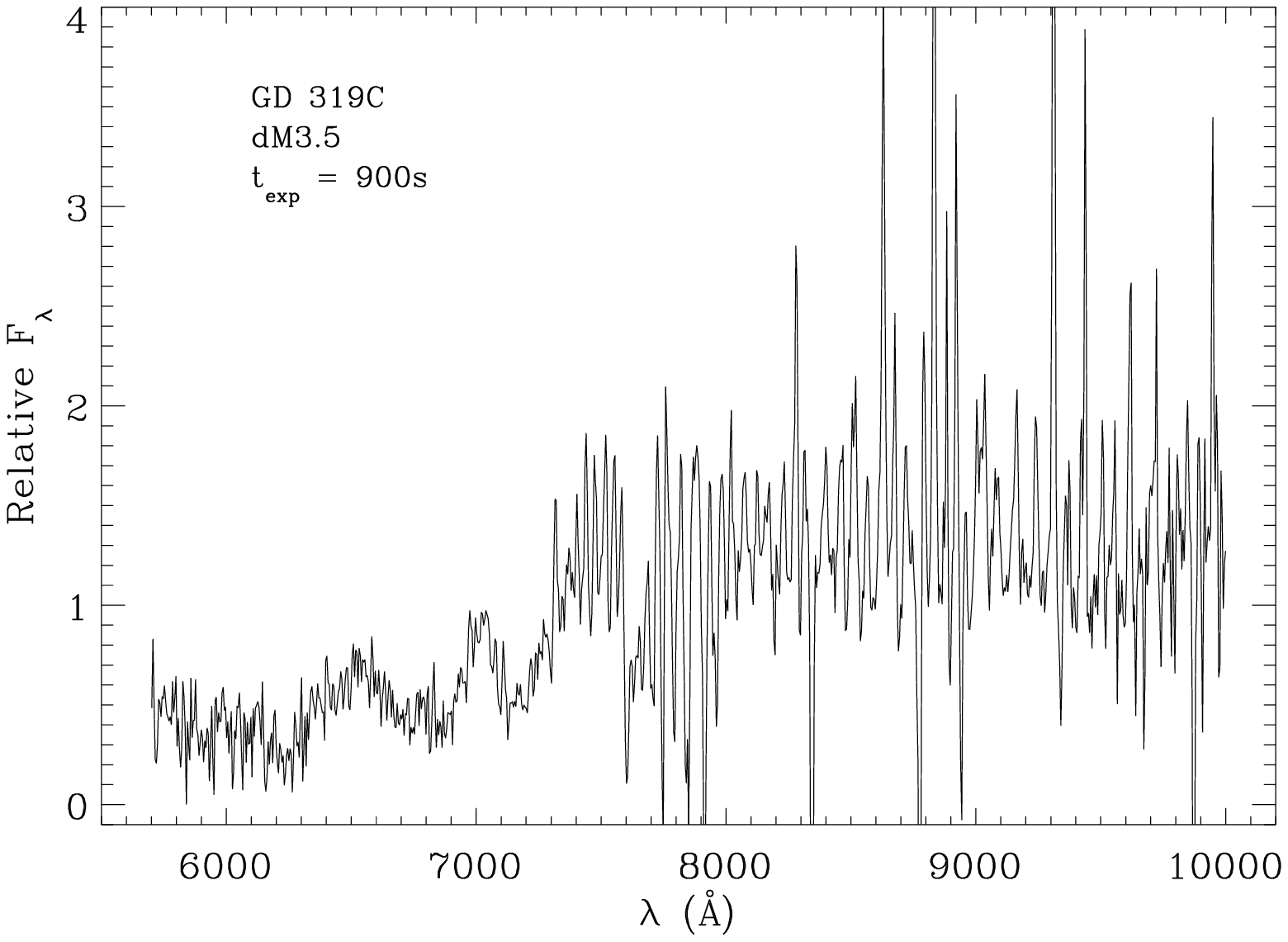}
\caption{Red optical spectrum of GD 319C taken with the Kast Spectrograph
on the Shane 3 meter telescope in August 2002.
\label{fig36}}
\end{figure}

\begin{figure}
\plotone{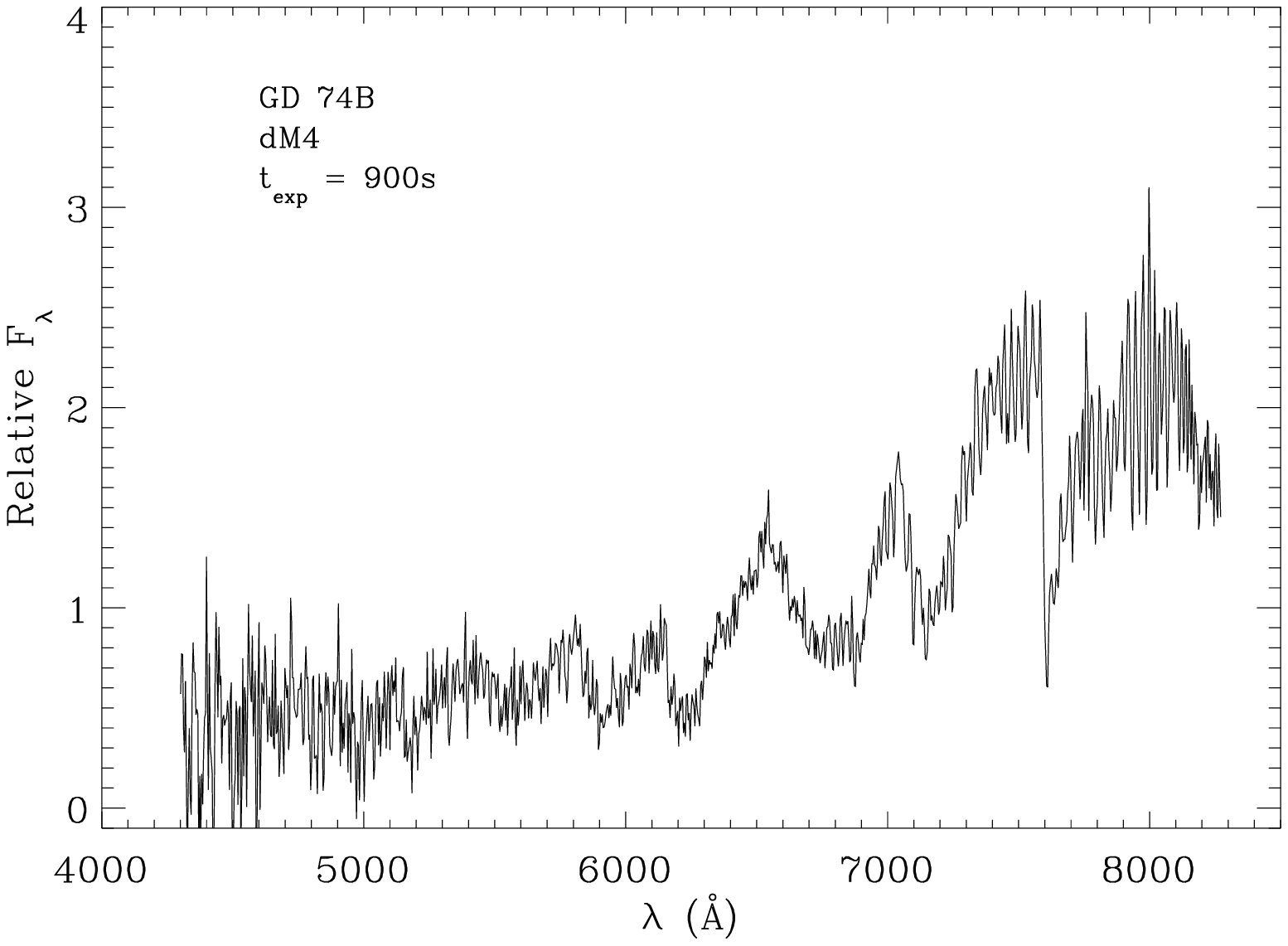}
\caption{Optical spectrum of GD 74B taken with the Boller \&
Chivens Spectrograph on the Bok 2.3 meter in April 2003.
\label{fig37}}
\end{figure}

\begin{figure}
\plotone{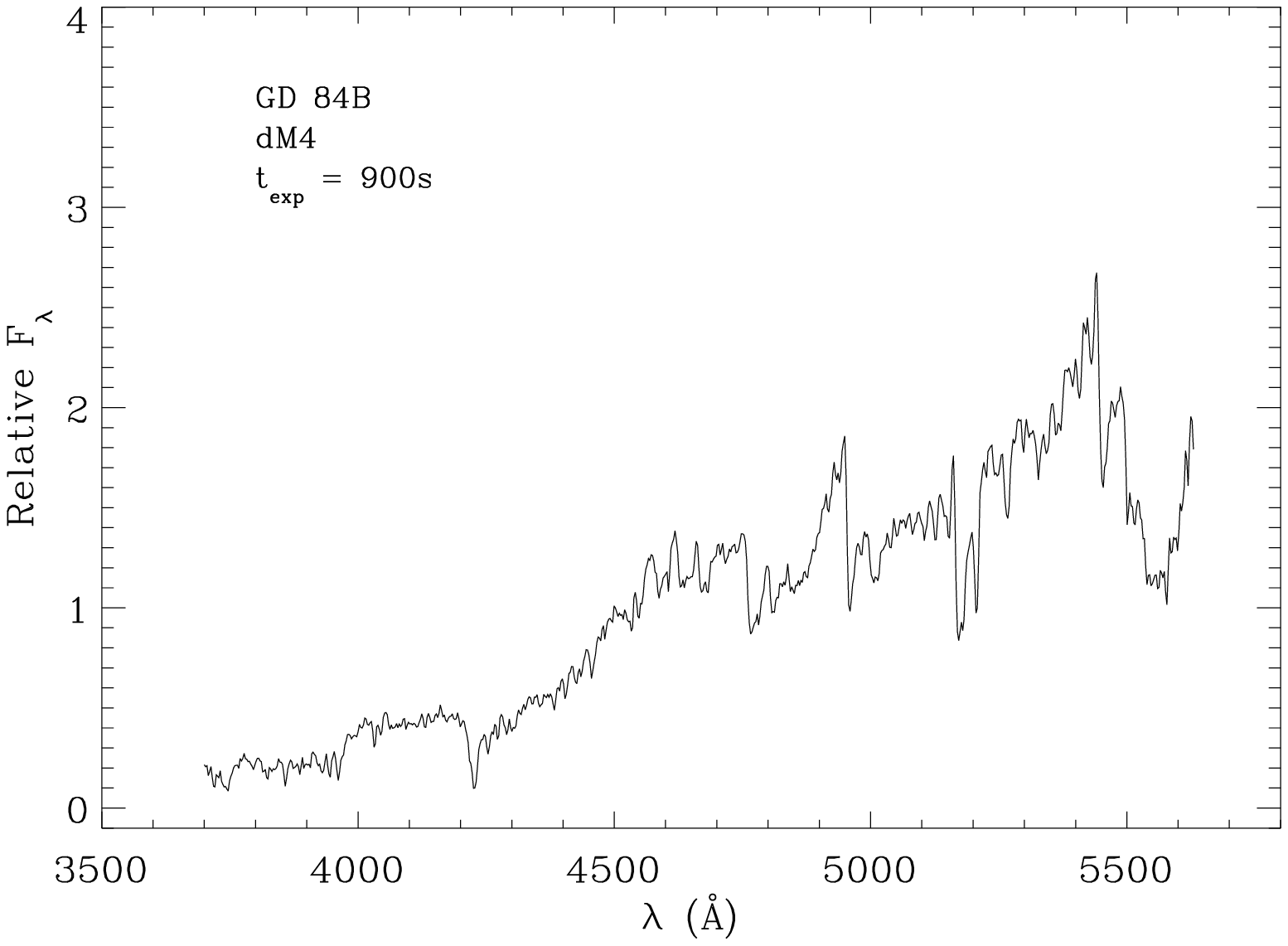}
\caption{Blue optical spectrum of GD 84B taken with the Kast Spectrograph
on the Shane 3 meter telescope in February 2002.
\label{fig38}}
\end{figure}

\begin{figure}
\plotone{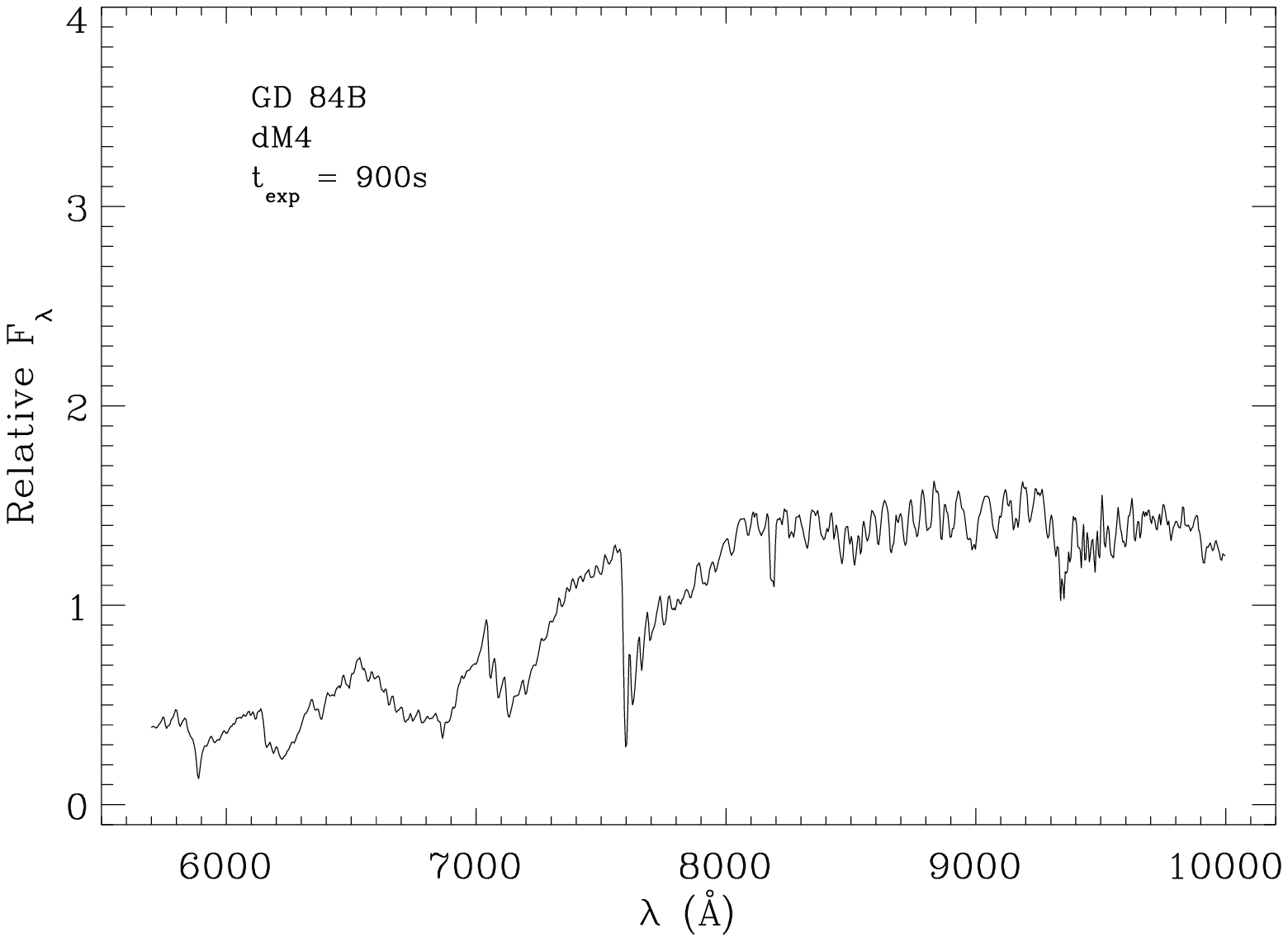}
\caption{Red optical spectrum of GD 84B taken with the Kast Spectrograph
on the Shane 3 meter telescope in February 2002.
\label{fig39}}
\end{figure}

\begin{figure}
\plotone{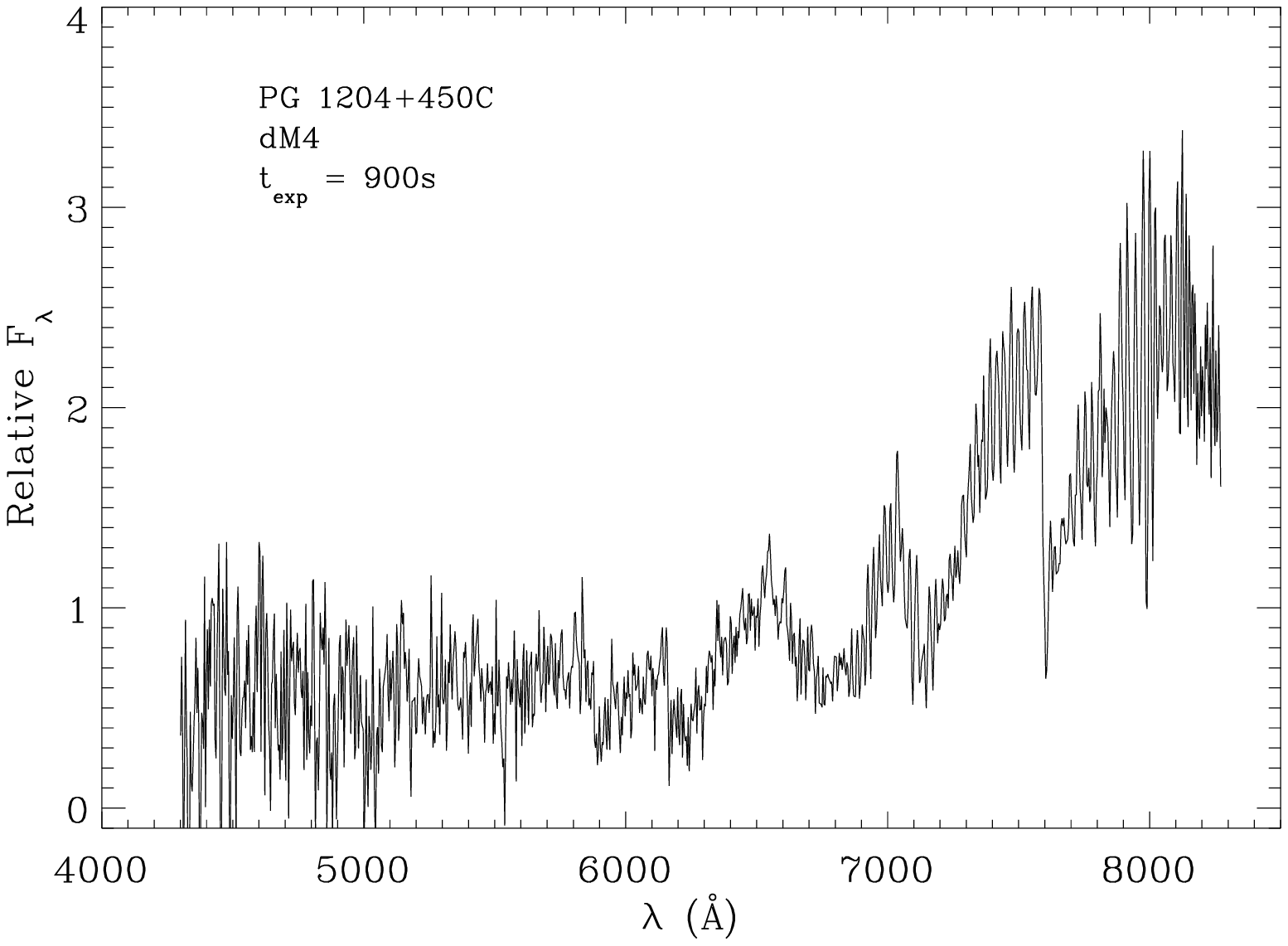}
\caption{Optical spectrum of PG 1204+450C taken with the Boller \&
Chivens Spectrograph on the Bok 2.3 meter in April 2003.
\label{fig40}}
\end{figure}

\begin{figure}
\plotone{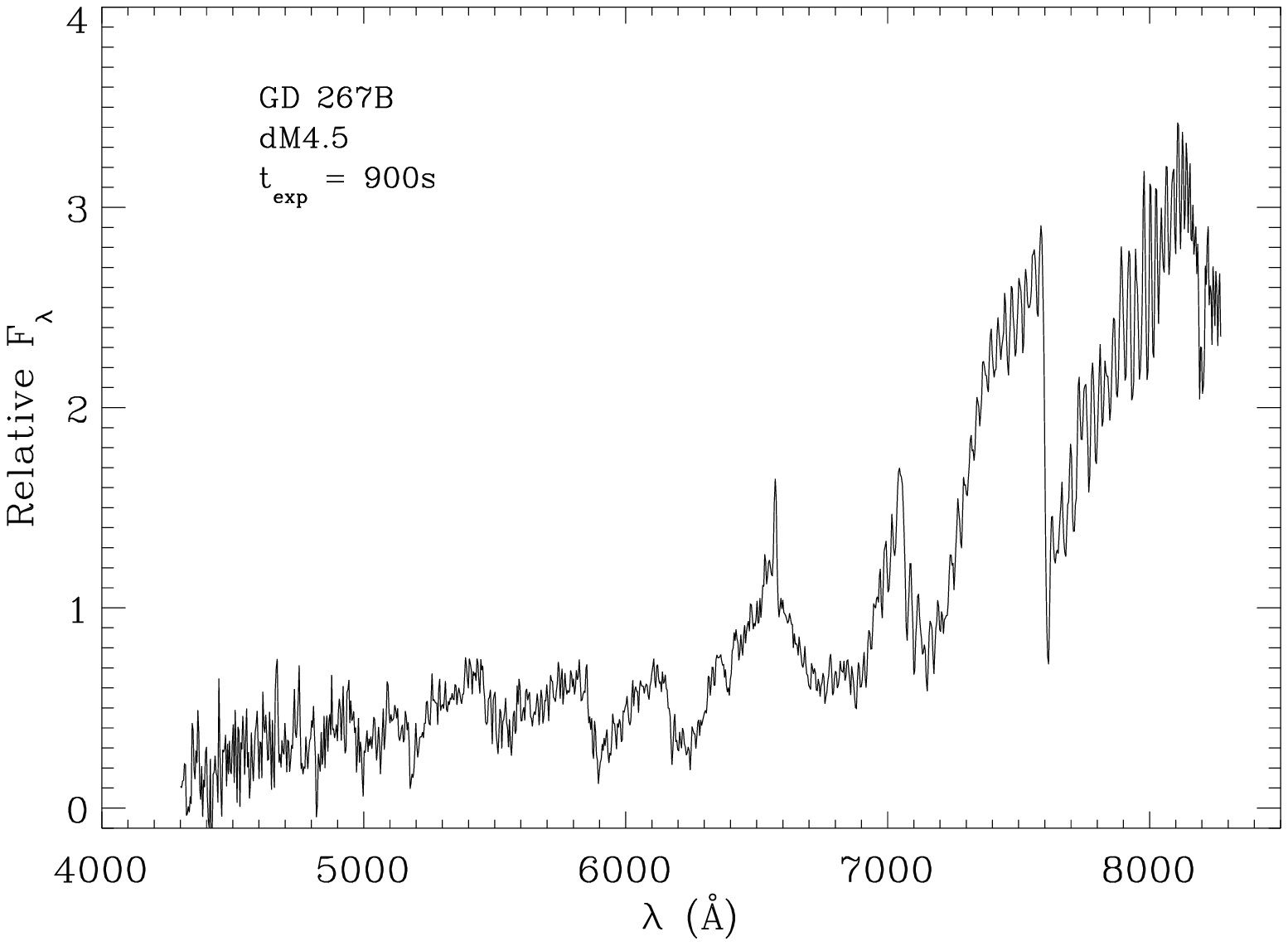}
\caption{Optical spectrum of GD 267B taken with the Boller \&
Chivens Spectrograph on the Bok 2.3 meter in April 2003.
\label{fig41}}
\end{figure}

\begin{figure}
\plotone{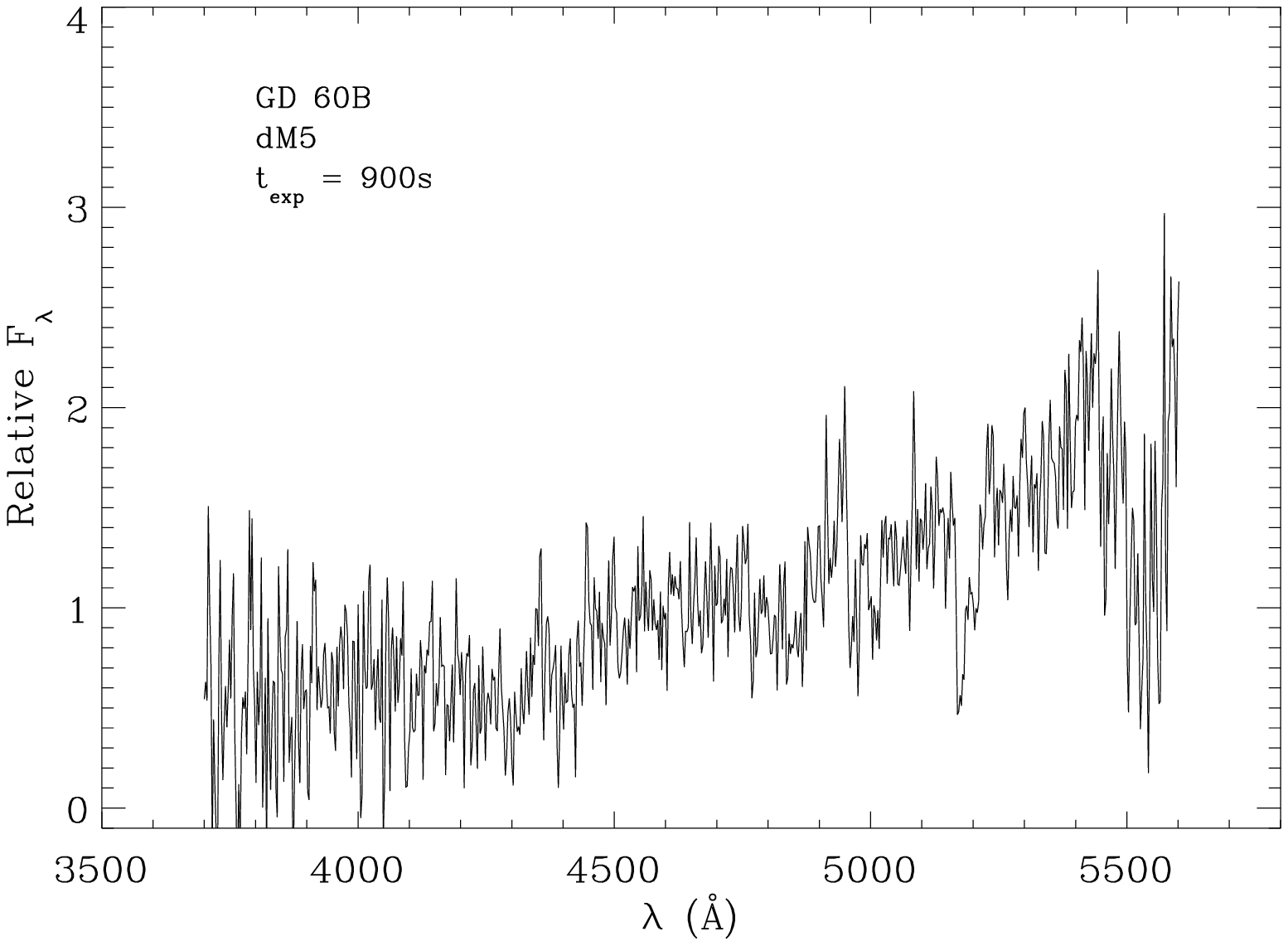}
\caption{Blue optical spectrum of GD 60B taken with the Kast Spectrograph
on the Shane 3 meter telescope in February 2002.
\label{fig42}}
\end{figure}

\begin{figure}
\plotone{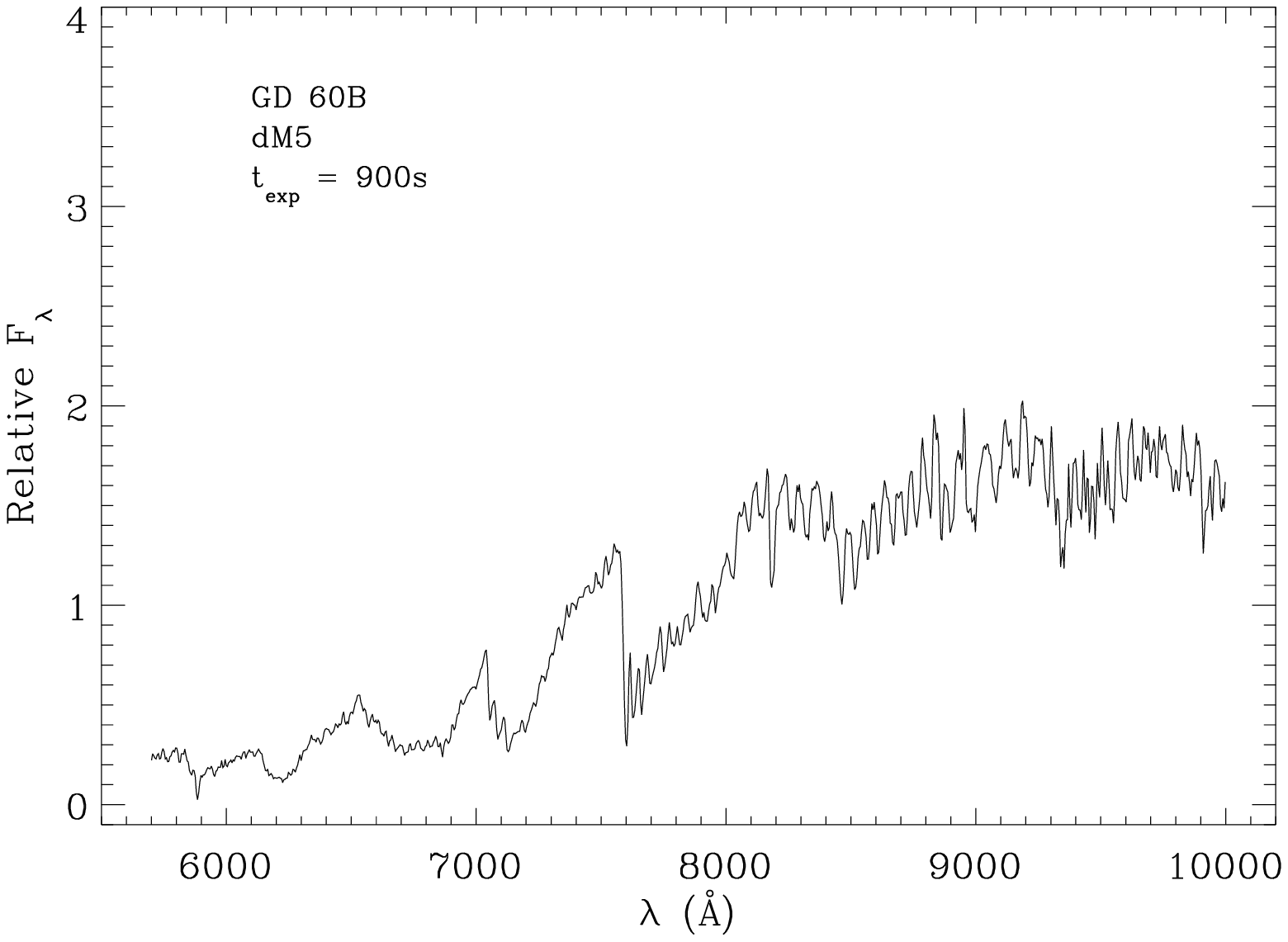}
\caption{Red optical spectrum of GD 60B taken with the Kast Spectrograph
on the Shane 3 meter telescope in February 2002.
\label{fig43}}
\end{figure}

\begin{figure}
\plotone{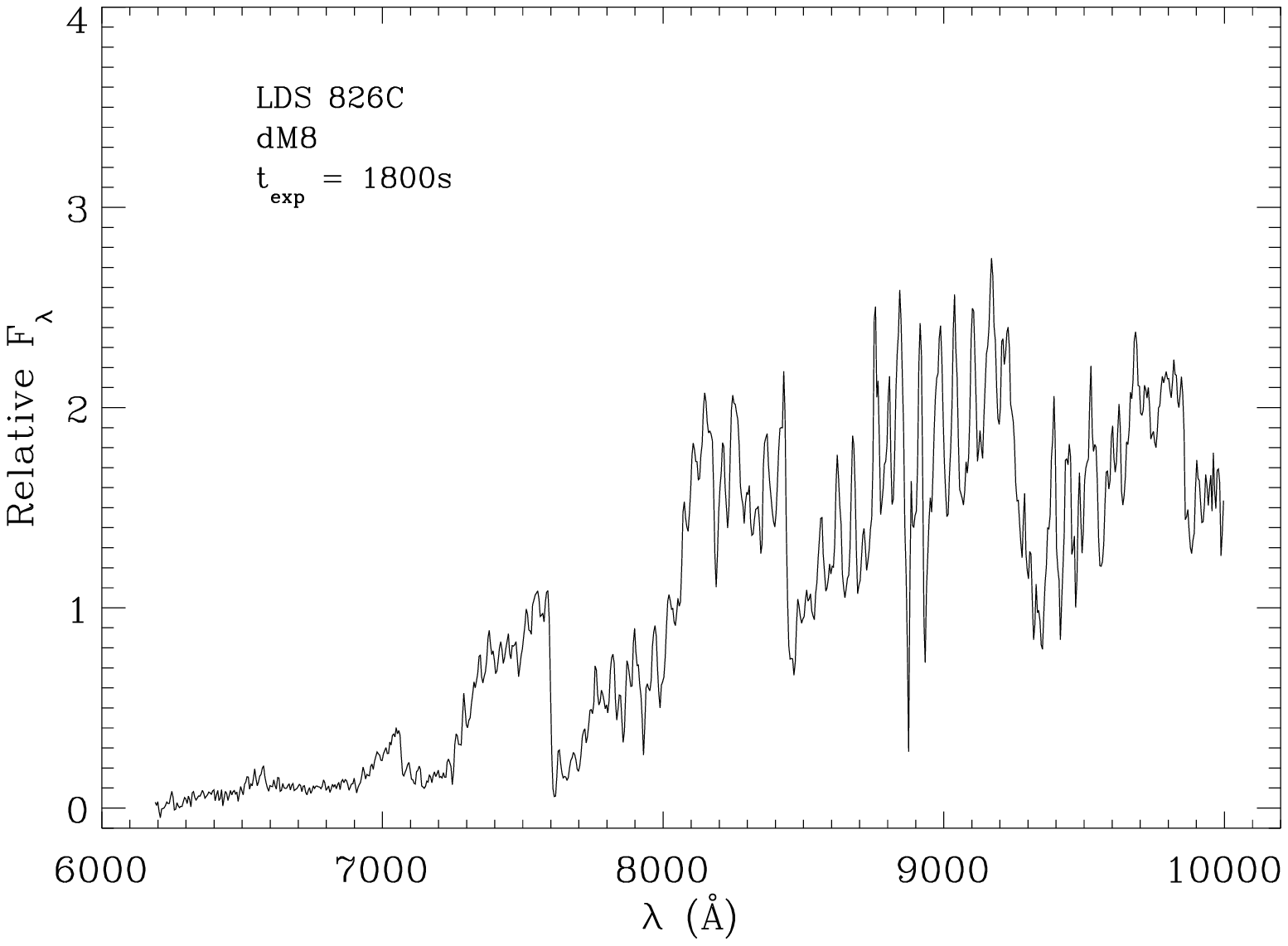}
\caption{Optical spectrum of LDS 826C taken with the Kast Spectrograph
on the Shane 3 meter telescope in August 2003.
\label{fig44}}
\end{figure}

\begin{figure}
\plotone{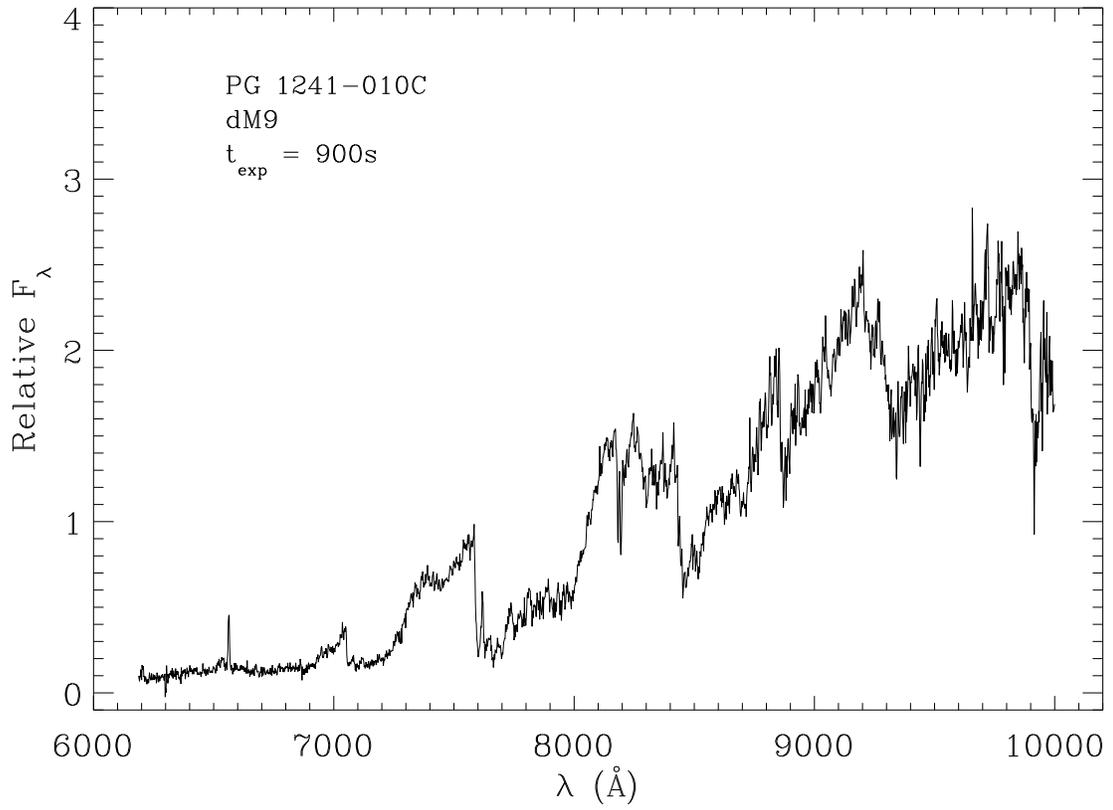}
\caption{Optical spectrum of PG 1241$-$010C taken with the Low Resolution
Imaging Spectrograph on the Keck I 10 meter telescope in May 2003.
\label{fig45}}
\end{figure}

\begin{figure}
\plotone{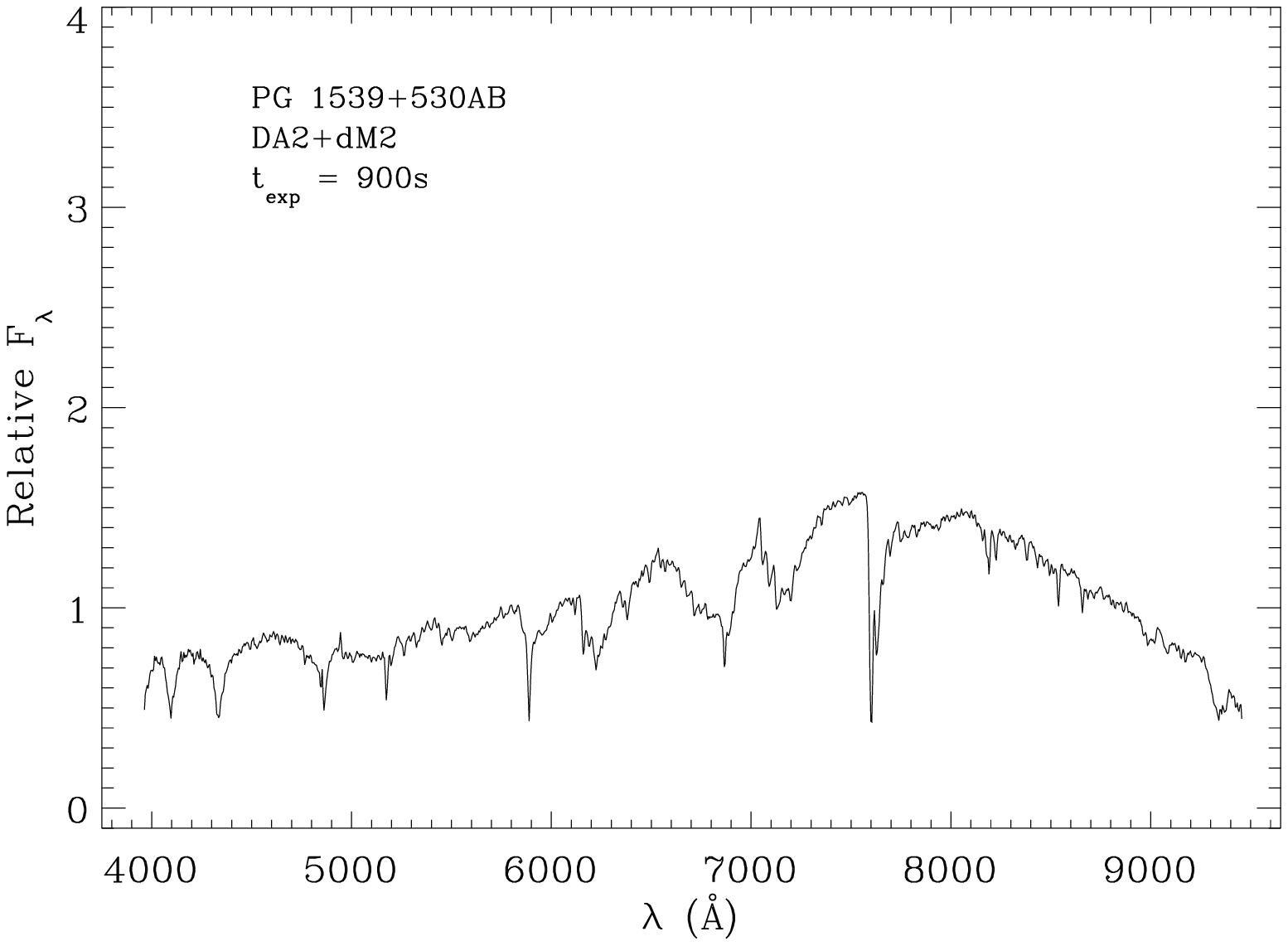}
\caption{Optical spectrum of PG 1539+530AB taken with the Kast Spectrograph
on the Shane 3 meter telescope in August 2003.
\label{fig46}}
\end{figure}

TS 392A is a very hot DA white dwarf that is not
well studied. The only published paper containing 
data on TS 392A is \citet{gre79}, who notes the white
dwarf is a ``very hot'' narrow lined DA.  The coordinates
in \citet{mcc99} are inaccurate by almost $3'$.  Accurate
coordinates are given in Figure \ref{fig23} and have been
checked against a photographic finding chart provided by
J. Greenstein (B. Zuckerman 2002, private communication).

The near-infrared excess of TS 392 was noticed in
January 1992 at the IRTF, but remained unpublished
until now.  \citet{wac03} were the first to publish
near-infrared magnitudes for TS 392 which indicate
the presence of a cool stellar companion.

\begin{figure}
\plotone{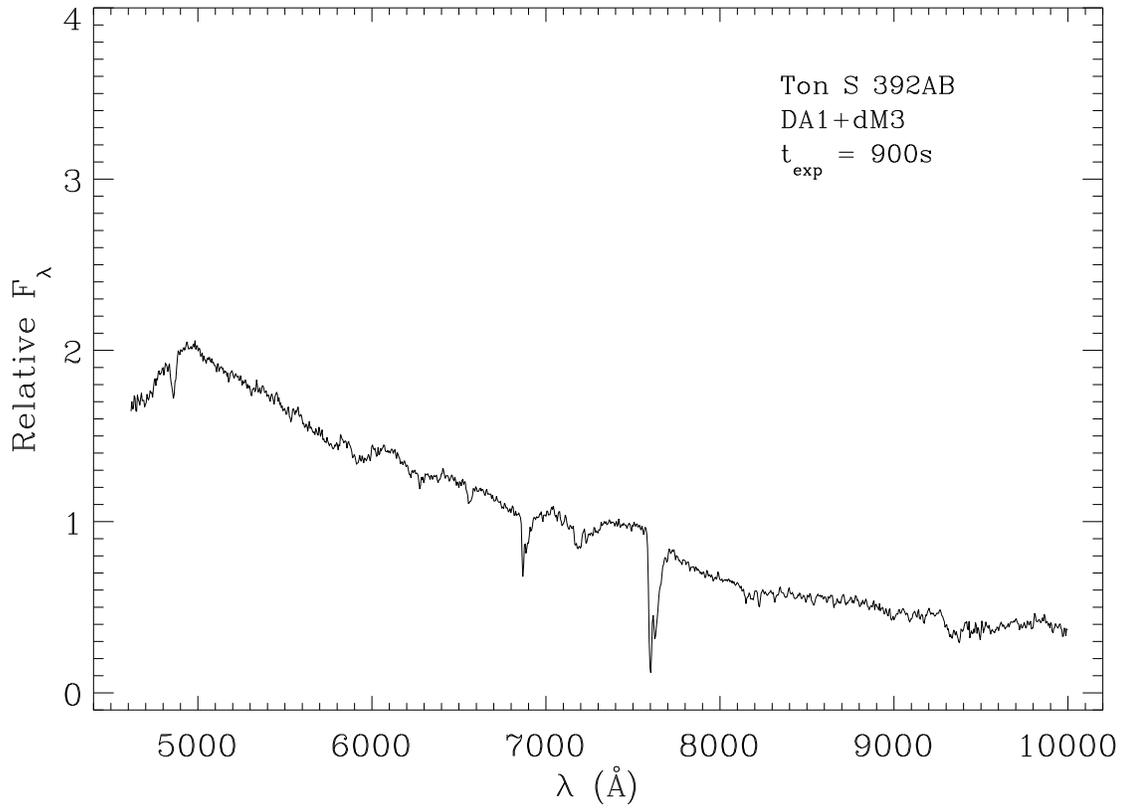}
\caption{Optical spectrum of Ton S 392AB taken with the Kast Spectrograph
on the Shane 3 meter telescope in August 2003.
\label{fig47}}
\end{figure}

\begin{figure}
\plotone{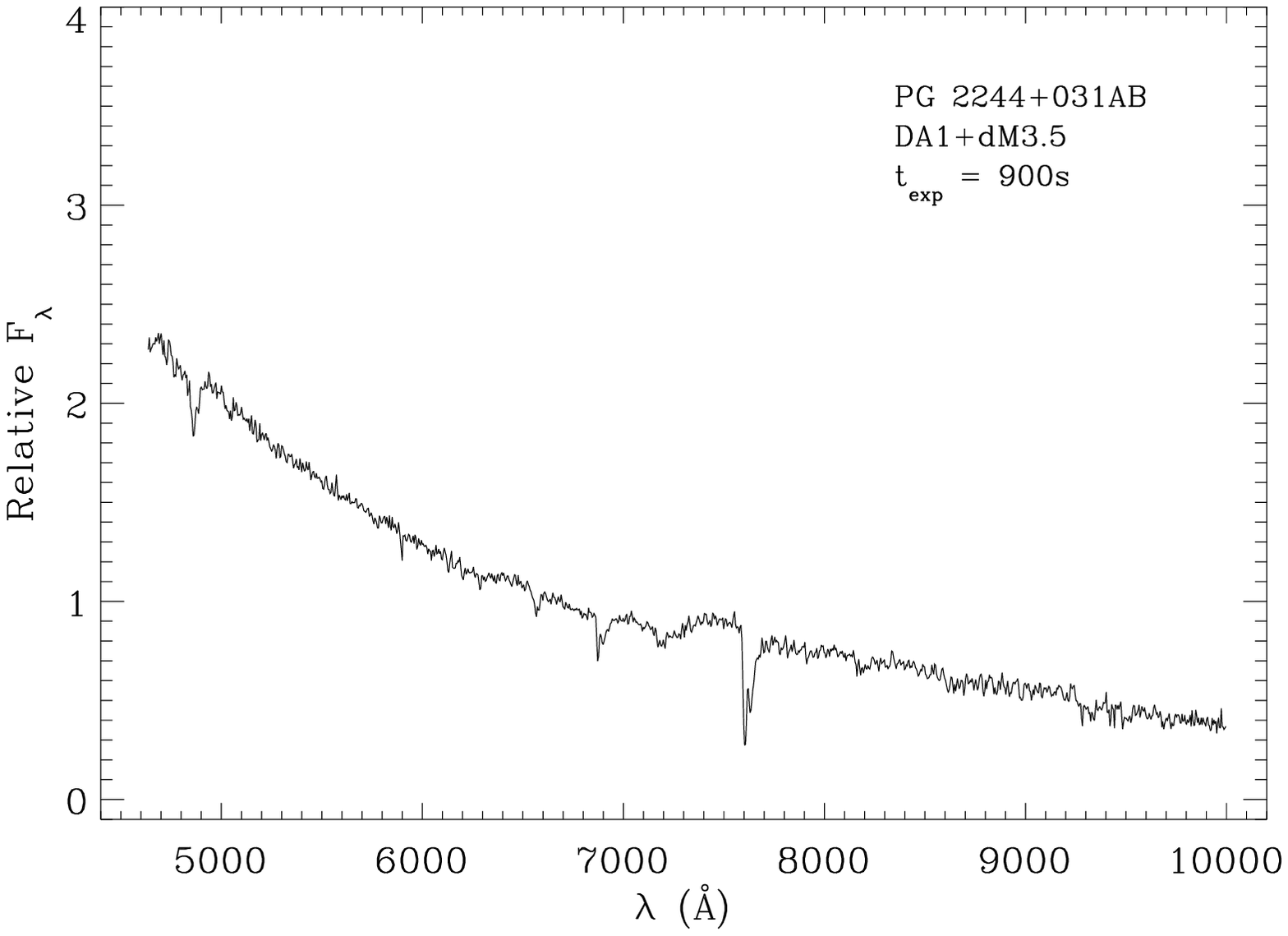}
\caption{Optical spectrum of PG 2244+031AB taken with the Kast Spectrograph
on the Shane 3 meter telescope in August 2003.
\label{fig48}}
\end{figure}

\begin{figure}
\plotone{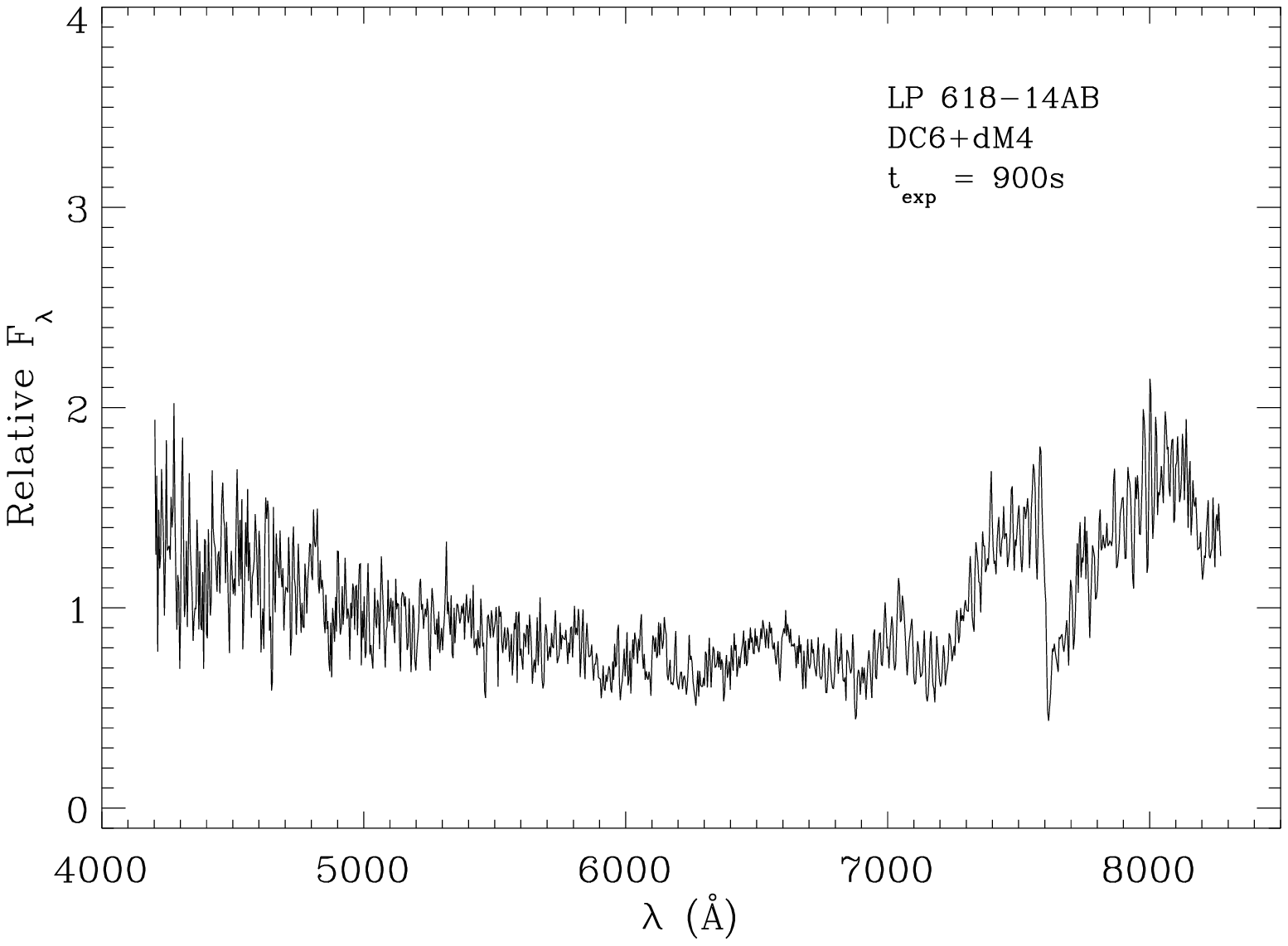}
\caption{Optical spectrum of LP 618-14AB taken with the Boller \&
Chivens Spectrograph on the Bok 2.3 meter in April 2003.
\label{fig49}}
\end{figure}

\begin{figure}
\plotone{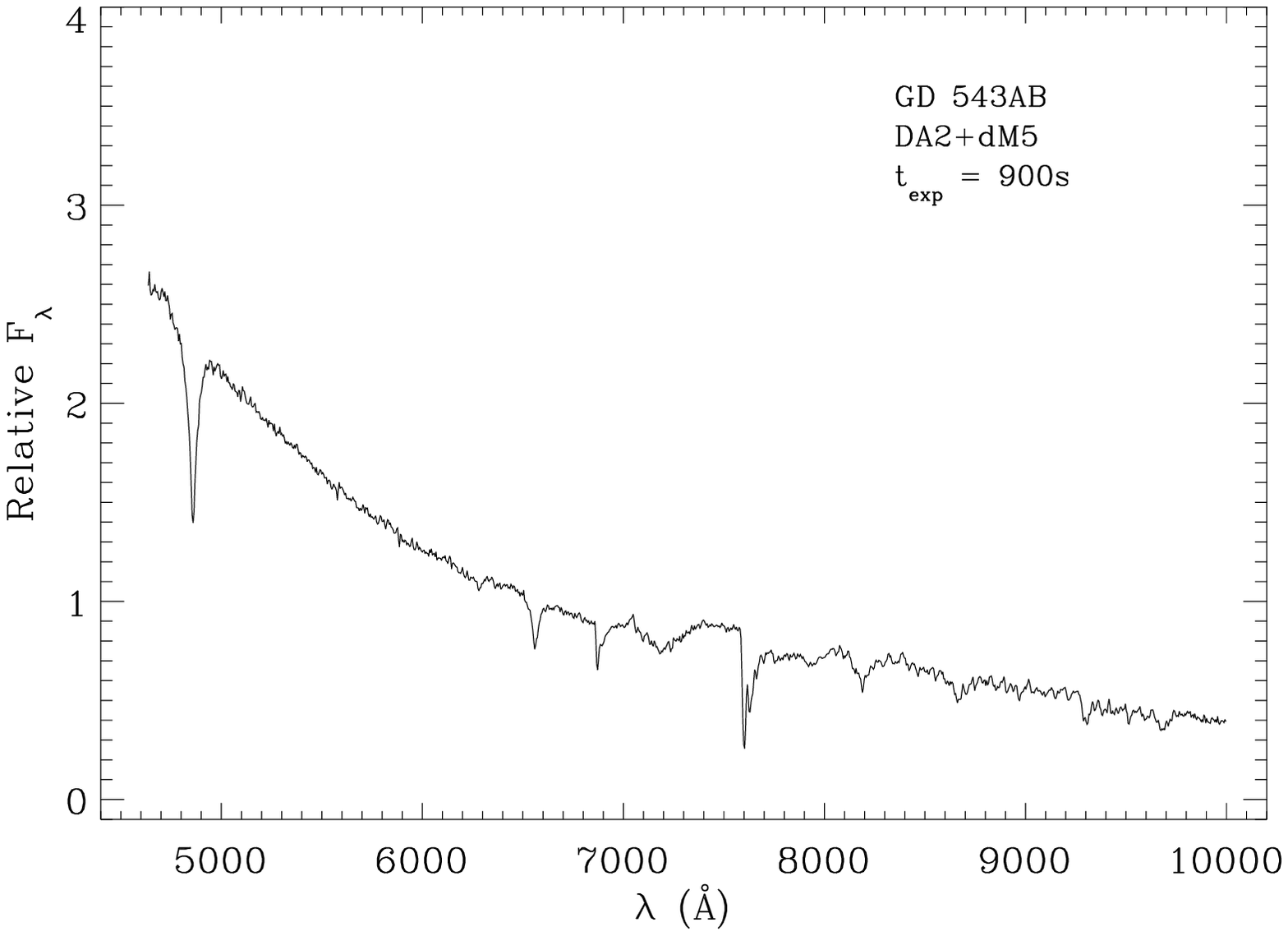}
\caption{Optical spectrum of GD 543AB taken with the Kast Spectrograph
on the Shane 3 meter telescope in August 2003.
\label{fig50}}
\end{figure}

\begin{figure}
\plotone{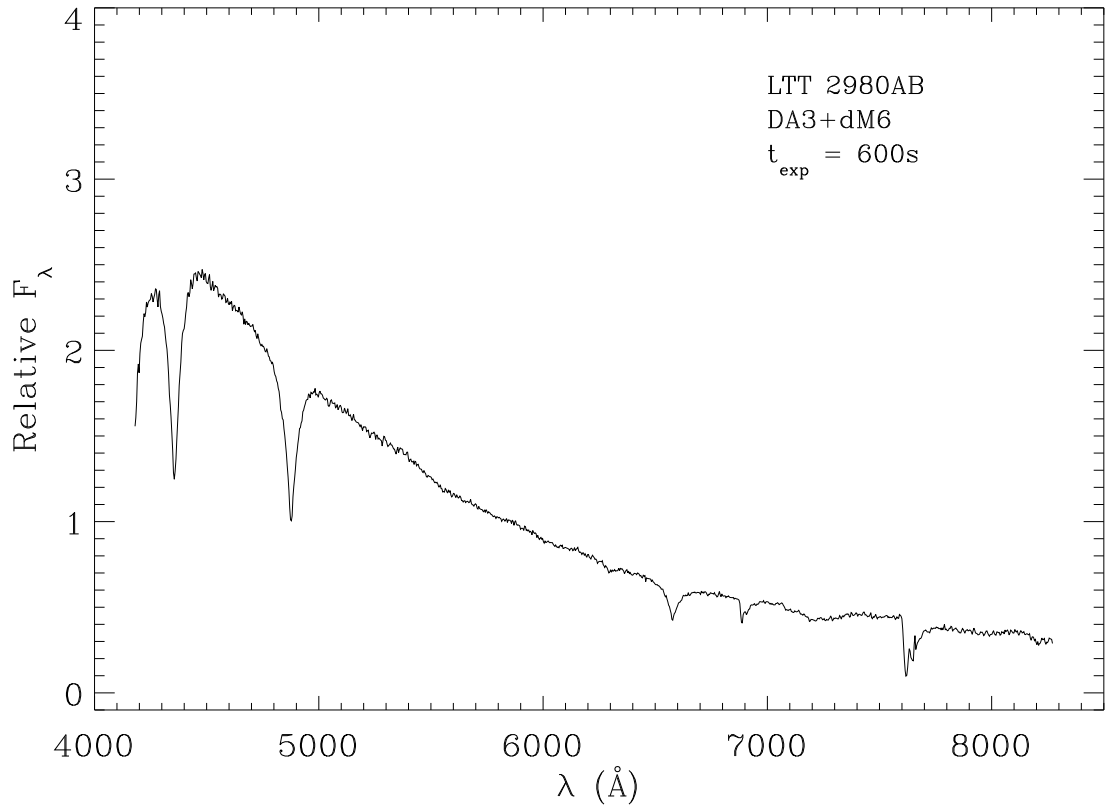}
\caption{Optical spectrum of LTT 2980AB taken with the Boller \&
Chivens Spectrograph on the Bok 2.3 meter in April 2003.
\label{fig51}}
\end{figure}

\begin{figure}
\plotone{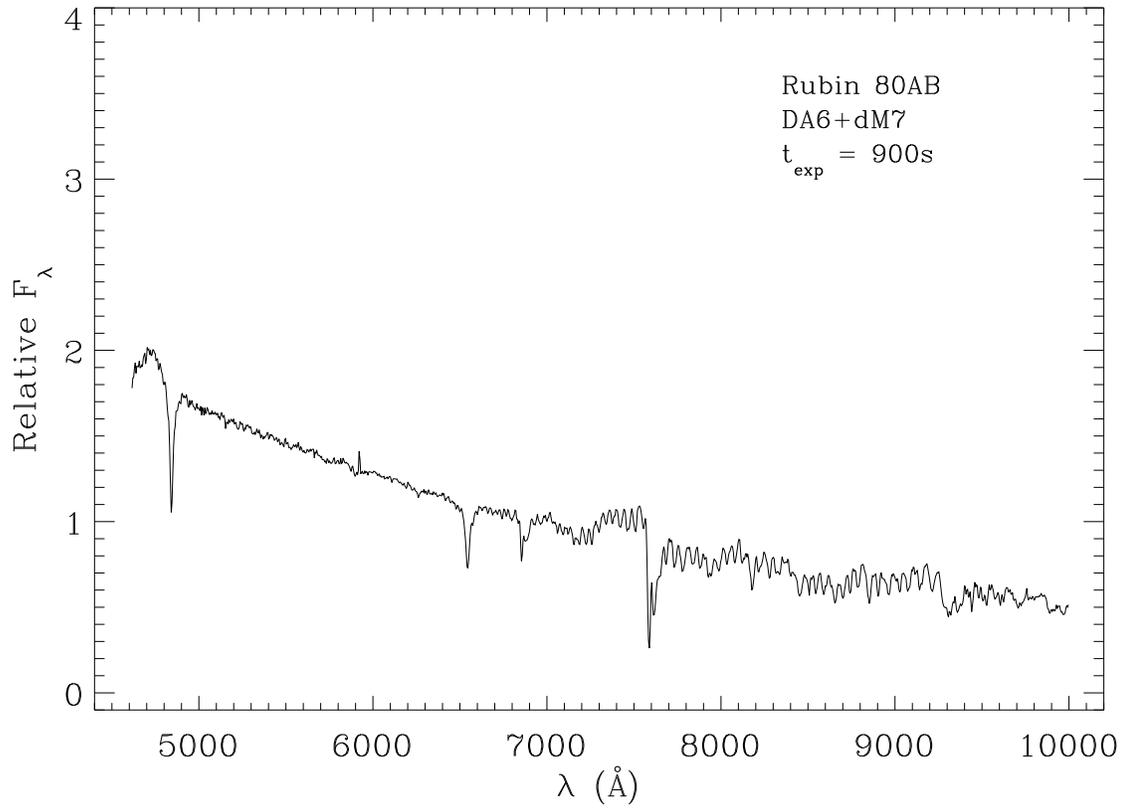}
\caption{Optical spectrum of Rubin 80AB taken with the Kast Spectrograph
on the Shane 3 meter telescope in August 2003.
\label{fig52}}
\end{figure}

\begin{figure}
\plotone{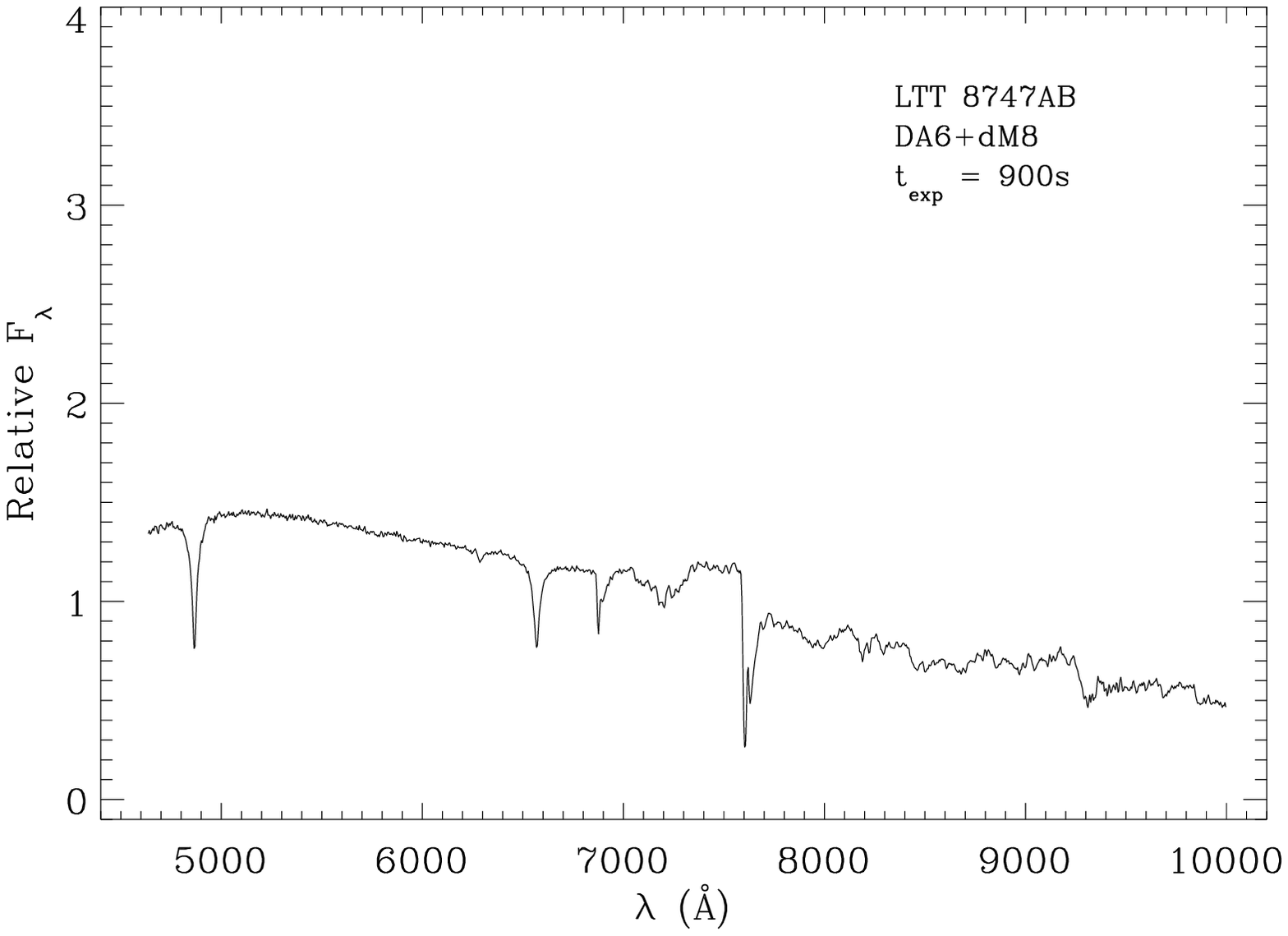}
\caption{Optical spectrum of LTT 8747AB taken with the Kast Spectrograph
on the Shane 3 meter telescope in August 2003.
\label{fig53}}
\end{figure}

\begin{figure}
\plotone{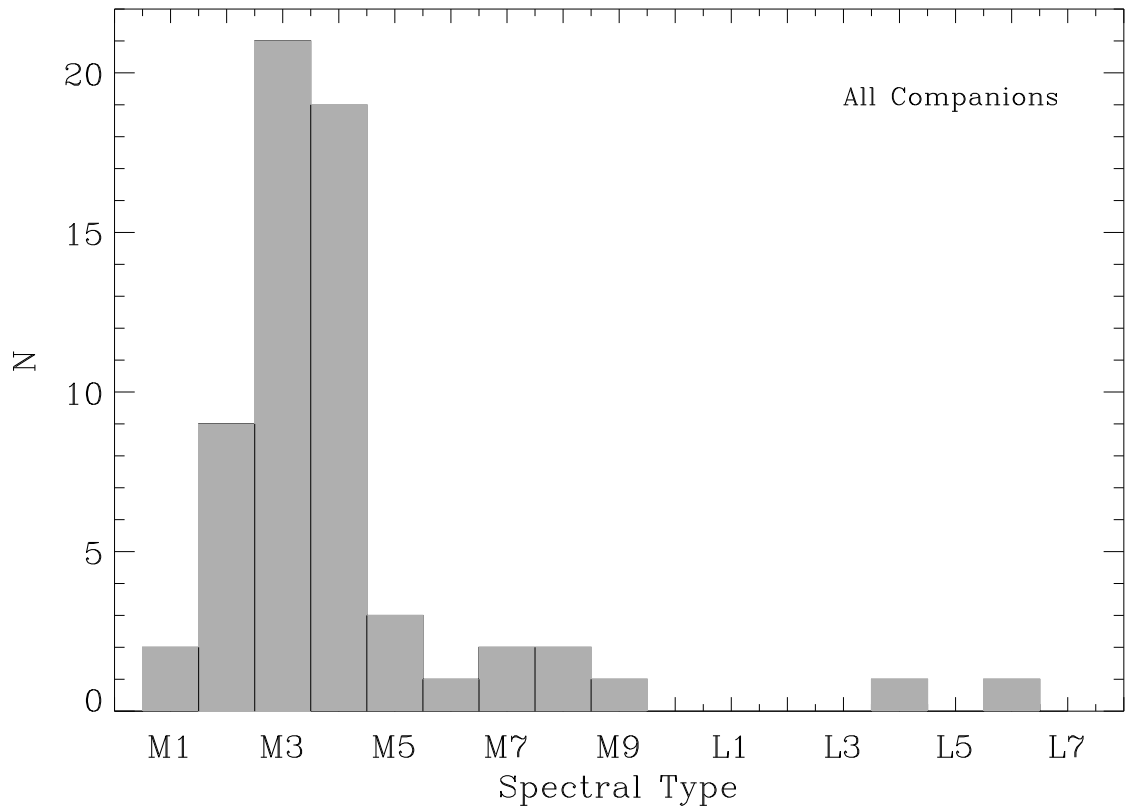}
\caption{The number of cool dwarf companions versus spectral type
for objects discovered and studied in this work.
\label{fig54}}
\end{figure}

\begin{figure}
\plotone{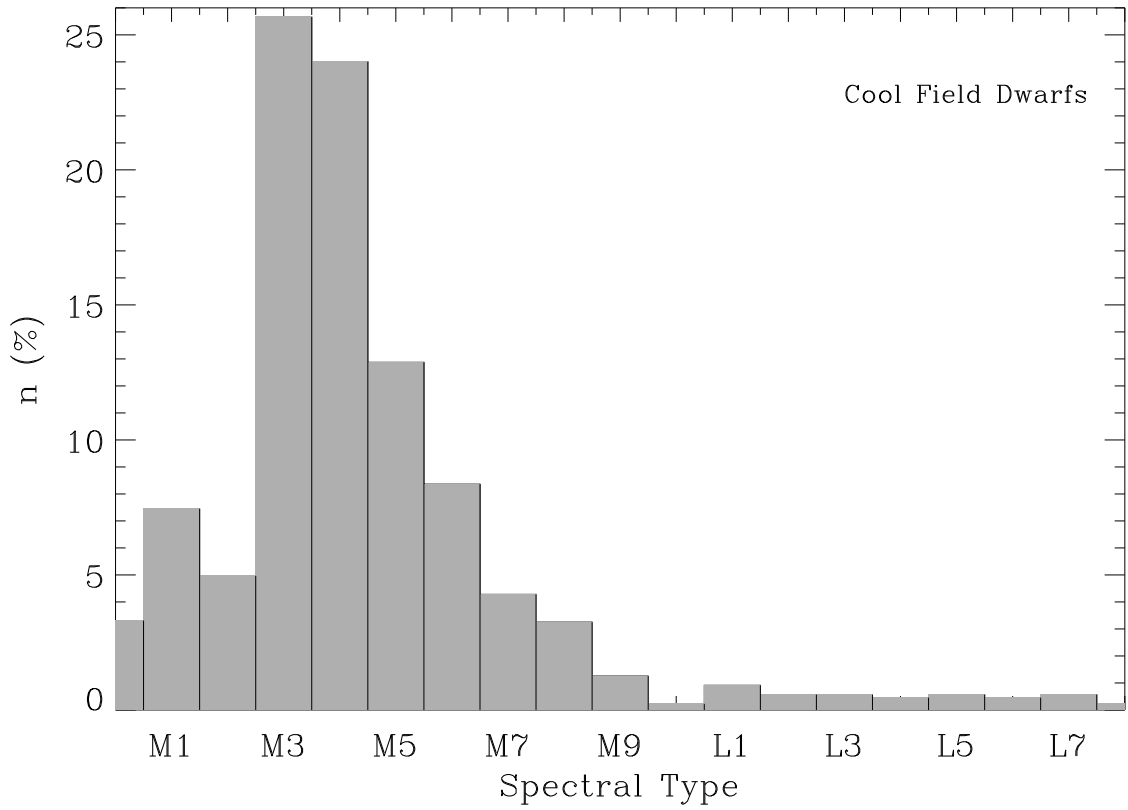}
\caption{The frequency of cool field dwarfs within $d=20$ pc versus
spectral type \citep{rei00,cru03}.  The data have been corrected
for volume, sky coverage, and estimated completeness.
\label{fig55}}
\end{figure}

\begin{figure}
\plotone{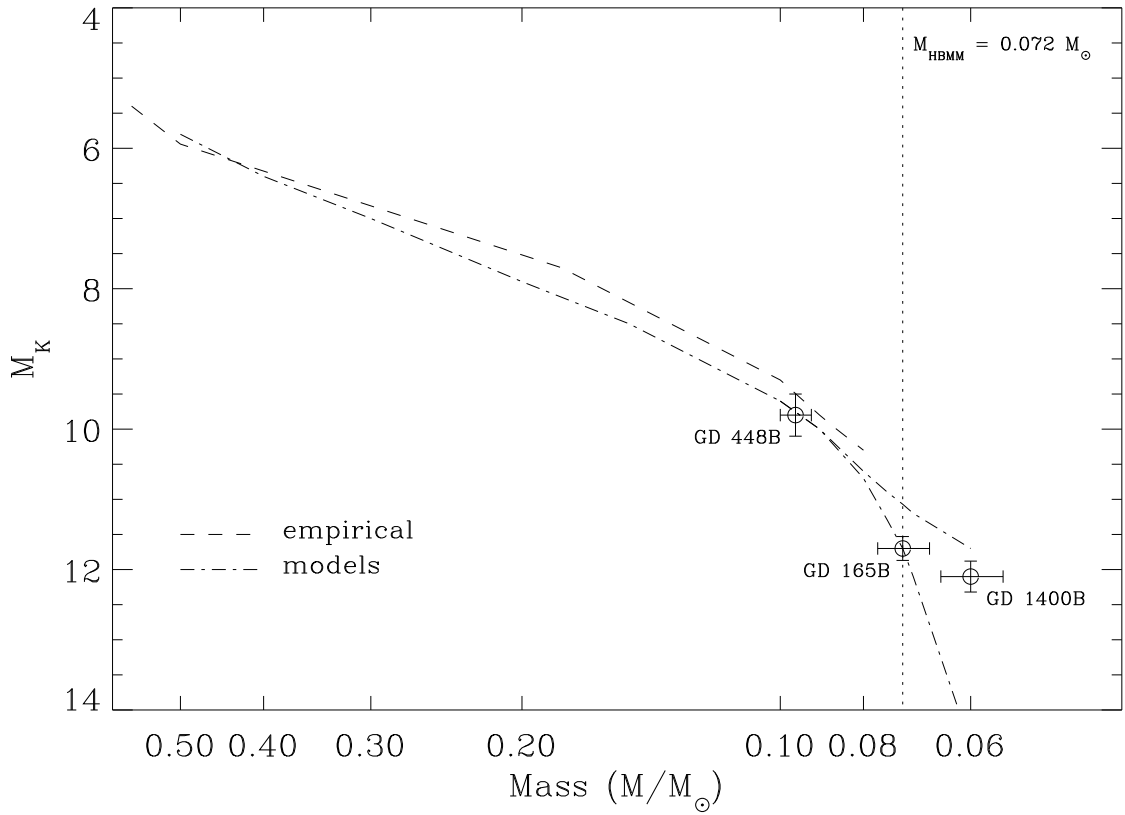}
\caption{Empirical and model relations between absolute $K$ band magnitude
and mass.  Three very cool companions to white dwarfs with mass estimates are
shown along with 1 and 5 Gyr brown dwarf model cooling tracks.
\label{fig56}}
\end{figure}

\begin{figure}
\plotone{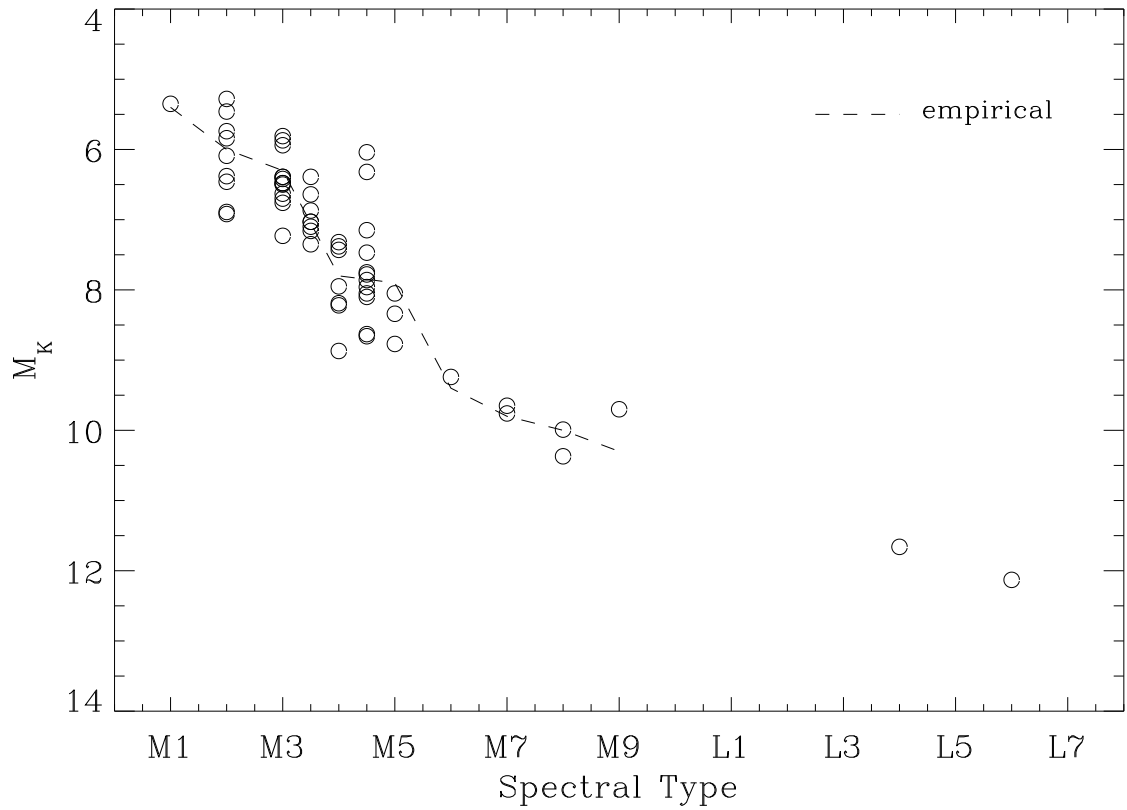}
\caption{Absolute $K$ magnitude versus spectral type for all low mass
companions based on the white dwarf distance.  The dashed line is the relation
of \citet{kir94}.
\label{fig57}}
\end{figure}

\begin{figure}
\plotone{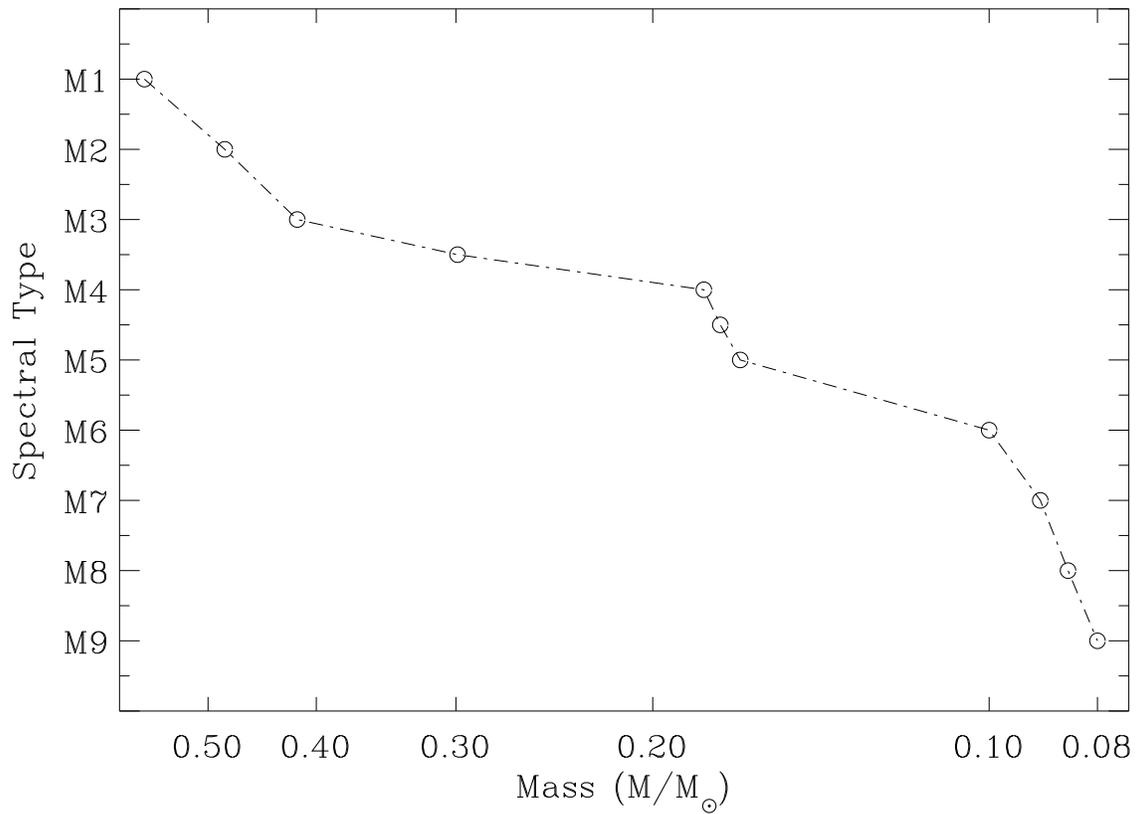}
\caption{Spectral type correlation with mass used for constructing the
companion mass function.  Data points represent the nodes in the constructed
correlation function.  These points are from the empirical and semiempirical
relations of \citet{hen93,kir94,dah02}, corrected for progress in the field
and the best available models \citep{bur97,cha00}.
\label{fig58}}
\end{figure}

\begin{figure}
\plotone{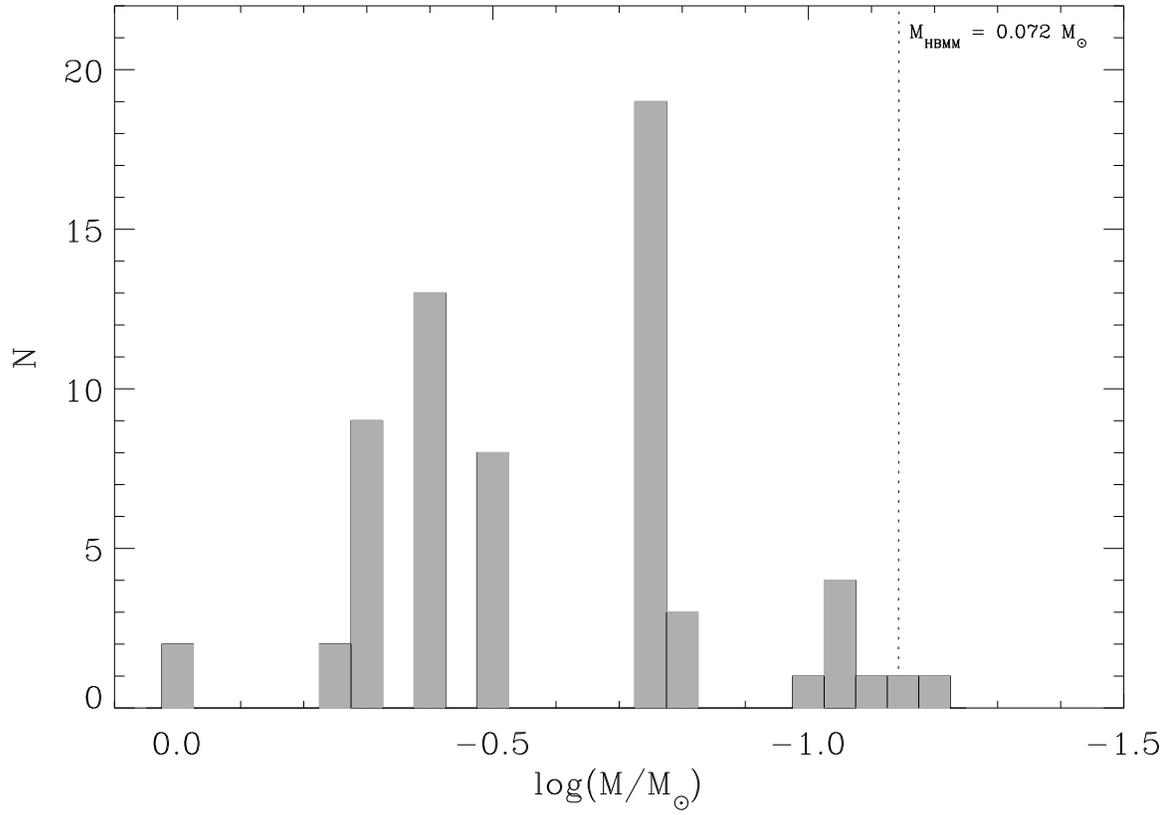}
\caption{Detected companion mass function.  The two wide G dwarf
companions are included. The empty bins are, for the most part, an
artifact of the discrete relations used between spectral type and
mass.
\label{fig59}}
\end{figure}

The spectrum in Figure \ref{fig47} has a turnover
near its blue end that is almost certainly not real.
Possible errors in flux calibration were searched
for unsuccessfully.  Attempting to correct the turnover
to the expected continuum resulted in a sensitivity
function that produced grossly incorrect shapes for
all other stars taken on the same night with the same 
instrument and setup.  In fact, all the other stars
similarly observed appear to have accurately flux
calibrated spectra.  TS 392 was observed at a high
airmass of 2.40 and this may be the source of the 
error.  In any case, the spectrum verifies a fairly
steep blue continuum from a hot DA white dwarf with
weak H$\alpha$ \& H$\beta$ absorption, plus TiO bands
from its $\sim1''$ distant M dwarf companion.

Virtually nothing was known or published about this
system until now, so distance estimates and parameter
determinations should be considered  somewhat preliminary.
Treating TS 392A as a $T_{\rm{eff}}=50,000$ K white dwarf,
the deconvolved magnitudes of TS 392B result in colors
consistent with an M3 dwarf, $\pm1$ spectral type, at a
distance of 409 pc.

\clearpage

\clearpage

\clearpage

\clearpage

\clearpage

\clearpage

\clearpage

\clearpage

\clearpage

\clearpage

\clearpage

\clearpage

\clearpage

\clearpage

\clearpage

\clearpage

\clearpage

\clearpage

\clearpage

\clearpage

\clearpage

\clearpage

\clearpage

\clearpage

\clearpage

\clearpage

\clearpage

\clearpage

\clearpage

\clearpage

\clearpage

\clearpage

\clearpage

\clearpage

\clearpage

\clearpage

\clearpage

\clearpage

\clearpage

\clearpage

\clearpage

\clearpage

\clearpage

\clearpage

\clearpage

\clearpage

\clearpage

\clearpage

\clearpage

\clearpage

\clearpage

\clearpage

\clearpage

\clearpage

\clearpage

\clearpage

\clearpage

\clearpage

\clearpage

\clearpage

\begin{deluxetable}{llcccccl}
\rotate
\tablecaption{Observed Targets\label{tbl-1}}
\tablewidth{0pt}
\tablehead{
\colhead{WD\#} 				&
\colhead{Name}				&
\colhead{SpT}				&
\colhead{$d$ (pc)} 			&
\colhead{$\mu$ ($''$ $\rm{yr}^{-1}$)}	&
\colhead{$\theta$ ($\arcdeg$)}		&
\colhead{($U,V,W$)\tablenotemark{*} (km $\rm{s}^{-1}$)} 	&
\colhead{Telescope}}

\startdata

0000$-$170      &G266-32	&DB4	&39.6	&0.246	&91.0	&($31,-10,-2$)	&K,I\\
0000$-$345      &LHS 1008	&DAH7	&13.2	&0.772	&169.2	&($-22,-35,11$)	&S\\
0002$+$729      &GD 408		&DBZ4	&34.7	&0.253	&57.4	&($19,-7,21$)	&S\\
0009$+$191\tablenotemark{\dag}  	
		&PG		&sdB	&\nodata&\nodata&\nodata&\nodata	&S\\
0009$+$501      &LHS1038	&DAP7	&11.0	&0.713	&219.8	&($-36,20,-17$)	&S\\

0011$+$000      &G31-35		&DA5	&30.4	&0.457	&114.2	&($30,-36,-14$)	&S\\
0013$-$241      &Ton 147	&DA3	&70.5	&0.149	&276.9	&($-49,41,14$)	&S\\
0016$-$220      &GD 597		&DA4	&72.8	&0.085	&230.1	&($-38,7,8$)	&S,K\\
\enddata

\tablenotetext{*}{$U,V,W$ calculated assuming $v_r=0$ with respect
		to the local standard of rest (\S3.2).}
\tablenotetext{\dag}{Not a white dwarf.}
\tablenotetext{\ddag}{Not listed in \citet{mcc99}.  The WD number is unofficial.}

\end{deluxetable}

\clearpage

\begin{deluxetable}{cc}
\tablecaption{Sample Kinematics\label{tbl-2}}
\tablewidth{0pt}
\tablehead{
\colhead{All 371 Stars}		&
\colhead{330 Stars with $\mu<0.5'' \ \rm{yr}^{-1}$}}

\startdata

$\langle U \rangle= -5$ km $\rm{s}^{-1}$	&$\langle U \rangle= -6$ km $\rm{s}^{-1}$\\
$\langle V \rangle= -6$ km $\rm{s}^{-1}$	&$\langle V \rangle= -3$ km $\rm{s}^{-1}$\\
$\langle W \rangle= \ 0$ km $\rm{s}^{-1}$	&$\langle W \rangle= \ 1$ km $\rm{s}^{-1}$\\
$\langle T \rangle= 37$ km $\rm{s}^{-1}$	&$\langle T \rangle= 33$ km $\rm{s}^{-1}$\\
$\langle\mu\rangle= 0.27'' \ \rm{yr}^{-1}$	&$\langle\mu\rangle= 0.16'' \ \rm{yr}^{-1}$\\
$\langle d \rangle= 56.6$ pc			&$\langle d \rangle= 61.7$ pc\\

$\sigma_{U}=32$ km $\rm{s}^{-1}$		&$\sigma_{U}=26$ km $\rm{s}^{-1}$\\
$\sigma_{V}=24$ km $\rm{s}^{-1}$		&$\sigma_{V}=20$ km $\rm{s}^{-1}$\\
$\sigma_{W}=20$ km $\rm{s}^{-1}$		&$\sigma_{W}=18$ km $\rm{s}^{-1}$\\
$\sigma_{T}=44$ km $\rm{s}^{-1}$		&$\sigma_{T}=38$ km $\rm{s}^{-1}$\\
$\sigma_{\mu}=0.45'' \ \rm{yr}^{-1}$	&$\sigma_{\mu}=0.11'' \ \rm{yr}^{-1}$\\
$\sigma_{d}=47.1$ pc				&$\sigma_{d}=47.6$ pc\
\enddata

\end{deluxetable}

\clearpage

\begin{deluxetable}{lccccccccl}
\rotate
\tablecaption{Summary of All Companion Systems\label{tbl-3}}
\tablewidth{0pt}
\tablehead{
\colhead{Companion} 	&
\colhead{SpT} 		&
\colhead{Primary}	&
\colhead{SpT}		&
\colhead{$a_s$\tablenotemark{1} ($''$)} 	&
\colhead{PA ($\arcdeg$)}	&
\colhead{$d$ (pc)}		&
\colhead{$a$ (AU)} 		&
\colhead{$M_{abs}$\tablenotemark{2} (mag)} 	&
\colhead{References}}

\startdata

GD 360B		&DA	&1713$+$332	&DA2	&close		&\nodata	&85.5	&$<0.1$	&\nodata&8\\
G1-45B		&DC	&0101$+$048	&DA5	&close		&\nodata	&13.5	&$<0.1$	&\nodata&5,9\\
G21-15B		&DC	&1824$+$040	&DA4	&close		&\nodata	&54.9	&$<0.1$	&\nodata&5,13\\
GD 429B		&DC	&0416$+$701	&DA4	&close		&\nodata	&44.7	&$<0.1$	&\nodata&9\\
PG 1241$-$010B	&DC	&1241$-$010	&DA2	&close		&\nodata	&88.7	&$<0.1$	&\nodata&8\\
PG 1428$+$373B	&DC	&1428$+$373	&DA5	&close		&\nodata	&96.8	&$<0.1$	&\nodata&14\\
PG 0922$+$162B	&DA2	&0922$+$162	&DA2	&4.4		&287.0 		&118.7	&522	&11.93	&1,22\\
PG 1204$+$450B	&DA3	&1204$+$450	&DA2	&close		&\nodata	&92.9	&$<0.1$	&\nodata&5,12\\
\enddata

\end{deluxetable}

\clearpage

\begin{deluxetable}{lccc}
\tablecaption{Measured Proper Motions\label{tbl-4}}
\tablewidth{0pt}
\tablehead{
\colhead{Star}						&
\colhead{$\mu_{\alpha}$ ($\rm{yr}^{-1}$)}		&
\colhead{$\mu_{\delta}$ ($\rm{yr}^{-1}$)}		&
\colhead{Reference}}

\startdata

G21-15AB	&$-0.25''$ &$-0.28''$ &1\\
G21-15C		&$-0.27''$ &$-0.28''$ &1\\
&	&		&\\

GD 60A		&$+0.08''$ &$-0.17''$ &1\\
GD 60B		&$+0.08''$ &$-0.17''$ &1\\
&	&		&\\

GD 74A		&$-0.01''$ &$-0.11''$ &1\\
GD 74B		&$-0.01''$ &$-0.08''$ &1\\
\enddata

\tablecomments{Uncertainties are generally $\sim0.01''
\ {\rm{yr}}^{-1}$ (\S4.3).}

\tablenotetext{\dag}{Candidate companion.}

\tablerefs{
(1) This work; \citealt{far04c}; (2) USNO B1.0 Catalog \citep{mon03}; 
(3) UCAC Catalogs \citep{zac00,zac04}; (4) Tycho 2
Catalog \citep{hog00}; (5) \citealt{rey02}.}

\end{deluxetable}

\clearpage

\begin{deluxetable}{lcccccccl}
\tablecaption{Optical \& Near-infrared Photometry of Resolved Pairs \label{tbl-5}}
\tablewidth{0pt}
\tablehead{
\colhead{Star}		&
\colhead{$B$}		&
\colhead{$V$}		&
\colhead{$R$}		&
\colhead{$I$}		&
\colhead{$J$}		&
\colhead{$H$}		&
\colhead{$K$}		&
\colhead{Reference}}

\startdata

G21-15AB	&14.03	&13.92	&13.94	&13.96	&14.10	&14.16	&14.16	&1,5\\
G21-15C		&20.05	&19.00	&18.44	&17.82	&17.53	&17.05	&16.88	&1\\
&	&	&	&	&	&	&	&\\

G130-5		&13.06	&12.92	&\nodata&\nodata&13.17	&13.20	&13.18	&2,4\\
G130-6		&13.25	&11.69	&10.59	&9.21	&7.86	&7.25	&7.04	&2,3\\
&	&	&	&	&	&	&	&\\

G148-7		&13.69	&13.64	&\nodata&\nodata&14.01	&13.98	&14.03	&2,4\\
G148-6		&15.72	&14.09	&\nodata&\nodata&9.96	&9.39	&9.14	&2,3\\
\enddata

\tablecomments{All entries are in magnitudes.  Uncertainties for photometry are
5\% or less with a few exceptions (\S4.5).}

\tablenotetext{\dag}{Candidate companion.}

\tablerefs{(1) This work; \citealt{far04c}; (2) 2MASS Catalog \citep{cut03};
(3) ARICNS \citep{gle91}; (4) \citealt{mcc99}; (5) \citealt{ber01}.}

\end{deluxetable}

\clearpage

\begin{deluxetable}{lcccccccl}
\tablecaption{Optical \& Near-infrared Photometry of Composite Pairs \label{tbl-6}}
\tablewidth{0pt}
\tablehead{
\colhead{Star}		&
\colhead{$B$}		&
\colhead{$V$}		&
\colhead{$R$}		&
\colhead{$I$}		&
\colhead{$J$}		&
\colhead{$H$}		&
\colhead{$K$}		&
\colhead{Reference}}

\startdata

GD 51AB		&15.27	&14.99	&14.64	&13.32	&12.05	&11.57	&11.30	&1,2\\
WD		&15.62	&15.67	&15.77	&15.90	&16.24	&16.31	&16.42	&3,9\\
RD		&\nodata&\nodata&\nodata&13.43	&12.07	&11.58	&11.31	&\\
&	&	&	&	&	&	&	&\\

GD 123AB	&14.21	&14.40	&14.42	&13.76	&12.56	&12.03	&11.75	&1,2\\
WD		&14.22	&14.41	&14.54	&14.71	&15.13	&15.23	&15.36	&3,4\\
RD		&\nodata&\nodata&\nodata&14.35	&12.67	&12.09	&11.79	&\\
&	&	&	&	&	&	&	&\\
\enddata

\tablecomments{All entires are in magnitudes.  Uncertainties for photometry are
5\% or less with a few exceptions (\S4.5).}

\tablerefs{(1) This work; \citealt{far04c}; (2) 2MASS Catalog \citep{cut03};
(3) \citealt{mcc99};
(4) \citealt{lie05}; (5) \citealt{zuc03}; (6) DENIS Catalog;
(7) \citealt{ber92}; (8) \citealt{bra95}; (9) \citealt{koe01};
(10) \citealt{sch72}; (11) \citealt{sch95b}; (12) \citealt{hin79};
(13) \citealt{fin97}; (14) \citealt{dow86}; (15) \citealt{gre74};
(16) \citealt{egg65}; (17) \citealt{kaw00}; (18) \citealt{mar97};
(19) \citealt{bon84}; (20) \citealt{kaw95}; (21) \citealt{ven99};
(22) \citealt{schw95}; (23) \citealt{nap99}; (24) \citealt{gre79}.}

\end{deluxetable}

\clearpage

\begin{deluxetable}{lcccccc}
\tablecaption{Survey Completeness for $d=57$ pc, $\tau=3$ Gyr \label{tbl-7}}
\tablewidth{0pt}
\tablehead{
\colhead{Survey}		&
\colhead{$a_{in}$ (AU)}		&
\colhead{$a_{out}$ (AU)}	&
\colhead{$m_{abs}$ (mag)}	&
\colhead{SpT}			&
\colhead{$M$ $(M_{\odot})$}	&
\colhead{N}}

\startdata

IRTF		&0	&700	&$M_K=12.2$	&L6	&0.065	&84\\

Steward		&110	&4700	&$M_J=14.2$	&L7	&0.060	&261\\

Keck		&55	&1100	&$M_J=17.2$	&T9\tablenotemark{\dag}
							&0.030	&86\\

All		&0	&110	&$M_H=13.5$	&L8	&0.058	&371\\

\enddata

\tablecomments{This table presents only average separations
and sensitivities.  The actual values depend on each individual
white dwarf distance and age.  The ``All'' entry refers to
detection in 2MASS of an $H$ band excess above a white dwarf
photosphere (\S5.3).}

\tablenotetext{\dag}{No objects are known with spectral type later
than T8.  However, the average limiting magnitude of the Keck survey probed
$\sim1.5$ magnitudes deeper than that of any known brown dwarf
\citep{vrb04,leg02}.}

\end{deluxetable}

\clearpage

\begin{deluxetable}{cccc}
\tablecaption{Median Spectral Type for Low Mass Companions\label{tbl-8}}
\tablewidth{0pt}
\tablehead{
\colhead{Subgroup}		&
\colhead{Description}		&
\colhead{Median SpT}		&
\colhead{N}}

\startdata

0		&All			&M3.5		&62\\
1		&Resolved		&M3.5		&38\\
2		&Close			&M4.5		&9\\
3		&Unresolved / not close	&M3.5		&15\\
4		&Groups $1+3$		&M3.5		&47\\
5		&Groups $2+3$		&M4.5		&24\\

\enddata

\end{deluxetable}

\clearpage

\begin{deluxetable}{lccccccc}
\rotate
\tablecaption{High Velocity Background Star Data \label{tbl-9}}
\tablewidth{0pt}
\tablehead{
\colhead{Object}					&
\colhead{$a$}						&
\colhead{PA}						&
\colhead{$V$ (mag)}					&
\colhead{$V-K$ (mag)}						&
\colhead{$\mu_{\alpha}$ ($\rm{yr}^{-1}$)}		&
\colhead{$\mu_{\delta}$ ($\rm{yr}^{-1}$)}		&
\colhead{SpT}}

\startdata

GD 248		&	&		&15.1	&$+0.0$	&$-0.05''$ &$-0.11''$	&DC5\\
background star	&$93''$	&$9\arcdeg$	&19.6	&$+3.4$	&$-0.06''$ &$-0.13''$	&sdK/M\\
&	&	&	&		& 	&	&\\

GD 304		&	&		&15.3	&$-0.8$	&$+0.06''$ &$-0.01''$	&DA2\\
background star	&$104''$&$33\arcdeg$	&20.5	&$+4.7$	&$+0.05''$ &$-0.01''$	&dM\\
&	&	&	&		& 	&	&\\
	
PG 1038$+$633	&	&		&14.8	&$-1.0$	&$-0.08''$ &$-0.04''$	&DA2\\
background star	&$37''$	&$37\arcdeg$	&18.9	&$+3.2$	&$-0.08''$ &$-0.04''$	&dK/M\\
&	&	&	&		& 	&	&\\

PG 1026$+$002	&	&		&13.8	&$+2.9$	&$+0.06''$ &$-0.08''$	&DA3+dM\\
background star	&$114''$&$21\arcdeg$	&19.7	&$+3.5$	&$+0.05''$ &$-0.07''$ 	&sdK/M\\

\enddata

\end{deluxetable}

\clearpage

\begin{deluxetable}{lcc}
\tablecaption{Candidate Binary Proper Motions \label{tbl-10}}
\tablewidth{0pt}
\tablehead{
\colhead{Object}				&
\colhead{$\mu_{\alpha}$ ($\rm{yr}^{-1}$)}	&
\colhead{$\mu_{\delta}$ ($\rm{yr}^{-1}$)}}

\startdata

GD 84A		&$-0.11\pm0.02''$ &$-0.16\pm0.02''$\\
GD 84B 		&$-0.06\pm0.02''$ &$-0.21\pm0.02''$\\
&	&\\

GD 683A		&$-0.04\pm0.01''$ &$-0.06\pm0.01''$\\
GD 683B		&$-0.03\pm0.01''$ &$-0.06\pm0.01''$\\
&	&\\

PG 0933$+$729A	&$-0.06\pm0.01''$ &$-0.04\pm0.01''$\\
PG 0933$+$729B	&$-0.06\pm0.01''$ &$-0.05\pm0.01''$\\

\enddata

\end{deluxetable}

\clearpage

\begin{deluxetable}{cccccccc}
\tablecaption{Possible Parameters for Rubin 80AB \label{tbl-11}}
\tablewidth{0pt}
\tablehead{
\colhead{Object}		&
\colhead{$T_{\rm eff}$ (K)}	&
\colhead{log $g$}		&
\colhead{$V$}			&
\colhead{$I$}			&
\colhead{$J$}		 	&
\colhead{$K$}			&
\colhead{$d$ (pc)}}

\startdata

WD	&9000		&8.0	&15.62	&15.38	&15.36	&15.33	&41.1\\
RD	&\nodata&\nodata&\nodata	&15.96	&13.83	&12.83	&40.7\\
	&		&	&	&	&	&	&\\

WD	&8000		&8.0	&15.58	&15.21	&15.08	&14.97	&32.6\\
RD	&\nodata&\nodata&\nodata	&16.33	&13.91	&12.88	&41.7\\
	&		&	&	&	&	&	&\\

WD	&8000		&7.6	&15.58	&15.21	&15.08	&14.97	&41.7\\
RD	&\nodata&\nodata&\nodata	&16.33	&13.91	&12.88	&41.7\\

\tablecomments{$VIJK$ in magnitudes.}

\enddata

\end{deluxetable}

\end{document}